\documentclass{emulateapj}
\usepackage{graphicx}
\usepackage{apjfonts}
\usepackage{amsfonts}
\usepackage{amsmath}
\usepackage[usenames]{color}
\usepackage{natbib}
\def\rund#1{\left( #1 \right)}
\def\ave#1{\langle #1 \rangle}
\def\eck#1{\left\lbrack #1 \right\rbrack}

\def\kms{\,km\,s$^{-1}$}
\def\msun{M$_\odot$\,}
\def\mo{M$_\odot$\,}

\def\Dwa{$\,$\uppercase\expandafter{\romannumeral5}$\,$}

\def\sles{\lower2pt\hbox{$\buildrel {\scriptstyle <}
   \over {\scriptstyle\sim}$}}

\def\sgreat{\lower2pt\hbox{$\buildrel {\scriptstyle >}
   \over {\scriptstyle\sim}$}}
\def\sharpnull#1{}

\begin{document}

\title{Neutrino signatures and the neutrino-driven wind in Binary Neutron Star Mergers}

\author{L. Dessart\altaffilmark{1,2},
C.D. Ott\altaffilmark{1}
A. Burrows\altaffilmark{2},
S. Rosswog\altaffilmark{3},
and E. Livne\altaffilmark{4},
}
\altaffiltext{1}{Department of Astronomy and Steward Observatory,
                 The University of Arizona, Tucson, AZ \ 85721;
                 luc@as.arizona.edu,cott@as.arizona.edu}
\altaffiltext{2}{Department of Astrophysical Sciences, Princeton University, Princeton, NJ 08544;
	        burrows@astro.princeton.edu}
\altaffiltext{3}{Jacobs University Bremen, Campus Ring 1, D-28759 Bremen, Germany;
s.rosswog@jacobs-university.de}
\altaffiltext{4}{Racah Institute of Physics, The Hebrew University,
Jerusalem, Israel; eli@phys.huji.ac.il}

\begin{abstract}

   We present VULCAN/2D multi-group flux-limited-diffusion radiation hydrodynamics
simulations of binary neutron star (BNS) mergers, using the Shen equation of state, 
covering $\sgreat$100\,ms, and starting from azimuthal-averaged 2D slices obtained from 3D SPH 
simulations of Rosswog \& Price for 1.4\,\mo (baryonic) neutron stars with no initial spins, 
co-rotating spins, and counter-rotating spins. Snapshots are post-processed at 10\,ms intervals 
with a multi-angle neutrino-transport solver. We find polar-enhanced neutrino luminosities, dominated 
by $\bar{\nu}_e$ and ``$\nu_\mu$'' neutrinos at peak, although $\nu_e$ emission may be 
stronger at late times. We obtain typical peak neutrino energies for $\nu_e$, $\bar{\nu}_e$, 
and ``$\nu_\mu$'' of $\sim$12, $\sim$16, and $\sim$22\,MeV.
The super-massive neutron star (SMNS) formed from the merger has a cooling timescale of $\sles$1\,s.
Charge-current neutrino reactions lead to the formation of a thermally-driven bipolar 
wind with <$\dot{M}$>$\sim$10$^{-3}$\,\mo s$^{-1}$, baryon-loading the polar regions,
and preventing any production of a $\gamma$-ray burst prior to black-hole formation.
The large budget of rotational free energy suggests magneto-rotational
effects could produce a much greater polar mass loss.
We estimate that $\sles$10$^{-4}$\,\mo of material with electron fraction 
in the range 0.1-0.2 become unbound during this SMNS phase as a result of neutrino heating.
We present a new formalism to compute the $\nu_i\bar{\nu}_i$ annihilation rate based on moments 
of the neutrino specific intensity computed with our multi-angle solver.
Cumulative annihilation rates, which decay as $t^{-1.8}$, decrease over our 100\,ms window
from a few $\times$10$^{50}$ to $\sim$10$^{49}$\,erg\,s$^{-1}$, equivalent 
to a few $\times$10$^{54}$ to $\sim$10$^{53}$ $e^-e^+$ pairs per second.

\end{abstract}


\section{Introduction}

   Coalescing neutron stars are one of the primary progenitor
candidates for short-duration (i.e., $\sles$\,2\,s) $\gamma$-ray bursts (GRBs; 
\citealt{paczynski:86,goodman:86,eichler:89,narayan:92,mochkovitch:93}).
In-spiral of the two neutron star components occurs due to energy and angular momentum
loss through gravitational radiation \citep{taylor:82,weisberg:05}, which is emitted
as the system evolves towards coalescence. 
Modern $\gamma$-ray and X-ray satellites have considerably improved our understanding
of short-hard GRBs \citep{nakar:07}, and have more strongly associated them 
with binary neutron star (BNS) merger events \citep{lee:07}.
Gravitational wave detectors such as LIGO \citep{abramovici:92} will most likely improve our
understanding of these events in the coming decade.

  The engine powering such short-hard GRBs and their associated high-Lorentz-factor
ejecta is thought to be linked to mass accretion from a torus onto a central black hole, 
formed when the super-massive neutron star (SMNS) born from the coalescence eventually collapses.
In recent years, models of GRB production from a high-mass, magnetar-like, neutron star have also received
some attention \citep{usov:92,usov:94,kluzniak:98,rosswog:03c,dai:06,metzger:07,dessart:07}. We will address the
likelyhood that the SMNS produces a GRB prior to black-hole formation in \S~\ref{sect:dynamics}.
A fraction of the gravitational energy in the disk around this black hole 
is radiated as neutrinos which power a disk wind. A fraction of these neutrinos and antineutrinos
annihilate into $e^-e^+$. Magnetic processes may, through their role in angular momentum transport,
be decisive in setting the time scale between merger and collapse and, in addition, 
they may extract rotational energy from the central black hole \citep{blandford:77}. 
The ejecta may be confined by the magnetic field morphology or by the intense 
neutrino-driven baryon-loaded wind thought to accompany coalescence and black hole mass accretion
\citep{levinson:00,rosswog:03b,aloy:05}.

Numerical simulations of BNS mergers have a rich history, and have been considerably
refined over the years, moving from 2D to 3D; from Newtonian to post-Newtonian,  
conformally flat, and finally to full General Relativity (GR); from simple
polytropic to more sophisticated equations of state (EOS); from the neglect of neutrinos 
to approximate neutrino-trapping schemes. BNS merger studies
yield predictions for 1) their associated gravitational wave signals
\citep{ruffert:96,calder:02,faber:02,shibata:02,shibata:03,shibata:05,oechslin:07a,anderson:08a}, 2)
the production of short-hard GRBs \citep{ruffert:97,ruffert:99,rosswog:03b,rosswog:03c,rosswog:05,janka:06,birkl:07},
and 3) the pollution of the environment by r-process nuclei 
\citep{lattimer:74,lattimer:76,symbalisty:82,eichler:89,davies:94,freiburghaus:99,rosswog:99a,surman:08}.

Different modeling ingredients and approaches have been employed.
3D Newtonian simulations without neutrino transport were performed by 
\cite{oohara:89,oohara:90}, \cite{nakamura:89,nakamura:91}, \cite{zhuge:94,zhuge:96}, 
and \cite{rosswog:00}. \cite{ruffert:96,ruffert:97} included neutrinos using a grid-based code, 
and this was followed by \cite{rosswog:02a} and \cite{rosswog:03a} using a 3D SPH code.
Recently, \cite{setiawan:04,setiawan:06} applied such a neutrino-leakage scheme 
in their study of torus-disk mass accretion around a black hole formed in a BNS merger, and addressed 
neutrino emission and annihilation. Magnetic fields in BNS merger evolution were introduced 
in 3D Newtonian simulations by \cite{price:06} and \cite{rosswog:07a}.

In parallel, there have been improvements to the Newtonian approach to incorporate the 
effects of GR (although largely neglecting microphysics), 
which become important as the two neutron stars come closer
and eventually merge. Post-Newtonian simulations were performed 
by \cite{oohara:92}, \cite{faber:00,faber:02}, \cite{faber:01}, \cite{ayal:01}, and \cite{calder:02}. 
The conformally-flat approximation was introduced in 
\cite{wilson:96} and followed by \cite{oechslin:02}, \cite{faber:04}, \cite{oechslin:06,oechslin:07a},
and \cite{oechslin:07b}. Full 2D GR simulations, sometimes including magnetic fields,  have been performed 
for super-massive neutron stars (SMNS), possibly resulting from BNS 
mergers, by \cite{baumgarte:00}, \cite{morrison:04}, \cite{duez:04,duez:06}, and \cite{shibata:06},
while full 3D GR simulations of the merger were carried out by 
\cite{shibata:02,shibata:03,shibata:05,shibata:07}, \cite{anderson:08b}, and \cite{baiotti:08}.

\begin{figure}
\plotone{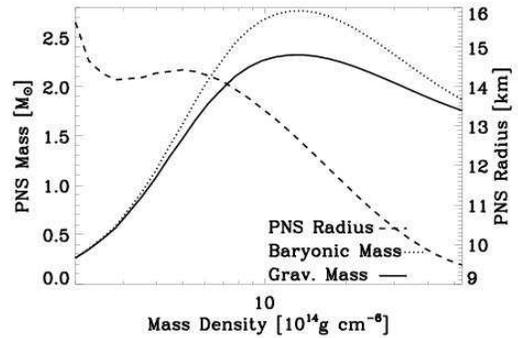}
\caption{Gravitational (solid line) and baryonic (dotted line) masses, as well
as the corresponding neutron star radius (dashed line), versus central density
derived for our Shen EOS and the Oppenheimer-Volkov equation 
(one form of the general relativistic equation of hydrostatic equilibrium), 
and assuming a non-rotating neutron star characterized by
a uniform electron fraction of 0.1 and a uniform temperature of 1\,MeV.
Here, the transition to a black hole will occur when the central density in the neutron star
reaches $\sim$1.3$\times$10$^{15}$\,g\,cm$^{-3}$, corresponding to a gravitational (baryonic)
mass of 2.32\,\msun\, (2.77\,\msun) and a neutron star radius of $\sim$13\,km.
\label{fig_BH}}
\end{figure}

All these simulations have been conducted at various levels of sophistication
for the thermodynamic properties of matter, ranging from simplistic and not so 
physically-consistent polytropic EOSs, to those employing
a detailed microphysical representation of nuclear matter at an arbitrary temperature
\citep{lseos:91,shen:98a,shen:98b}. Investigations performed with such detailed EOSs 
have been conducted by, e.g., \cite{ruffert:96,ruffert:97}, \cite{rosswog:99a}, and \cite{rosswog:02a},
and the dependency of BNS merger properties on the adopted EOS has been discussed by 
\cite{oechslin:07a} and \cite{oechslin:07b}.
Variations of a few times 10\% in the maximum allowed mass are seen, depending on the compressiblity 
of nuclear matter, but recent observations may 
suggest that neutron stars with a gravitational mass in excess of 2\,\mo do exist \citep{nice:05,freire:07},
supporting a stiff EOS for nuclear matter, such as the Shen EOS we employ here. 
In Fig.~\ref{fig_BH}, we show, as a function of the central density, the baryonic and gravitational 
neutron star masses that obtain for our implementation of the Shen EOS. 
The maximum occurs at $\sim$1.3$\times$10$^{15}$\,g\,cm$^{-3}$, 
corresponding to a gravitational (baryonic) mass of 2.32\,\msun\, 
(2.77\,\msun).\footnote{Note that the values given in \citealt{dessart:08} are not exact.}
Moreover, in the context of differentially rotating (and possibly magnetized) SMNSs,
the maximum mass that can be supported by a given EOS may be increased by up to $\sim$50\% compared to
the equivalent non-rotating object 
\citep{baumgarte:00,morrison:04,duez:04,duez:06,shibata:06,sumiyoshi:07,kiuchi:08}.
Ultimately, understanding the mechanisms and timescales for angular momentum to be redistributed 
to lead to solid-body rotation is, therefore, important. One possible agency of redistribution is 
the magneto-rotational instability \citep{bh:91,akiyama:03,pessah:06}.
The implications are non-trivial because the efficiency of angular-momentum transport determines 
in part whether a black hole forms promptly, or after a short or a long delay.
It also determines how much mass can be accreted, i.e., whether there is  
$\sim$0.01 or $\sim$0.1\,\mo available in the torus surrounding the black hole after it has formed, 
and what the timescale is over which such accretion can take place to power relativistic ejecta.
Although indirect, the relevance to short-hard GRBs and their properties is obvious, and perhaps central.
In fact, this issue is also germane to the production of collapsars and long-duration GRBs 
\citep{dessart:08}.

In this work, we perform 2D multi-group, flux-limited-diffusion (MGFLD), radiation hydrodynamics  
of merged BNSs using the Shen EOS. The main goal of this work is to document in detail
the neutrino signatures from such merger events. Our approach is to solve the neutrino transport problem
self-consistently (although with a flux-limiter and assuming diffusion), in combination with the 
dynamics of the system (but assuming axisymmetry), over $\sgreat$100\,ms. 
At selected times, we post-process such models with the multi-angle, $S_n$, neutrino 
transport algorithm of \cite{livne:04} and described recently in \cite{ott:08}.
In particular, our investigation allows for the first time the computation of the neutrino-antineutrino
annihilation rate from a solution of the transport equation, rather than using an approximate
leakage scheme. Our investigation applies strictly to the neutron-star phase of such BNS mergers.

This paper is organized as follows. In \S~\ref{sect:model}, we present the initial models
we employ in our work, based on 3D SPH simulations of BNS mergers with different spin configurations 
(\S~\ref{sect:prog}). In \S~\ref{sect:vulcan}, we present the characteristics and procedures
we employ to evolve such initial conditions with the MGFLD radiation hydrodynamics code VULCAN/2D.
In \S~\ref{sect:results}, we present our results for the neutrino signatures (\S~\ref{sect:neutrinos}) 
and the dynamics of BNS mergers (\S~\ref{sect:dynamics}).
In \S~\ref{sect:annihilate}, we address the neutrino-antineutrino annihilation rate 
in our three BNS merger models. We first present results employing the approach used so far
\citep{ruffert:96,ruffert:97,rosswog:03a,setiawan:04,setiawan:06,birkl:07}, based on a leakage scheme
for the neutrino-flavor dependent opacities and emissivities and a summation of all paired grid 
cells contributing at any given location. In \S~\ref{annihil:sn}, we then present a new formalism
which uses moments of the neutrino specific intensity computed with a multi-angle, $S_n$, scheme. 
Finally, we present our conclusions in \S~\ref{sect:conclusion} and discuss 
the most striking implications for the powering of short-duration GRBs.  
For completeness, in the appendix, we apply our formalism 
to the computation of the annihilation rate in the context of slow- and fast-rotating single protoneutron
stars (PNSs).

\begin{figure*}
\epsscale{0.35}
\plotone{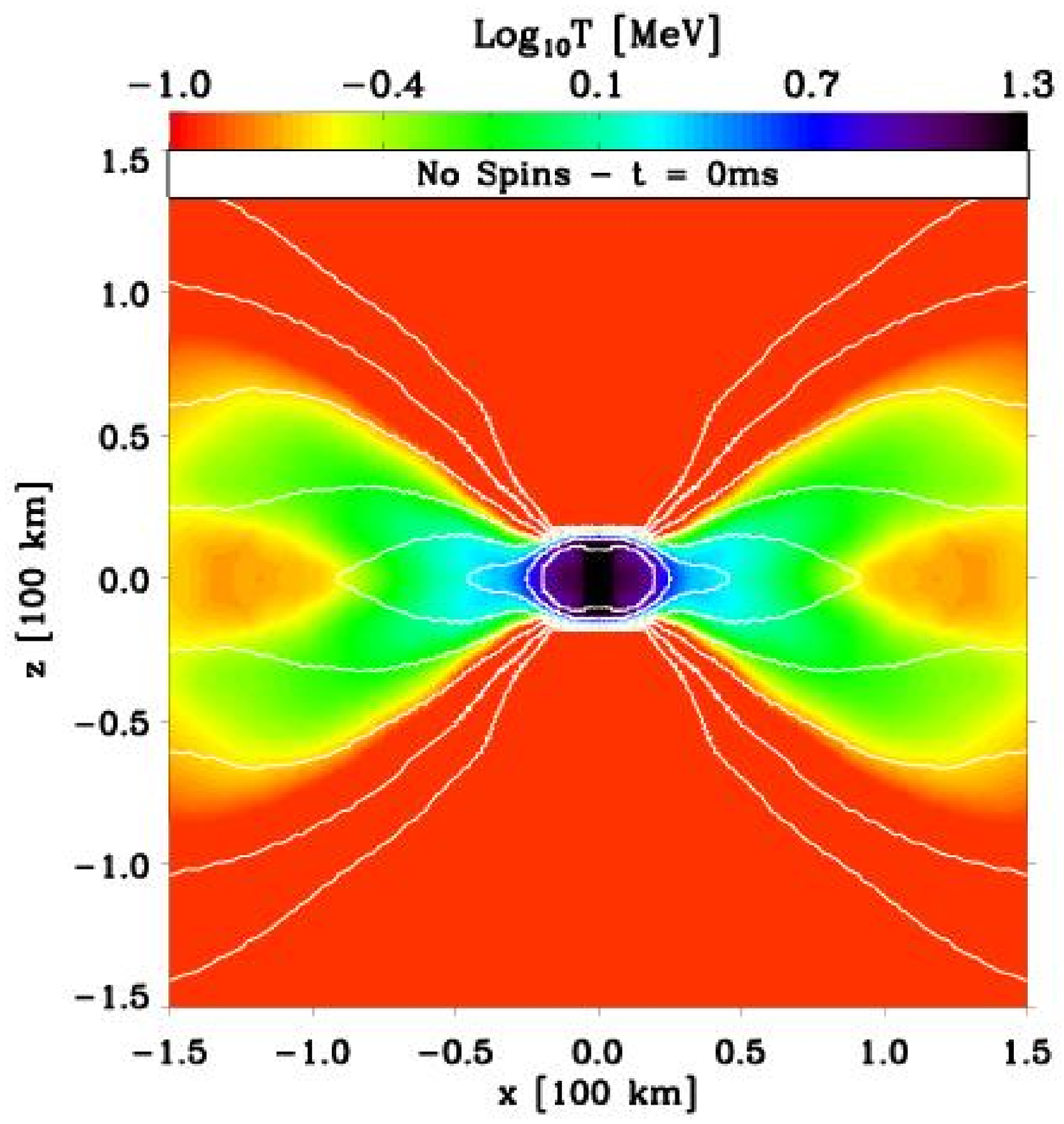}
\plotone{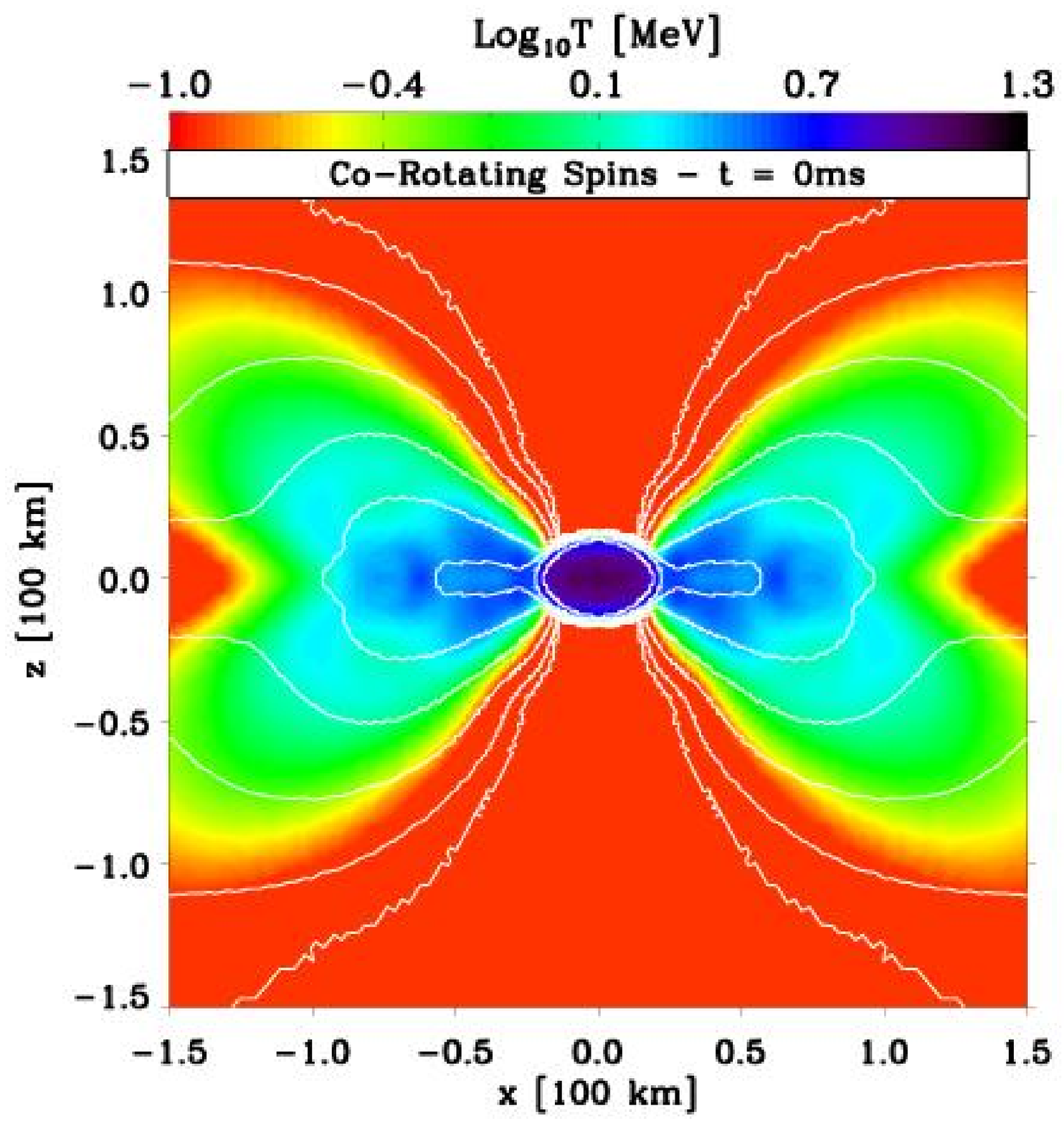}
\plotone{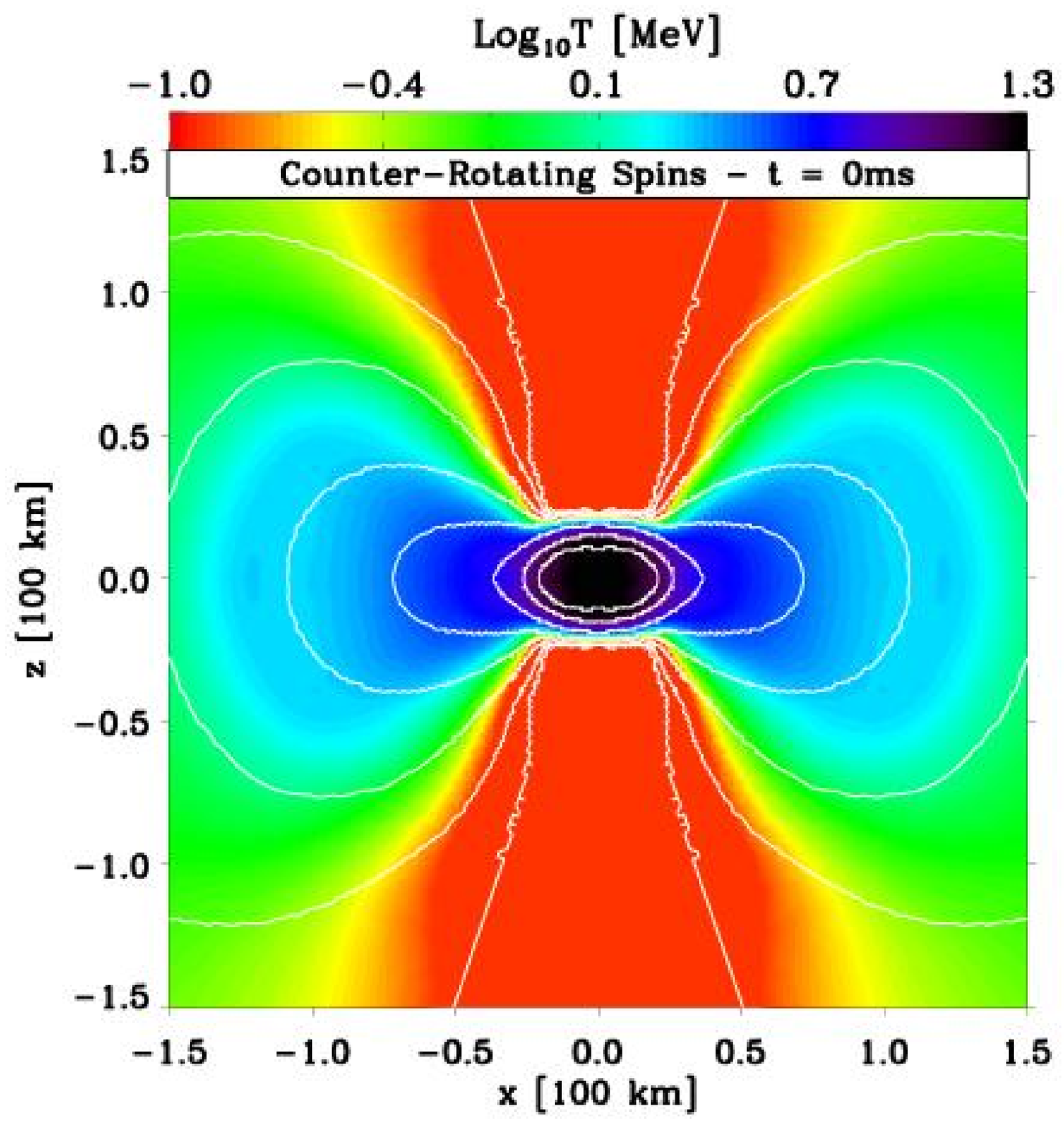}
\plotone{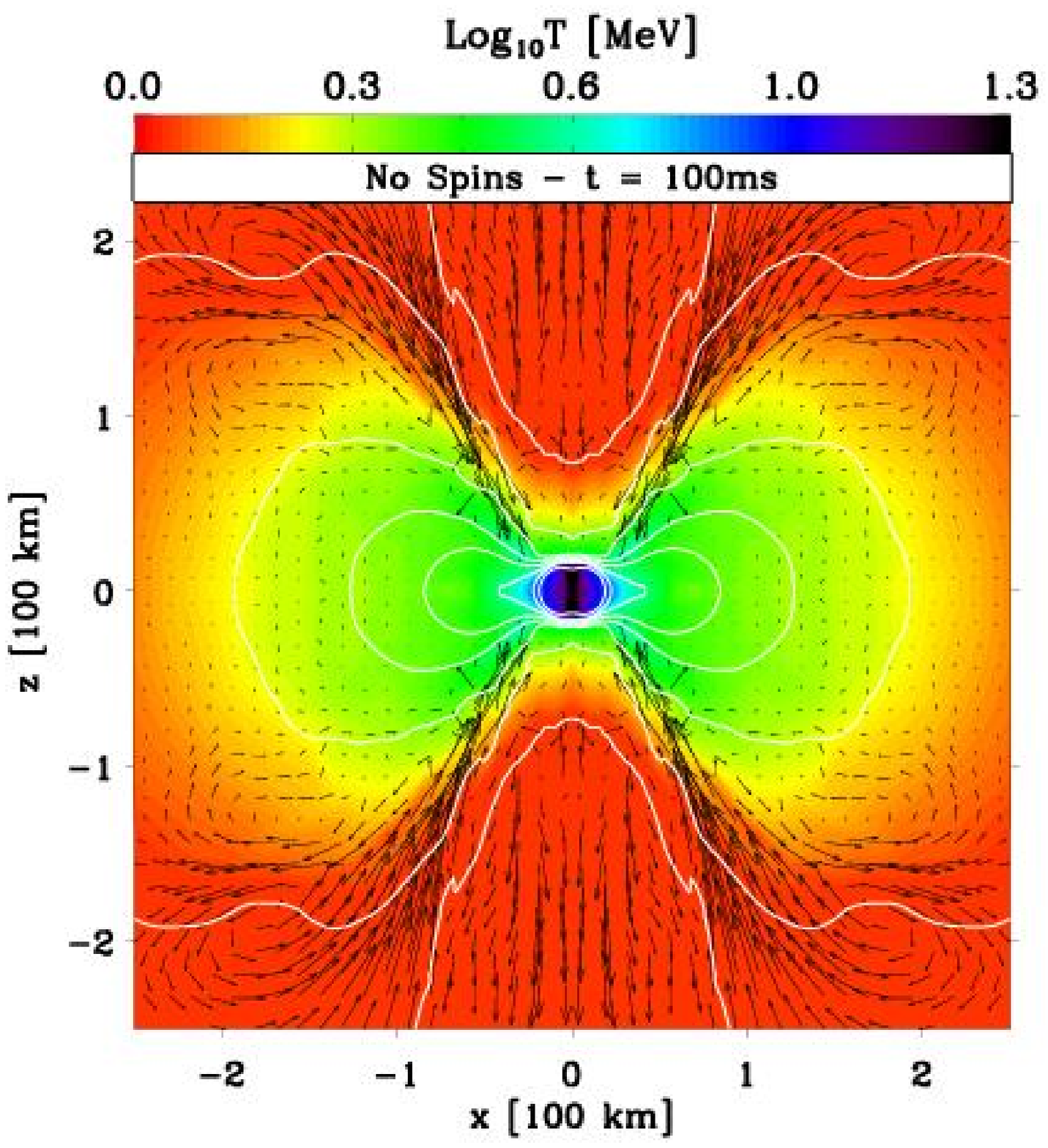}
\plotone{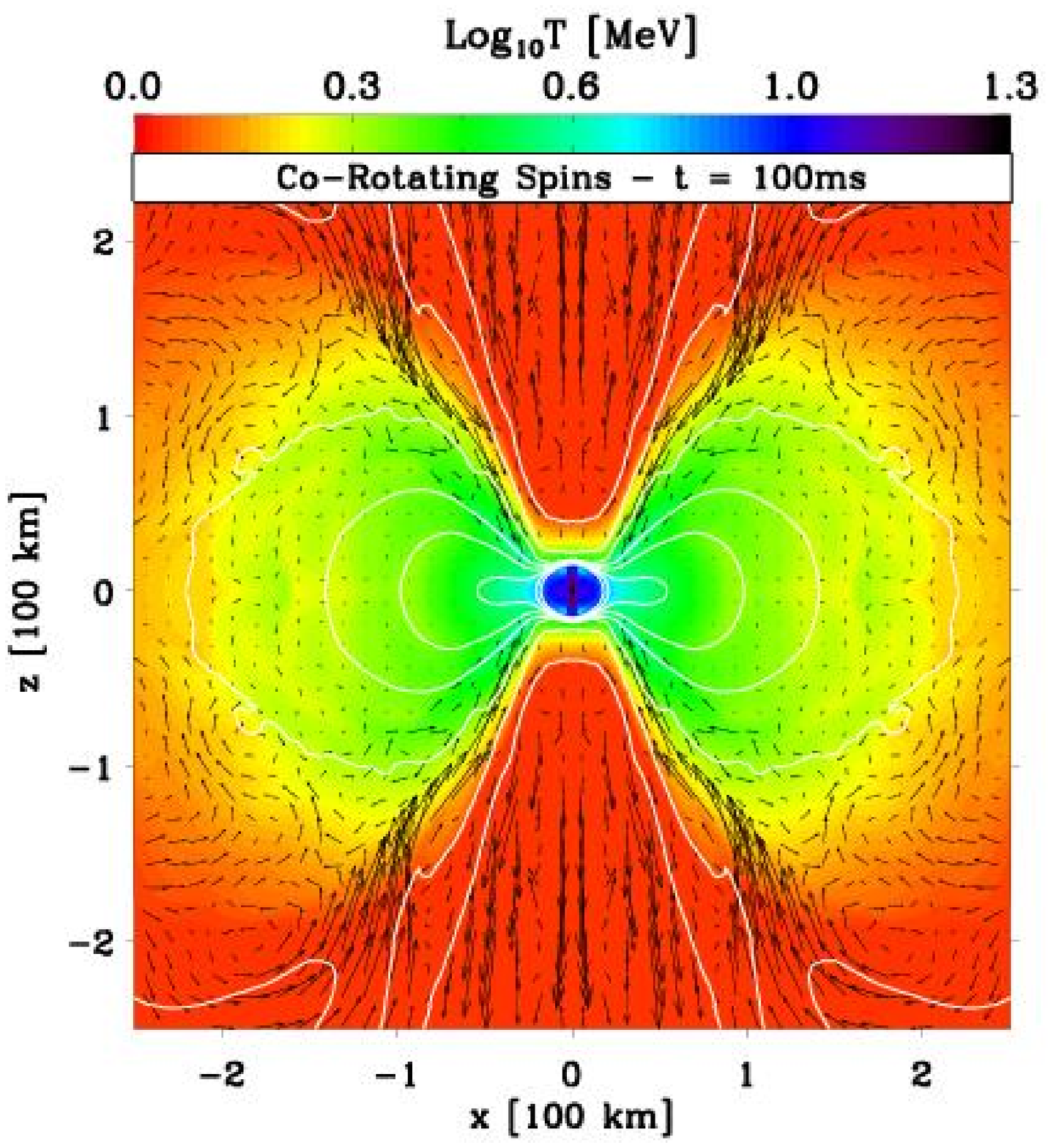}
\plotone{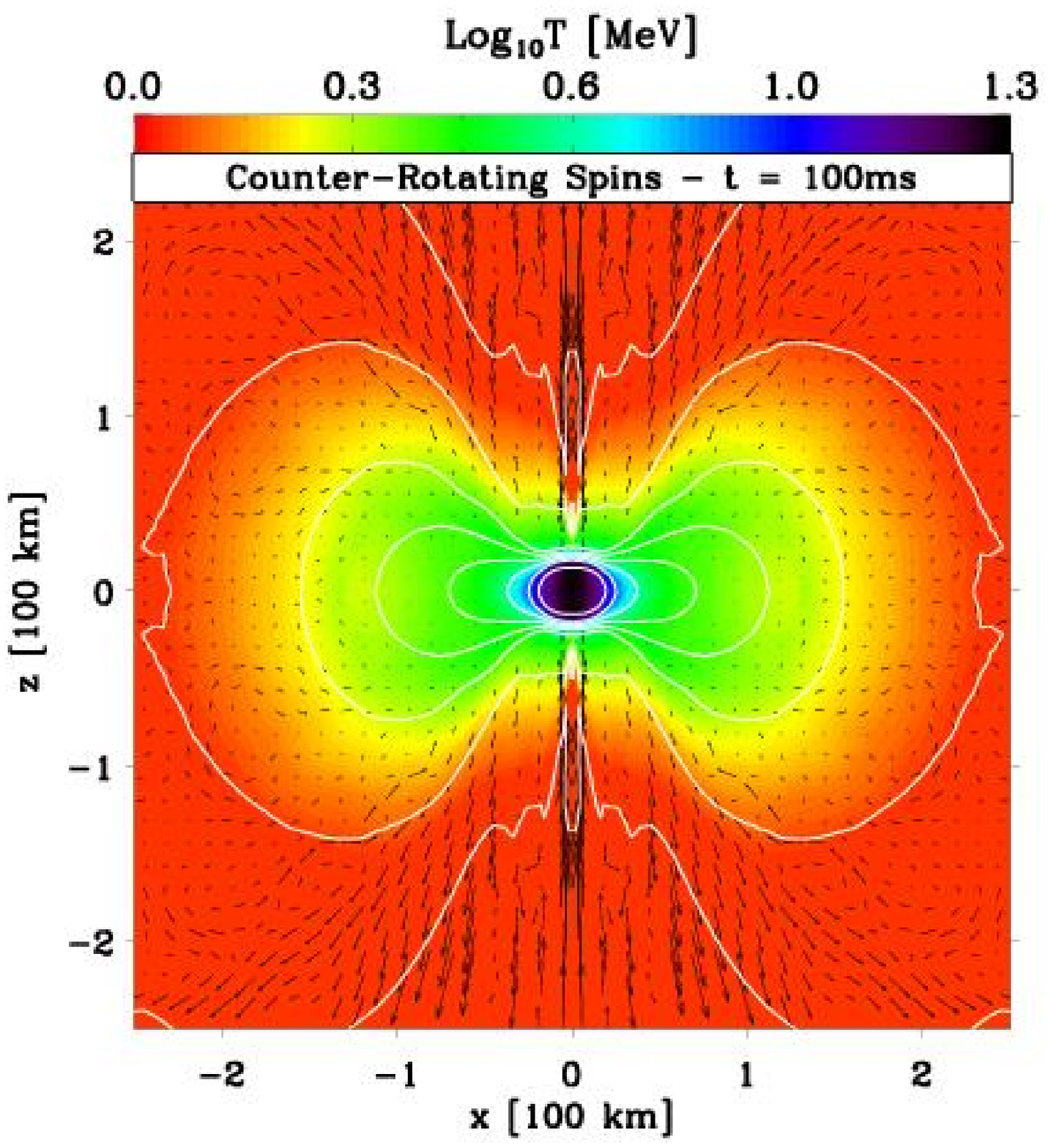}
\caption{
Colormap of the temperature (MeV; log scale) at 0 ({\it top row}; inner 150$\times$150\,km$^2$ region) 
and 100\,ms ({\it bottom row}; inner 250$\times$250\,km$^2$ region) after the start of the simulation. 
In the bottom panels, velocity vectors are added, with a saturation length at 15\% the width 
of the display and corresponding to a velocity of 30,000\,\kms. 
From left to right columns, we show the no-spin, co-rotating, and counter-rotating models.
We also overplot white density contours for every decade, 
starting at a maximum of 10$^{14}$\,g\,cm$^{-3}$. Note that the minimum of the colorbar changes
between the top and the bottom rows.
} 
\label{fig_temp_evol}
\end{figure*}

\section{Model description}
\label{sect:model}

    The main goal of this work is to understand the long-term (over many rotation periods), 
post-coalescence, evolution of BNS mergers, with particular attention to neutrino 
signatures (flavor dependence, angular distribution, and radial distribution).  
Starting from azimuthal-averaged slices constructed from 3D Smooth-Particle-Hydrodynamics (SPH) 
simulations with the MAGMA code \citep{rosswog:07a} and taken a few milliseconds after coalescence, 
we perform two-dimensional (axisymmetric) multi-group flux-limited-diffusion (MGFLD) radiation 
hydrodynamics simulations using the code VULCAN/2D \citep{livne:04,dessart:06a,burrows:07a,ott:08}.
Below, we present the properties of the three BNS merger configurations from which
we start (\S~\ref{sect:prog}), and then describe our approach with VULCAN/2D 
in more detail (\S~\ref{sect:vulcan}). 

\subsection{Initial conditions}
\label{sect:prog}

The MAGMA code \citep{rosswog:07a} is a state-of-the-art SPH code that contains 
physics modules that are relevant to compact binary mergers and on top implements 
a slew of numerical improvements over most ``standard'' SPH schemes. 
A detailed code-description can be found in \cite{rosswog:07a},
while results are presented in \cite{price:06} and \cite{rosswog:07c}.

We briefly summarize the most important physics modules.
For the thermodynamic properties of neutron star matter we use
a temperature-dependent relativistic mean-field equation of state 
\citep{shen:98a,shen:98b}. It can handle temperatures from 0 to 100
MeV, electron
fractions from $Y_{\rm e}$= 0 (pure neutron matter) up to 0.56 and densities
from about 10 to more than 10$^{15}$\,g\,cm$^{-3}$. No attempt is made to
include matter constituents that are more exotic than neutrons and
protons at high densities. For more details we refer the reader to \cite{rosswog:02a}.

The MAGMA code contains a detailed multi-flavor neutrino leakage scheme.
An additional mesh is used to calculate the neutrino opacities that are
needed for the neutrino emission rates at each particle position.
The neutrino emission rates are
used to account for the local cooling and the compositional changes due to
weak interactions such as electron captures. A detailed description of the
neutrino treatment can be found in \cite{rosswog:03a}.

The self-gravity of the fluid is treated in a Newtonian fashion.
Both the gravitational forces and the search for the
particle neighbors  are performed with a binary tree that is based
on the one described in \cite{benz:90}. These tasks are the
computationally
most expensive part of the simulations and in practice they completely
dominate the CPU-time usage.  Forces emerging from the emission of
gravitational waves are treated in a simple approximation. For more details,
we refer to the literature \citep{rosswog:00,rosswog:02a}

In terms of  numerical improvements over ``standard'' SPH techniques, the code
contains the following:

\begin{itemize}

\item To restrict shocks to a numerically resolvable width artificial viscosity
is used. The form of the artificial viscosity tensor is oriented at Riemann
solvers \citep{monaghan:97}. The resulting equations are similar to those constructed for
Riemann solutions of compressible fluid dynamics. In order to apply the
artificial viscosity terms only where they are really needed, i.e., close to
a shock, the numerical parameter that controls the strength of the
dissipative terms is made time dependent, as suggested by \cite{morris:97}. 
An extra equation is solved for this parameter which
contains a source term that triggers on the shock and a term causing
an exponential decay of the parameter in the absence of shocks. Tests can
be found in \cite{rosswog:00} and an illustration of the time-dependent 
viscosity parameter in the context of Sod's shock tube problem can be found 
in \citet[][their Fig.~1]{rosswog:08b}.

\item A self-consistent treatment of extra terms in the hydrodynamics equations
that arise from varying smoothing lengths (so-called ``grad-h''-terms;
\citealt{springel:02,monaghan:02,price:04,rosswog:07a}).

\item A consistent implementation of adaptive gravitational softening
lengths as described in \cite{price:07}.

\item The option to evolve magnetic fields via so-called Euler potentials
\citep{stern:70} that guarantee that the constraint $\nabla \cdot B = 0$ is fulfilled. 
The details can be found in \cite{rosswog:07a} and \cite{rosswog:08a}.

\end{itemize}

  Our work starts from 3D SPH simulations of BNS mergers
in an isolated binary system, made up of two individual neutron stars, each with a baryonic mass of
1.4\,\mo (equivalent to a gravitational mass of $\sim$1.3\,\mo), and separated by a distance of 48\,km. 
Their initial hydrostatic structure is computed by solving the Newtonian equations.
The properties of nuclear matter (pressure, energy, entropy etc.) are 
obtained by interpolation in the Shen EOS \citep{shen:98a,shen:98b},
assuming zero temperature and $\beta$ equilibrium. 
A large number of SPH particles ($N \approx 10^6$) are then mapped onto the resulting density profiles 
These particles are further relaxed to find their true equilibrium state (see, e.g., \citealt{rosswog:07a}).

\begin{figure*}
\epsscale{0.35}
\plotone{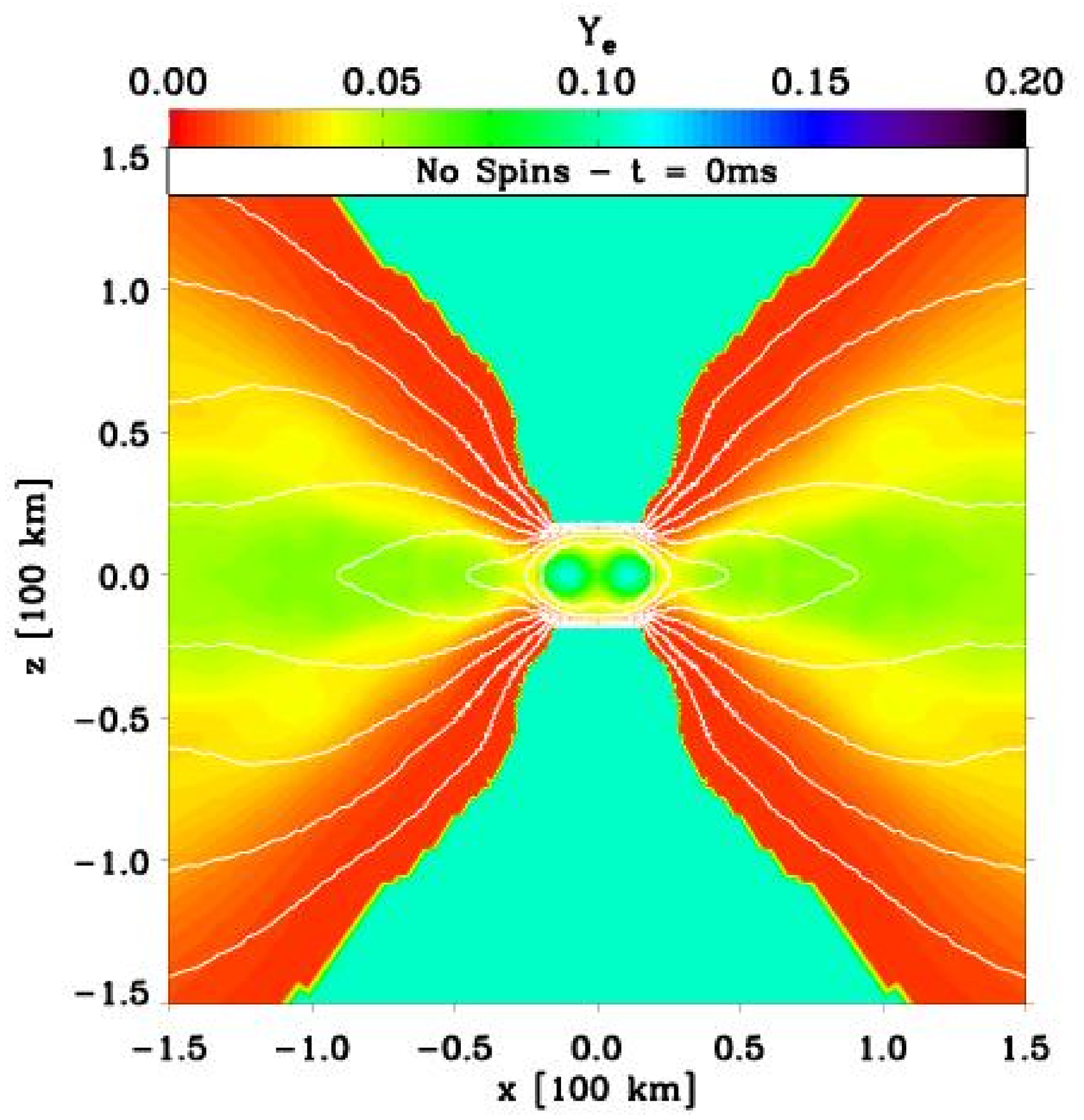}
\plotone{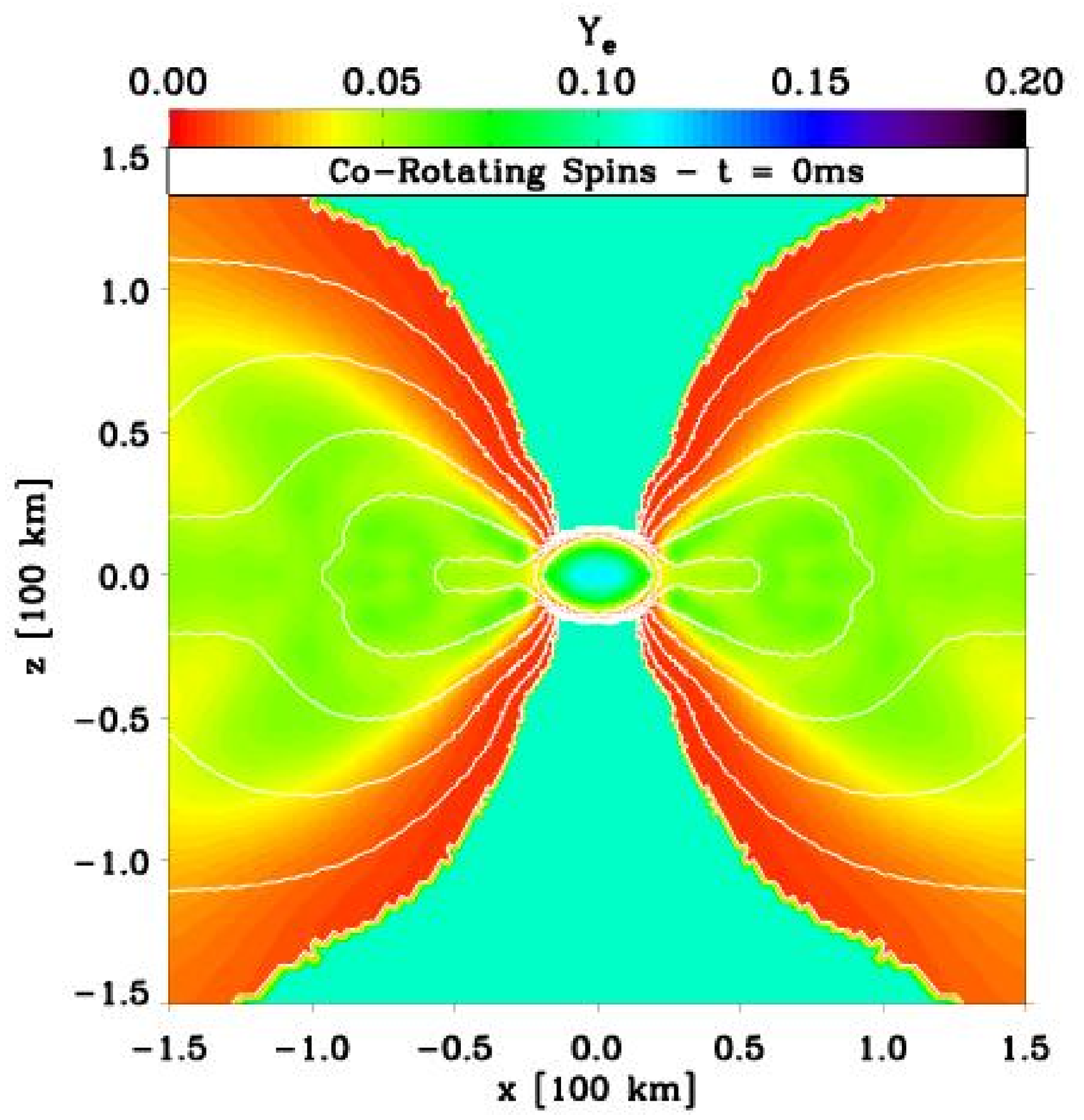}
\plotone{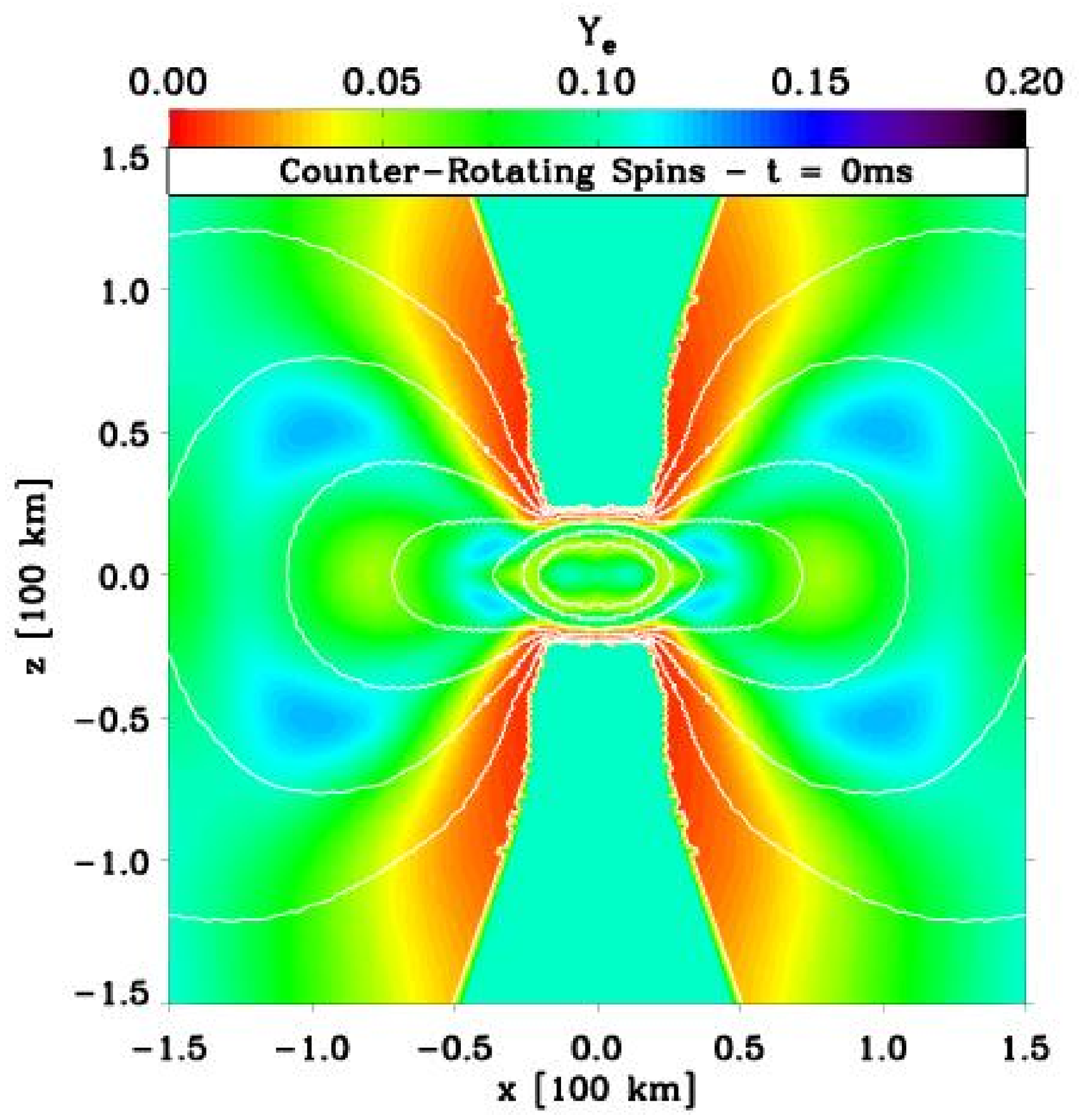}
\plotone{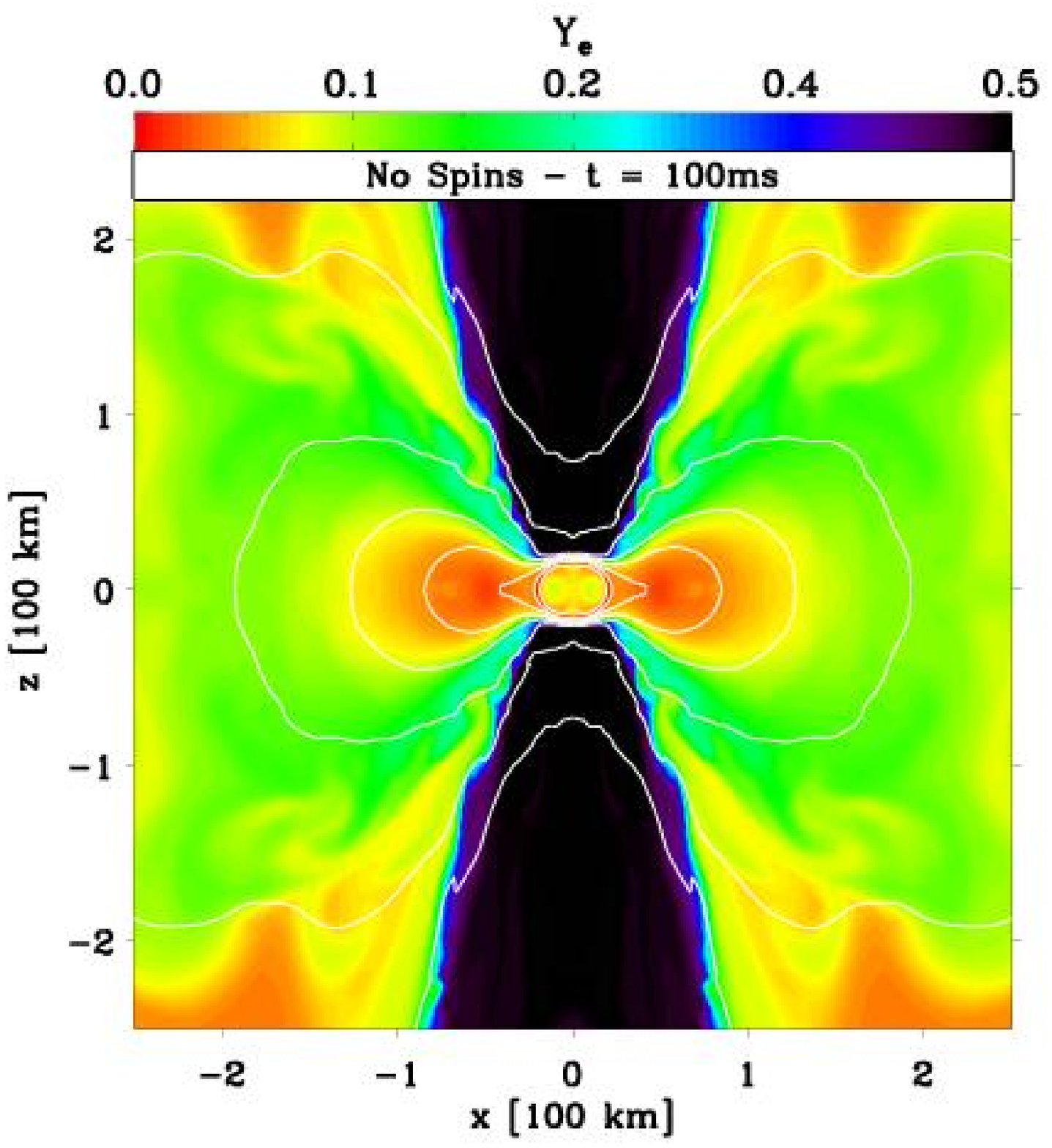}
\plotone{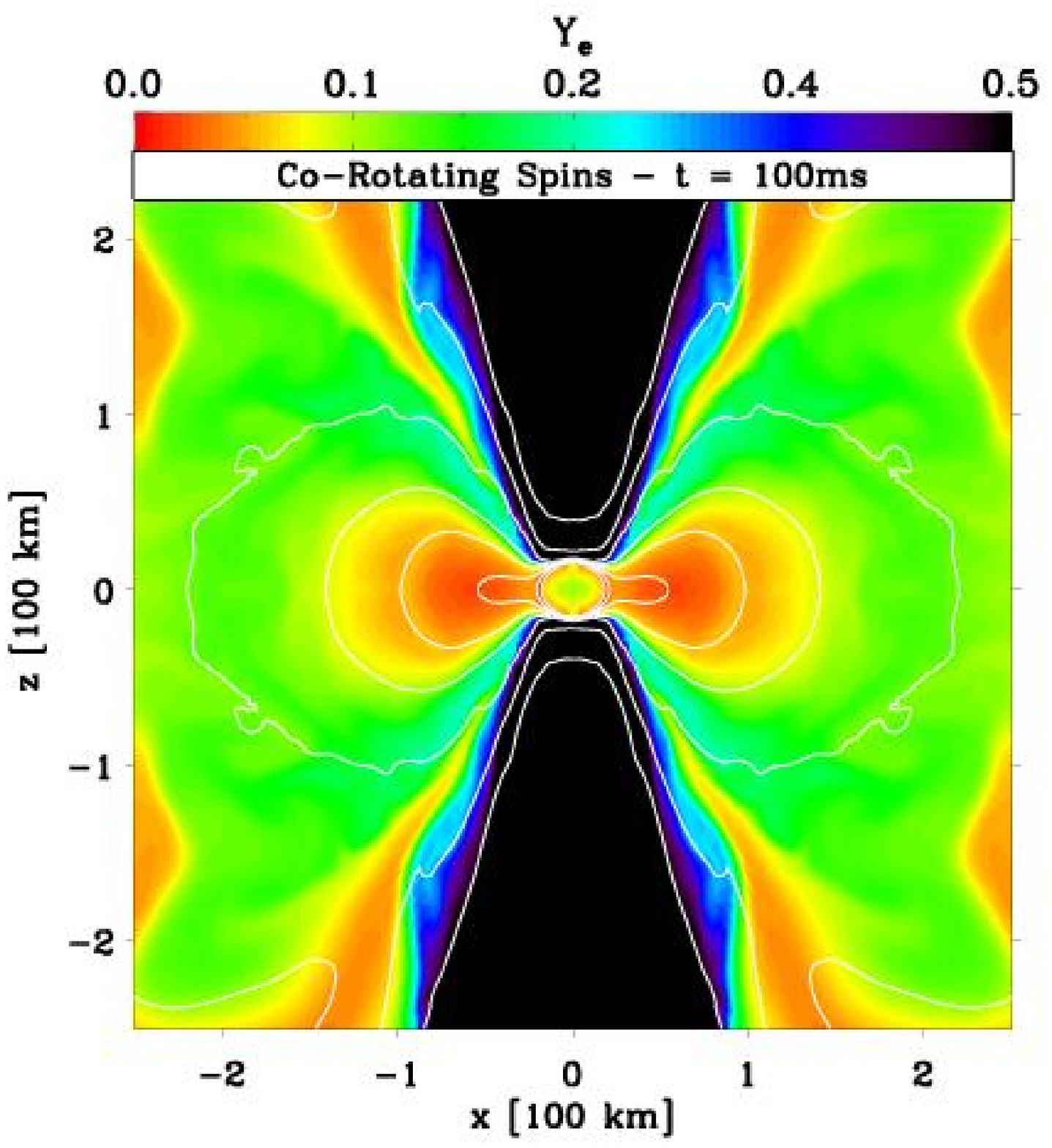}
\plotone{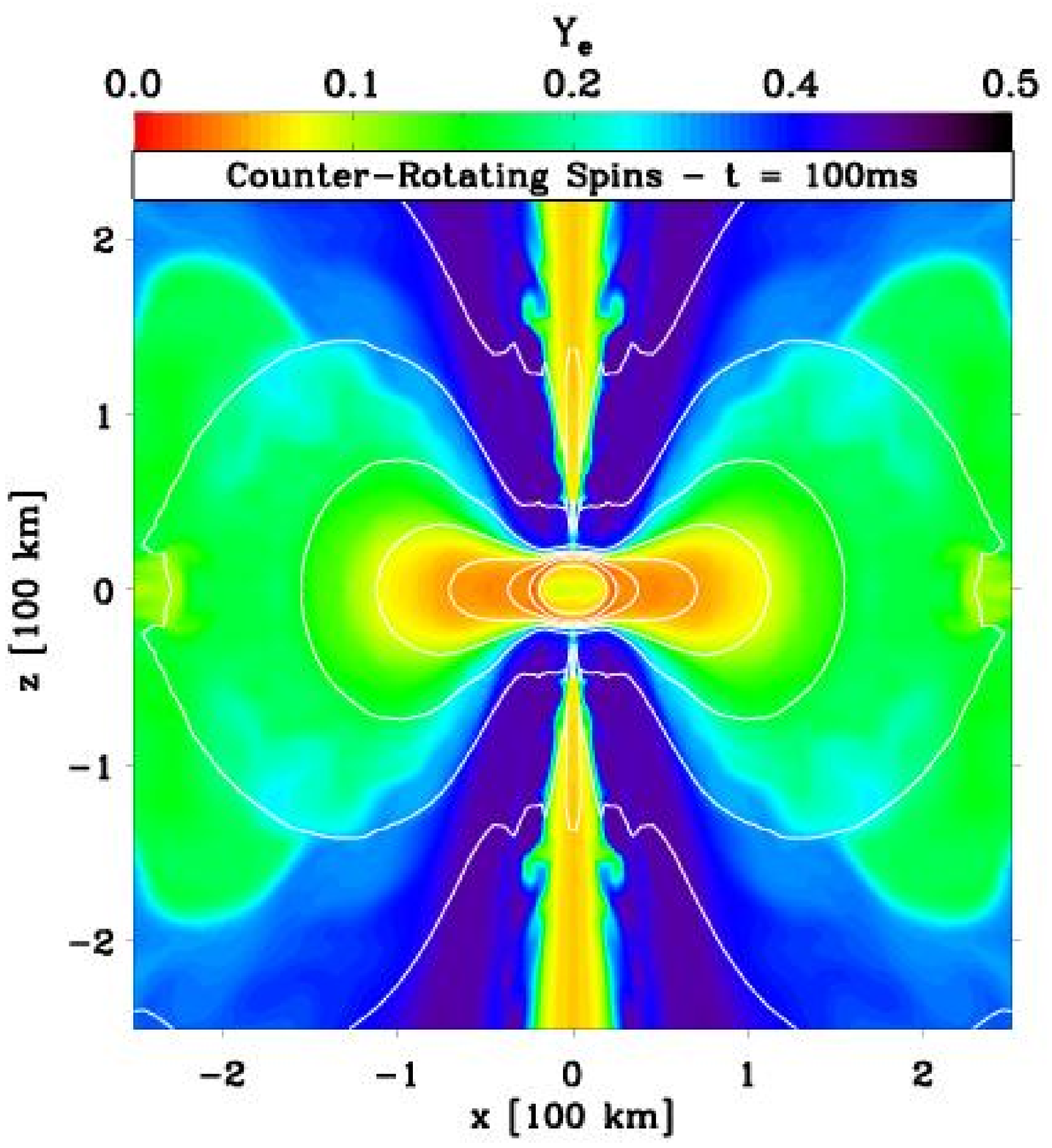}
\caption{
Same as for Fig.~\ref{fig_temp_evol}, but now for the electron fraction $Y_{\rm e}$.
Note that the extent of the displayed region and the extrema of the colorbar 
change between the top and the bottom rows.
In the counter-rotating spin model, an axis problem develops early on in the 
VULCAN/2D simulation (visible as a spurious low $Y_{\rm e}$ ``needle'' along the 
rotation axis), despite the high spatial resolution employed.
}
\label{fig_ye_evol}
\end{figure*}

  Because the tidal synchronization time 
exceeds the gravitational decay time, BNSs cannot be tidally locked during their
inspiral phase, and are thus expected to be close to irrotational (see, e.g. 
\citealt{bildsten:92} and \citealt{kochanek:92}.
However, past investigations of the coalescence of BNSs suggest that the intrinsic spin of 
each neutron star is an important parameter, as it determines, for example,
the width of the baryon-poor funnel above the forming central object
that is thought to play a decisive role in the launching of a GRB
(see, e.g. \citealt{zhuge:96,faber:01,oechslin:07b}).
Therefore, to encompass the range of possible outcomes, we
simulate three initial configurations: 1) no spins (no intrinsic neutron star spin), 
2) co-rotating spins (each neutron star has a period equal to that of the orbit, and rotates 
in the same direction), and 3) counter-rotating spins (same spin period as for co-rotating spins, 
but with rotation in a reverse direction to that of the orbit; the spins are counter-aligned).
In the rest of this paper, we broadly refer to these models as the no-spin,
the co-rotating spin, and the counter-rotating spin BNS models.
The co-rotating and counter-rotating cases are included here to bracket the range of possibilities,
but they represent unlikely extremes.
We leave to future work the consideration of higher mass BNS mergers 
\citep{rosswog:03b} or asymmetric systems \citep{rosswog:00}.

In practice, we post-process these 3D SPH simulations by constructing azimuthal-averages,
selecting a time after coalescence (when the two neutron stars come into contact) of 
10.0 (no-spins), 12.4 (co-rotating spins), and 20.4\,ms (counter-rotating spins),
corresponding in each VULCAN/2D simulations to the time origin.
At these times, these systems are evolving towards, but have not yet reached, complete axisymmetric 
configurations. This is a compromise made to model with VULCAN/2D the epoch of peak neutrino 
brightness. 
At these initial times, the amount of mass with a density greater than 10$^{13}$\,g\,cm$^{-3}$ is 2.5\,\mo in 
the no-spin BNS model, 2.31\,\mo in the co-rotating spin model, and 2.43\,\mo in the counter-rotating spin model.
In the same model order, and at these times, the mass contained inside the $\sim$13\,km radius of the 
highest mass neutron star allowed by the Shen EOS is only 1.11, 
1.26, and 1.03\,\mo. The SPH simulations employed to generate these models are Newtonian, and, thus, one would
expect more mass at small radii and systematically higher densities if allowance for GR 
effects were made, but the use of the highly-incompressible Shen EOS together with the significant amount of 
mass in quasi-Keplerian motion and on wide orbits suggests that the systems should survive at least for some time before the
general-relativistic gravitational instability occurs.

\subsection{VULCAN/2D}
\label{sect:vulcan}

  Starting from fluid variables constructed with azimuthal averages of the 3D SPH simulations 
described above, we follow the evolution of three BNS merger configurations with the
MGFLD radiation-hydrodynamics code VULCAN/2D \citep{livne:04,dessart:06a,burrows:07a}.
The gravitational potential is assumed Newtonian and computed using a multipole solver \citep{dessart:06a}.
Due to the difficulty of handling the superluminal Alfv\'en speeds prevailing in 
the low-density material surrounding the BNS merger for even modest field strengths, 
we do not investigate the dependence on magnetic field strength and morphology.
However, our study presents a self-consistent treatment of neutrino emission, scattering,
and absorption in a multi-species MGFLD context.
Associated neutrino-matter coupling
source terms are included in the momentum and energy conservation equations
and their relevance to the merger evolution is, thus, computed explicitly,
for a typical duration of $\sgreat$100\,ms.
For these dynamical calculations, we include all neutrino processes described in \cite{brt:06}, 
but neglect the secondary processes of neutrino-electron scattering.
The transport solution in the MGFLD scheme we employ is solved for the three neutrino species 
$\nu_e$, $\bar{\nu}_e$, and ``$\nu_\mu$,'' (which groups together the $\mu$ and $\tau$ neutrinos)
and at eight neutrino energies: 2.50, 6.87, 12.02, 21.01, 
36.74, 64.25, 112.36, 196.48\,MeV. The energy spacing is constant in the log.

  The choice of spatial grid is determined primarily by the very aspherical density distribution
of the merger and the very steep density drop-off at the neutron star surface in the polar direction. 
We present in Figs.~\ref{fig_temp_evol}-\ref{fig_ye_evol} ({\it top row}) the initial temperature 
(logarithmic scale and in MeV), electron fraction, and the density distribution 
(white line contours overplotted for every decade between 10$^7$ and 10$^{14}$\,g\,cm$^{-3}$) 
for each BNS merger configuration
that we map onto the VULCAN/2D grid. Note that fast rotation in the inner region, 
although sub-Keplerian, displaces the density maximum by 8\,km from the center and 
along the equator in the no-spin BNS model. These density peaks also 
correspond to extrema in the electron fraction of $\sim$0.1 at this time.
To setup the hybrid VULCAN/2D grid, we position the transition radius between
the inner Cartesian and the outer spherical polar regions at 12\,km. 
The density at this location is at all
times in excess of 10$^{14}$\,g\,cm$^{-3}$, and thus -- due to the stiffness of the Shen EOS in 
that regime -- away from the regions of large density gradient. 
The resolution in this inner region is typically 200$\times$200\,m$^2$, corresponding
to $12/0.2=60$ zones in each (cylindrical) direction $r$ and $z$ from the center to the transition radius.
Beyond the transition radius, we use a logarithmically increasing radial grid spacing with $\Delta r / r$ = 1.85\%,
using 301 zones to 3000\,km. The grid covers one hemisphere, from the rotation axis to the
equator (both treated as reflecting boundaries), and the computation assumes axisymmetry, 
i.e., the azimuthal gradients are zero. We fill the grid outside the merger with material 
having a low density 
of 5$\times$10$^3$\,g\,cm$^{-3}$ and a low temperature of 4$\times$10$^8$\,K. 
Below $Y_{\rm e}=$0.05, we compute thermodynamic variables, as well as 
opacities/emissivities, by adopting an electron fraction of 0.05. Given the smooth 
variation for the corresponding quantitites (pressure, entropy, mean-free path, etc.) 
at this level of neutron richness, this approximation is expected to be quite good.

\begin{figure*}
\epsscale{0.35}
\plotone{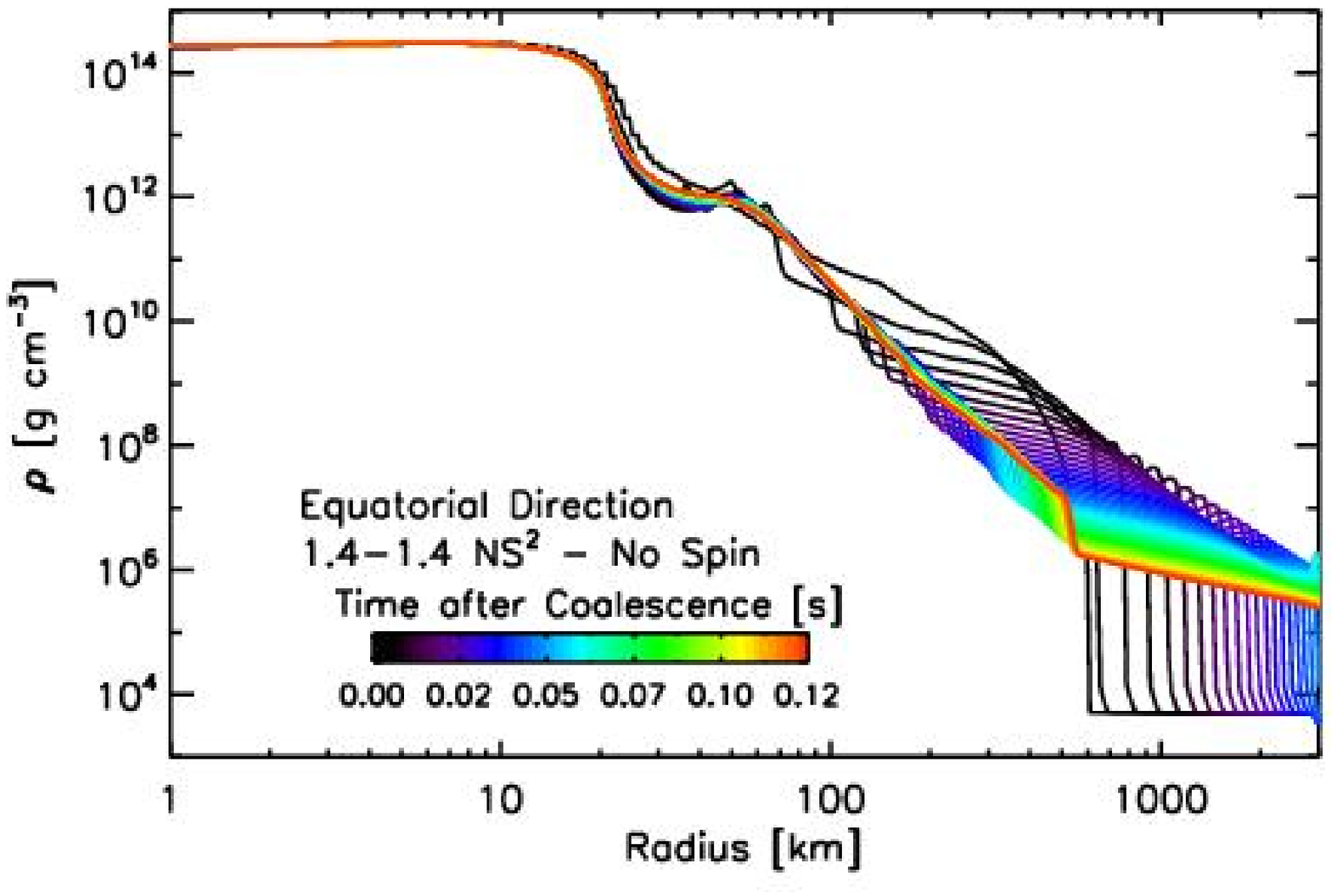}
\plotone{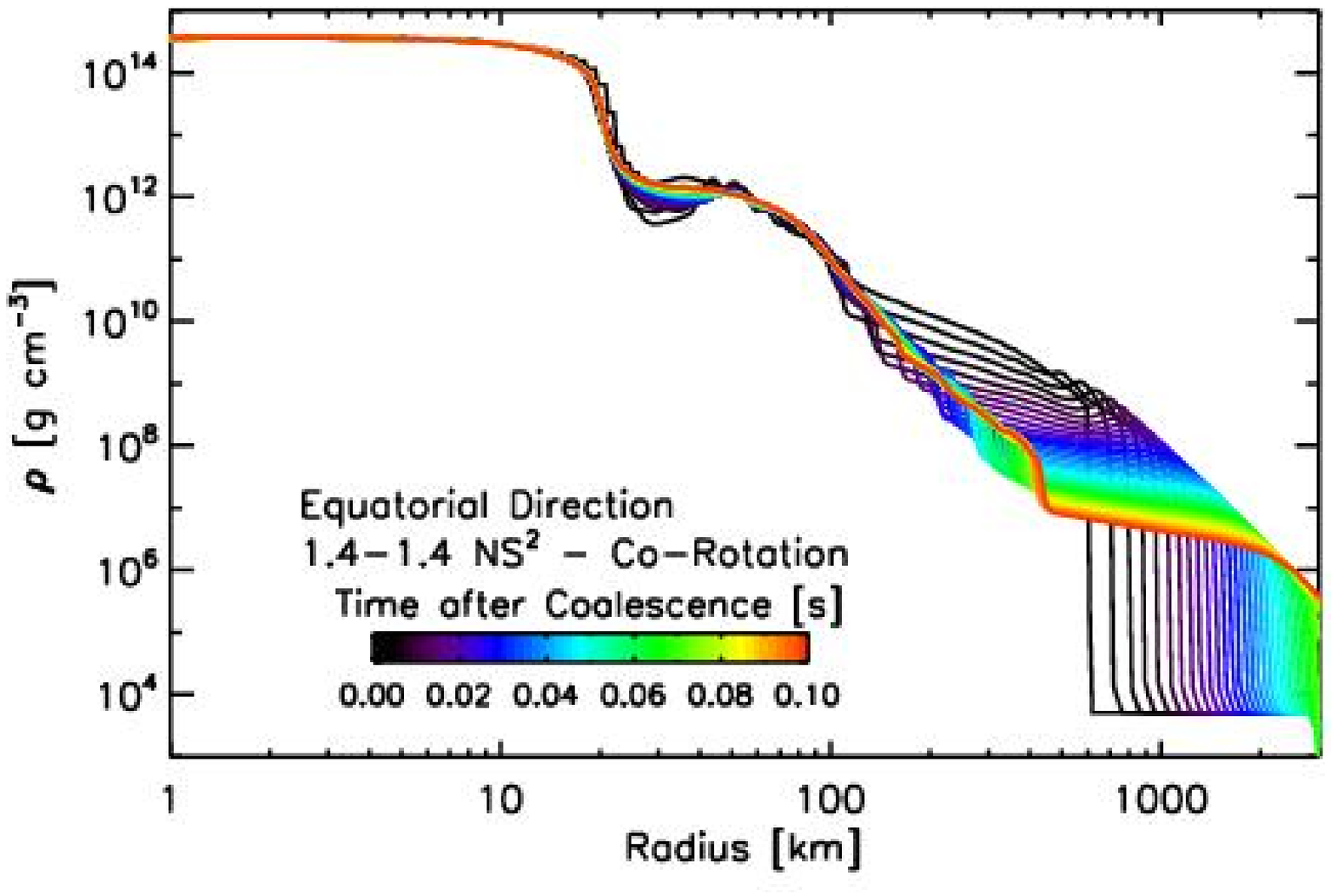}
\plotone{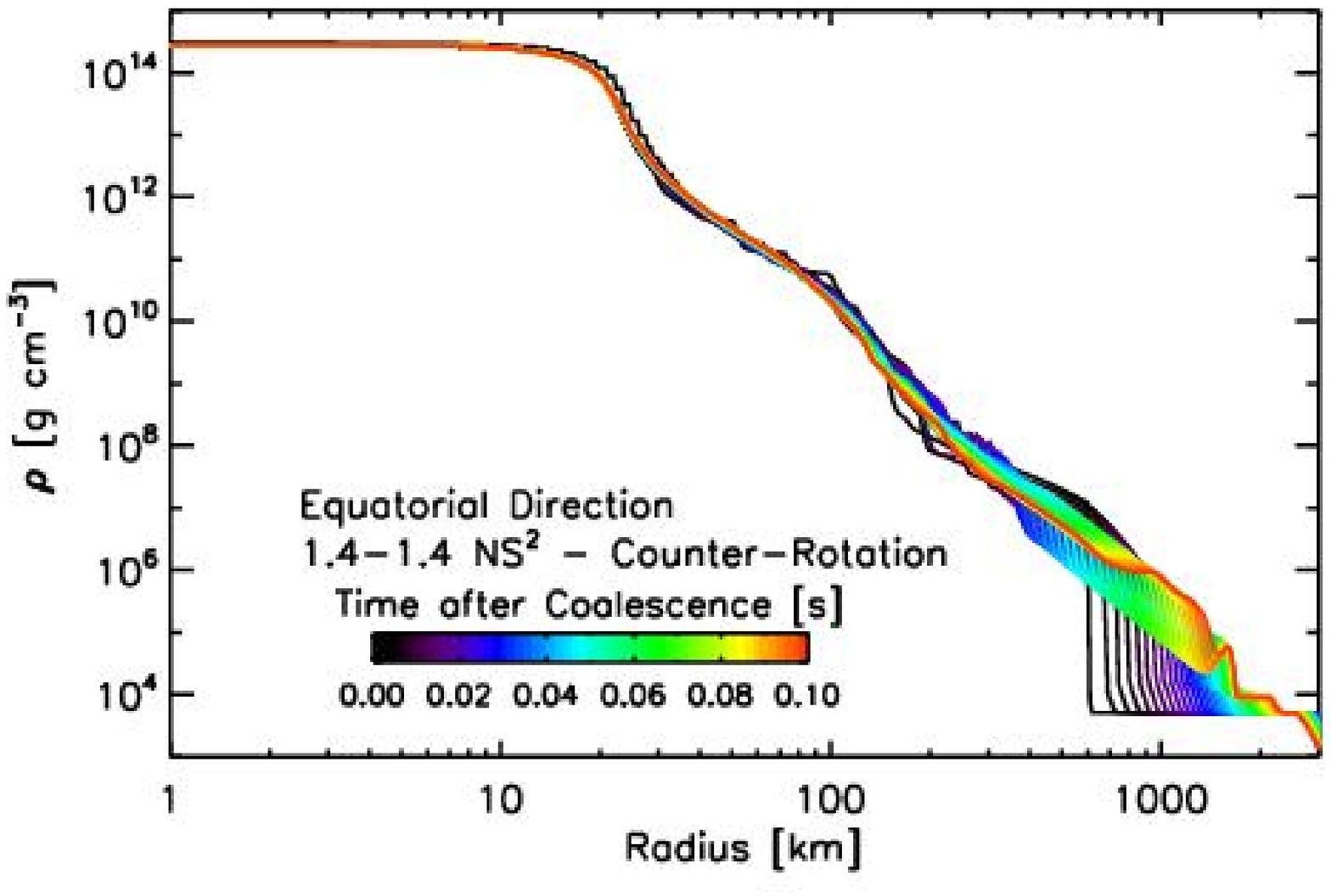}
\plotone{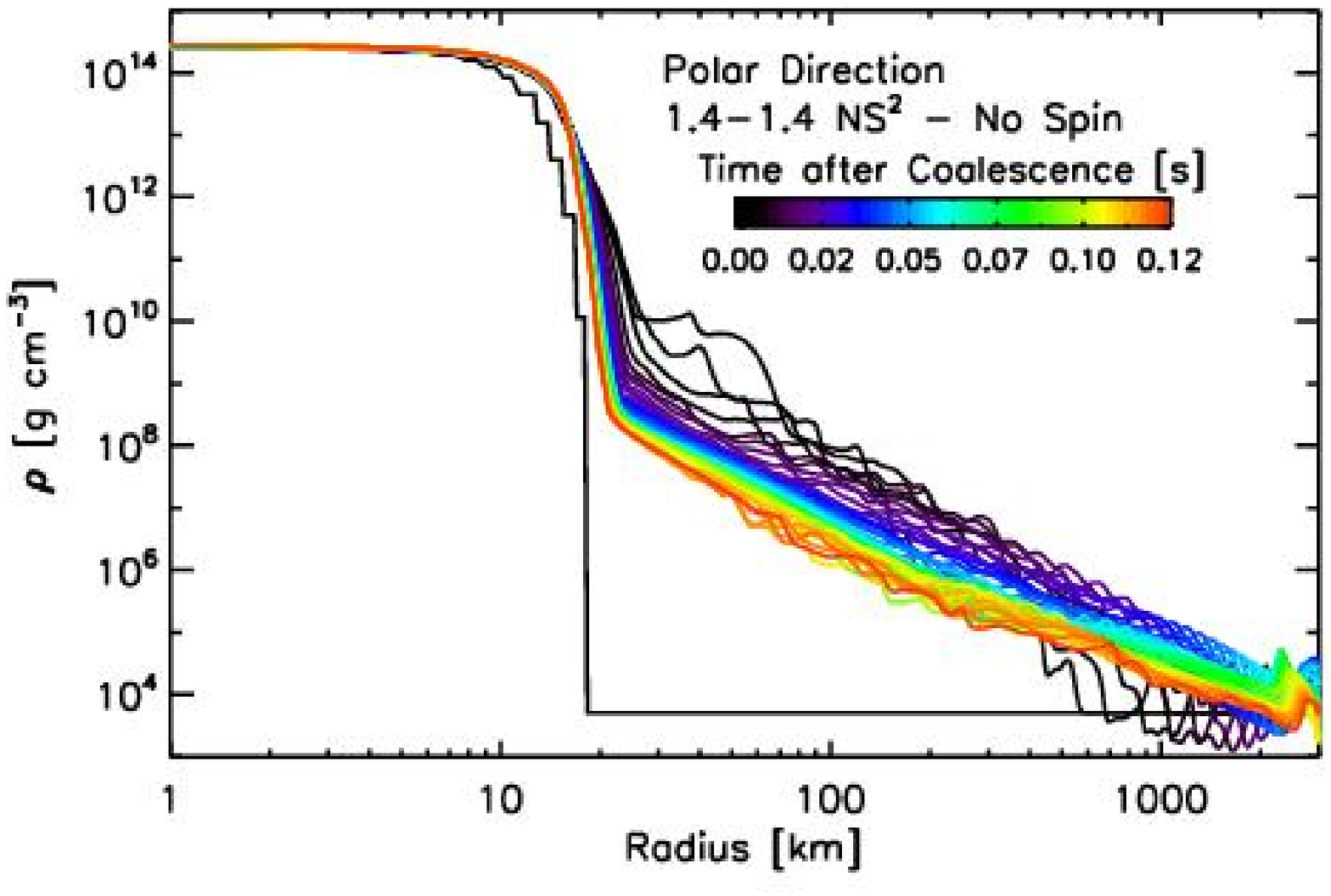}
\plotone{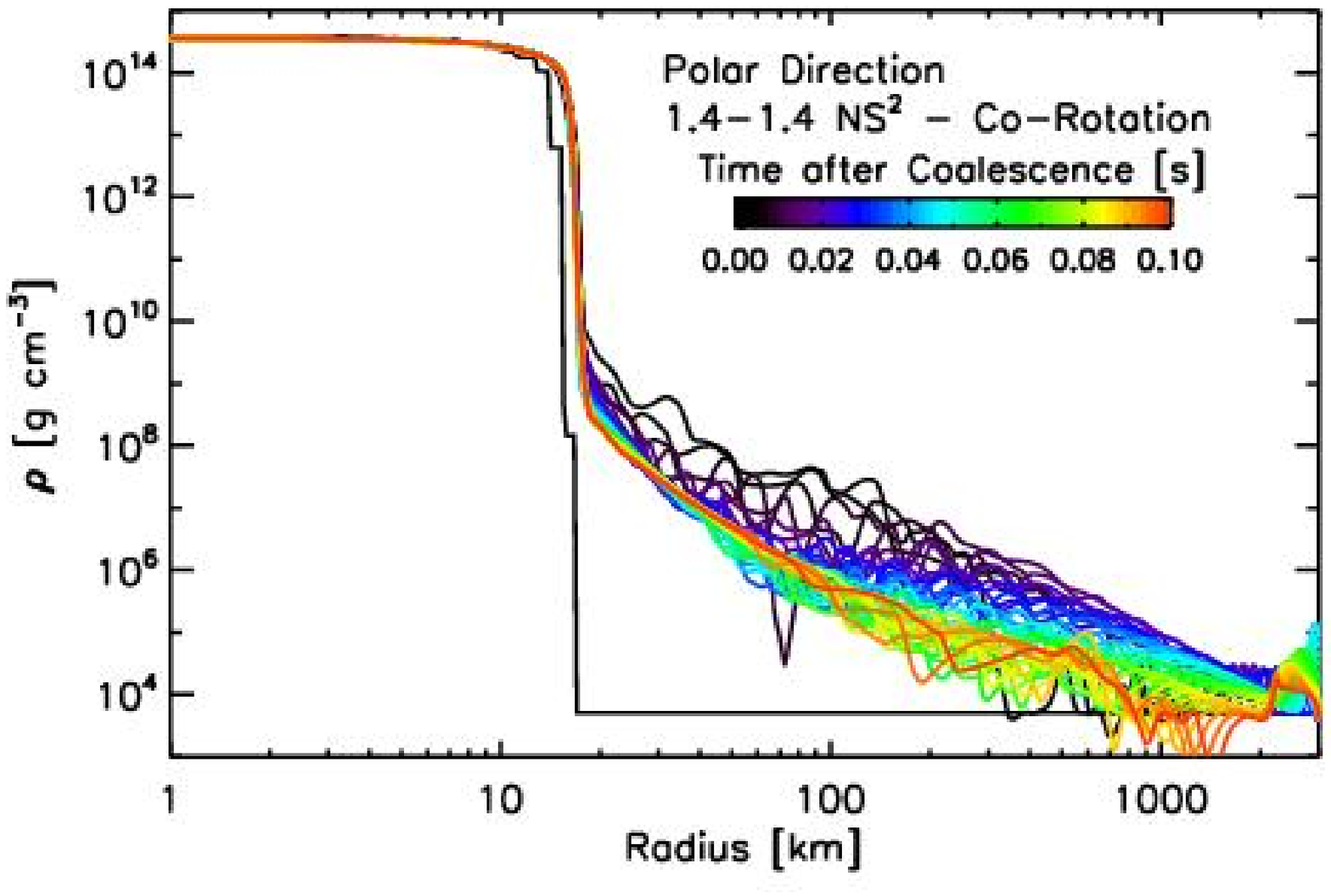}
\plotone{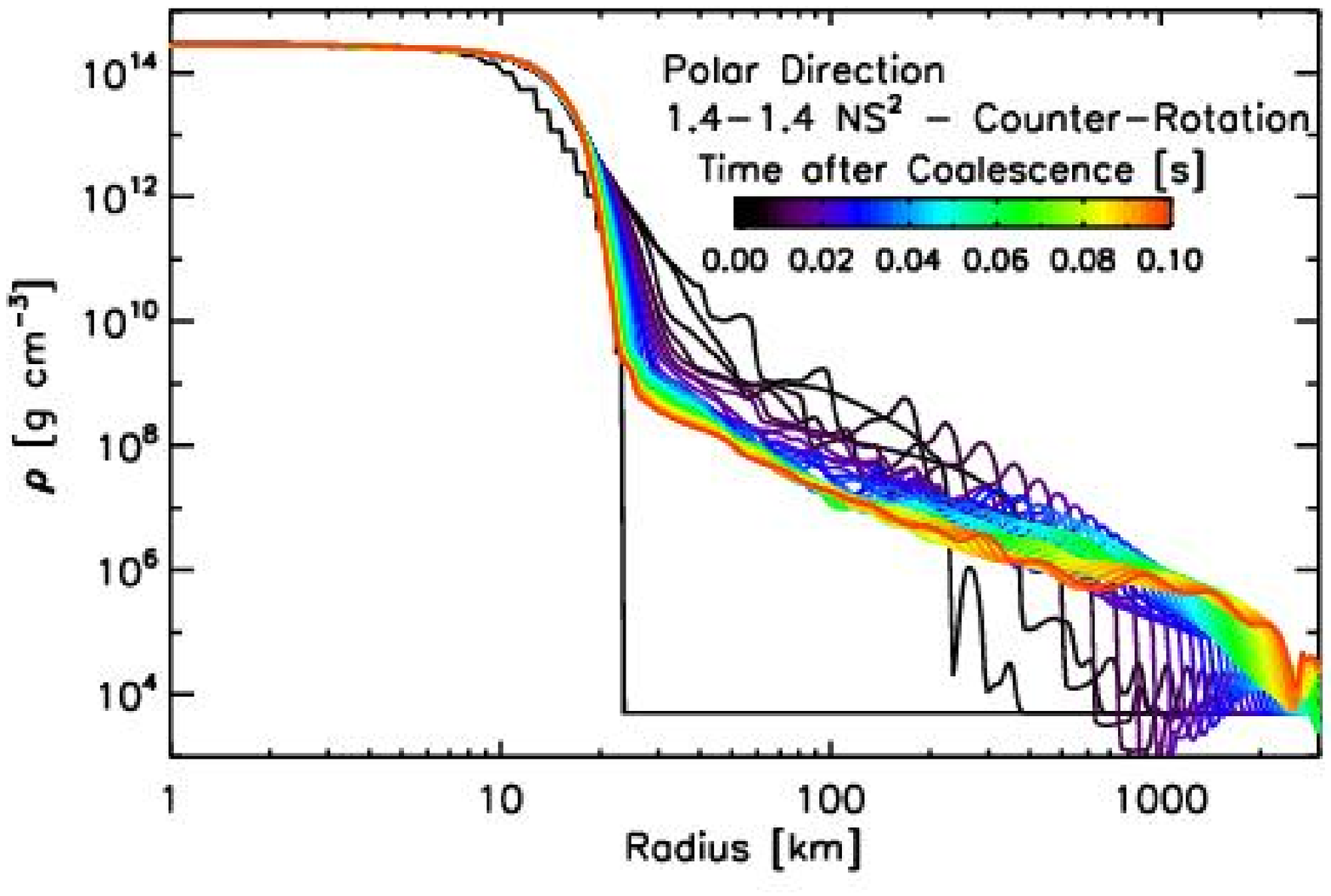}
\caption{
Time evolution of the density distribution along the equatorial ({\it top row}) and 
the polar ({\it bottom row}) directions 
for the BNS merger models with initially no spins ({\it left}), 
co-rotating spins {\it middle}), and counter-rotating spins ({\it right}).
}
\label{fig_rho_slice}
\end{figure*}

After the MGFLD radiation-hydrodynamics simulations are completed, we post-process individual 
snapshots keeping the hydrodynamics frozen, and relaxing only the radiation quantities. 
This operation is performed for each model at 10\,ms intervals over 
the whole evolution using the multi-angle, $S_n$, radiation-transport module discussed 
in \cite{livne:04} and described more recently in \cite{ott:08}. We refer the reader to those
papers for details on the method and the nomenclature. 
This $S_n$ variant yields the explicit angular dependence of the neutrino specific intensity,
and, thus, represents a more accurate modeling of the strongly anisotropic neutrino luminosity
in such highly aspherical systems \citep{ott:08}.
The $S_n$ calculation is done with the same number of energy groups (i.e., eight)
and on the same spatial grid. In the $S_n$ algorithm, the transition at depth to MGFLD
described in \cite{ott:08} is natural here, since it occurs at 12\,km, and, thus, in regions where 
the densities are nuclear. However, the spatial resolution is not fully satisfactory 
at the neutron star surface and along the 
polar direction, a region where the density decreases by few orders of magnitude in
just a few kilometers. This, thus, represents a challenge for radiation transport on our Eulerian grid.
In the first instance, we relax the MGFLD results with $S_n$ adopting 8 $\vartheta$-angles. 
In the $S_n$ approach, $n$ such $\vartheta$-angles translate into $n(n+2)/2$ directions
mapping the unit sphere uniformly.
Once converged, we perform an accuracy check by remapping 
the angle-dependent neutrino radiation field from 8 to 16 $\vartheta$-angles, and then relaxing 
this higher angular resolution simulation.
Due to the increased computational costs, we use the high (16 $\vartheta$-angles) 
angular-resolution configuration only for snapshots at 10, 60, and 100\,ms, 
but for all three BNS merger configurations
studied here. Importantly, the explicit angle-dependence of the neutrino-radiation field
allows us to compute its various moments, which we employ in a novel formulation
of the neutrino-antineutrino annihilation rate (\S~\ref{annihil:sn}). We find that these results 
compare favorably with the simpler approach of \citet{ruffert:96,ruffert:97},
although the multi-angle simulations offer a number of interesting new insights, which we explore
in \S~\ref{sect:annihilate}.

\section{BNS merger evolution and neutrino signatures} 
\label{sect:results}

\begin{figure*}
\epsscale{0.35}
\plotone{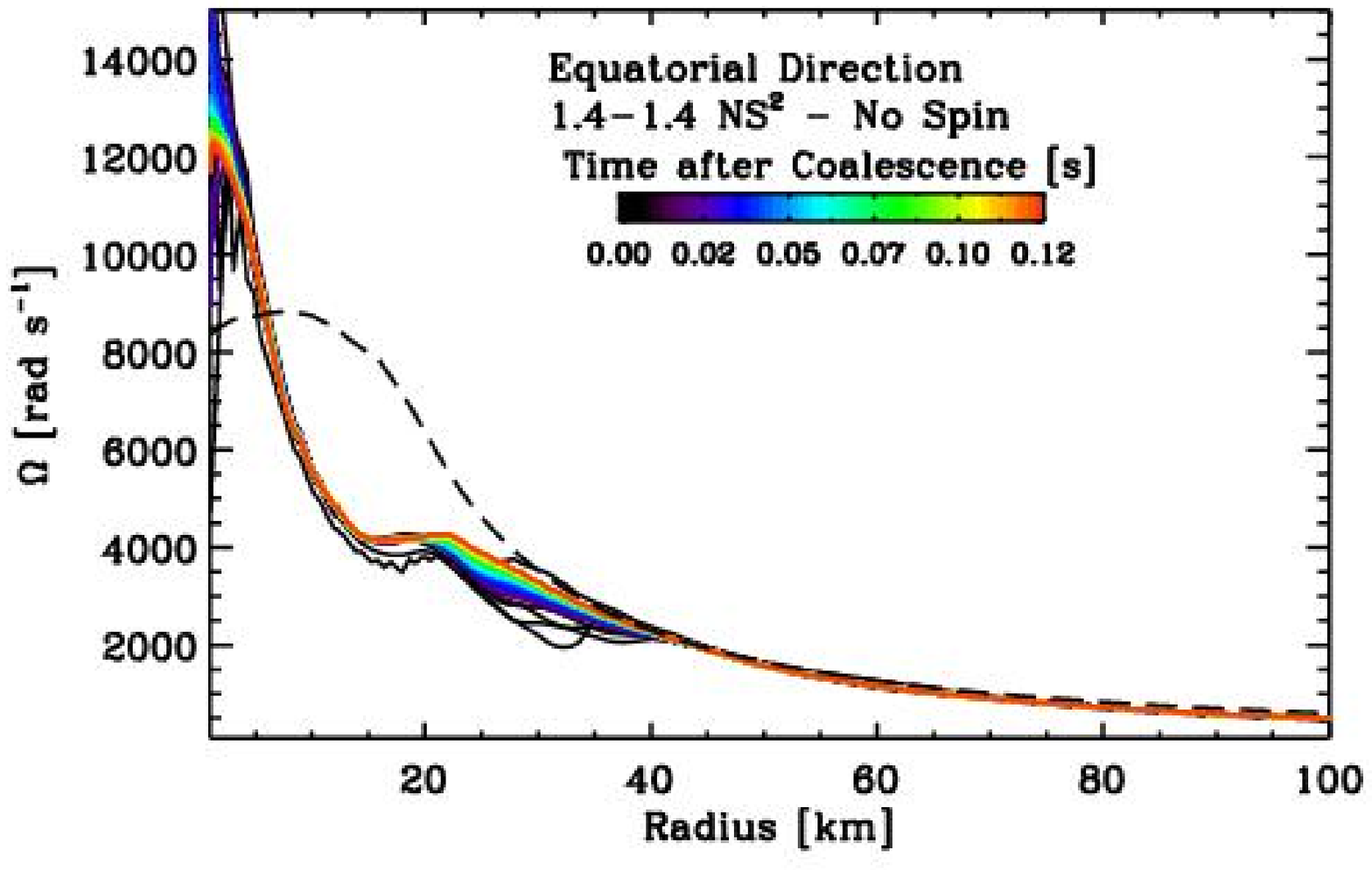}
\plotone{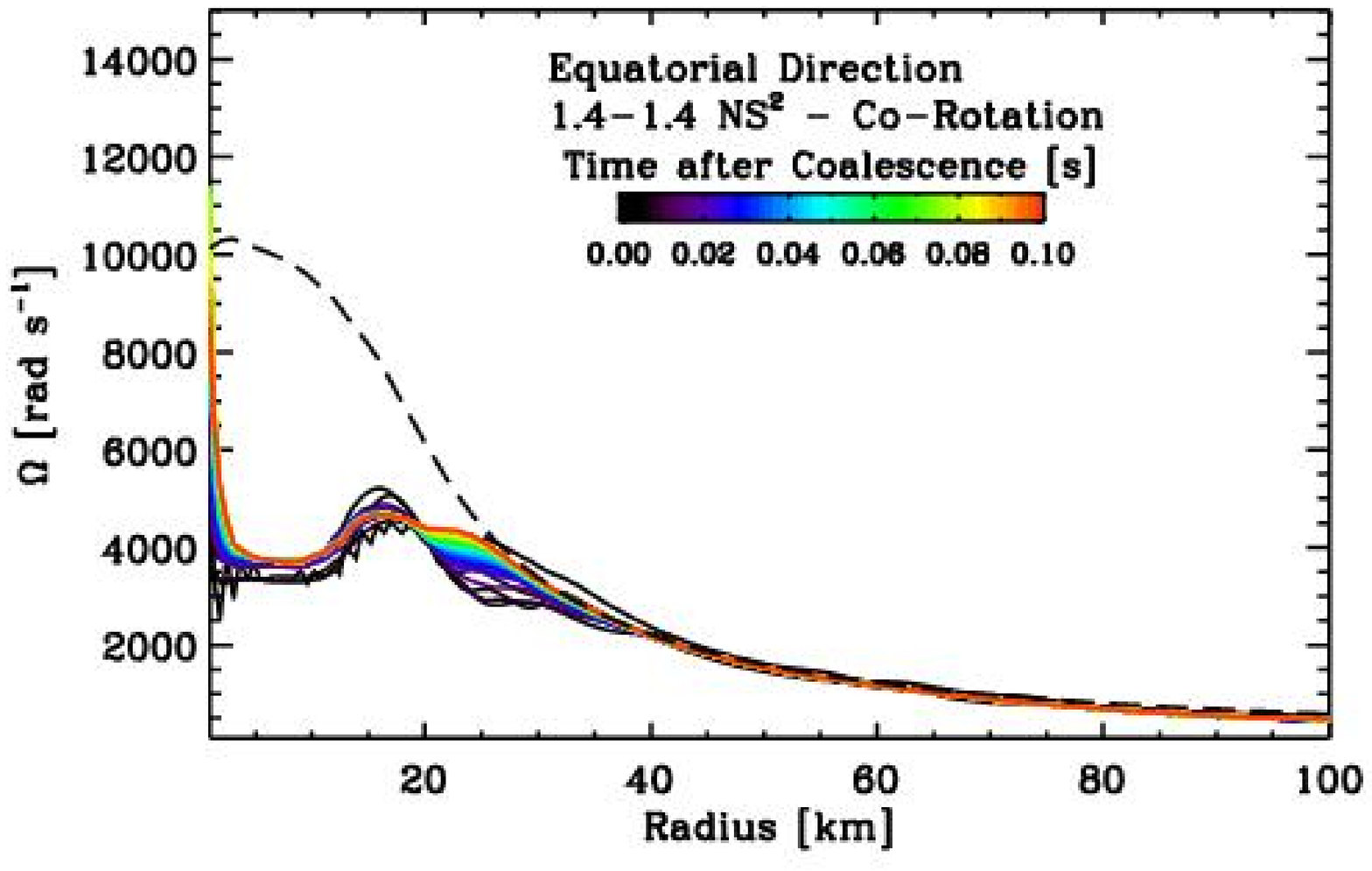}
\plotone{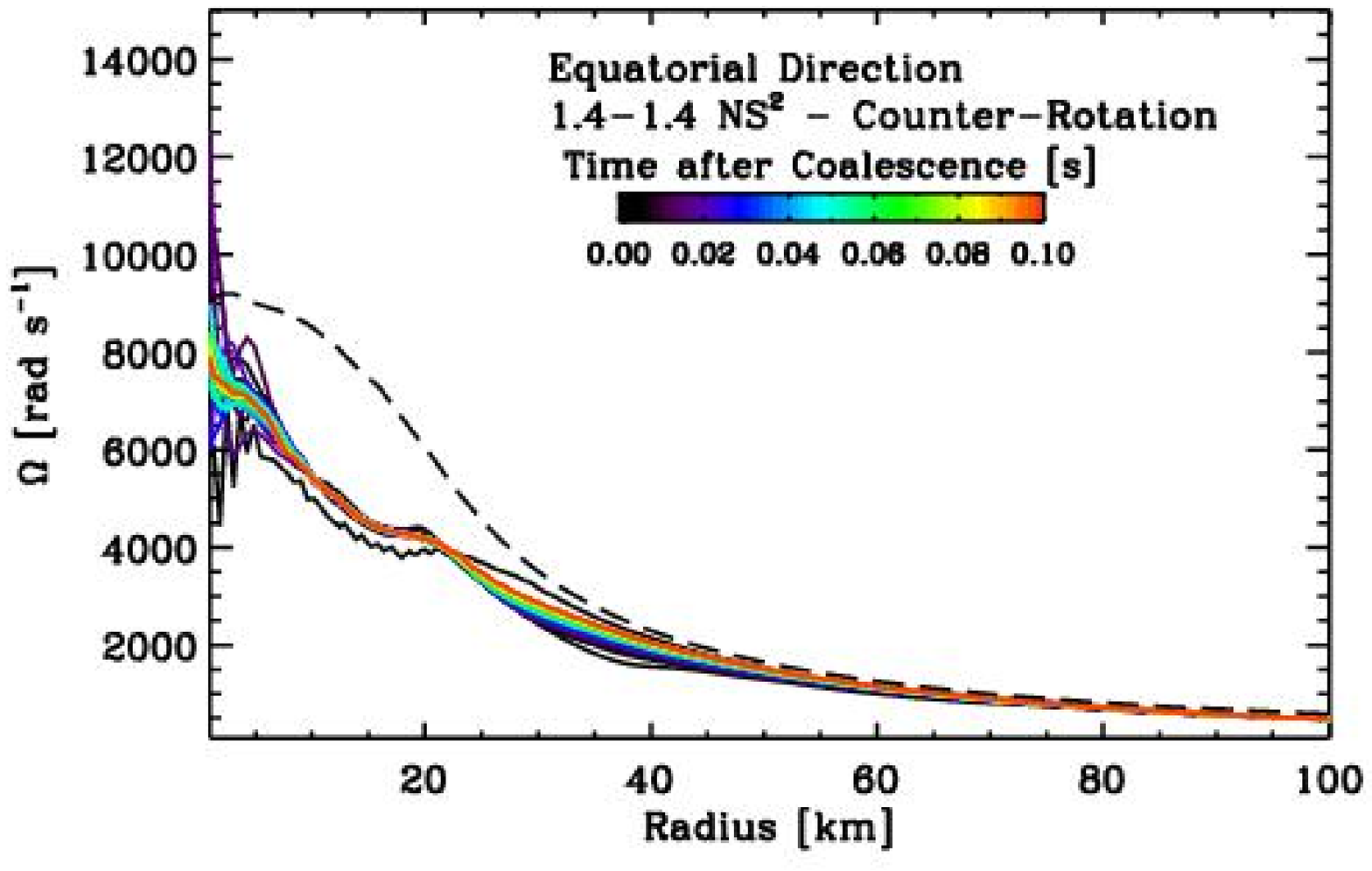}
\caption{
Time evolution of the angular velocity distribution along the equatorial 
direction for the BNS merger models with initially no spins ({\it left}), 
co-rotating spins {\it middle}), and counter-rotating spins ({\it right}).
The dashed line gives the local Keplerian angular velocity (adopting the 
mass of the spherical volume interior to a given radius), suggesting that 
the material at low latitudes and located between $\sim$30 and $\sim$100\,km forms 
a quasi-Keplerian disk, with densities on the order of 10$^{12}$\,g\,cm$^{-3}$.
Note also the strong degree of differential rotation in the no-spin
and counter-rotating spin cases, which is preserved throughout the VULCAN/2D simulation. 
}
\label{fig_omega_slice}
\end{figure*}

\begin{figure}
\plotone{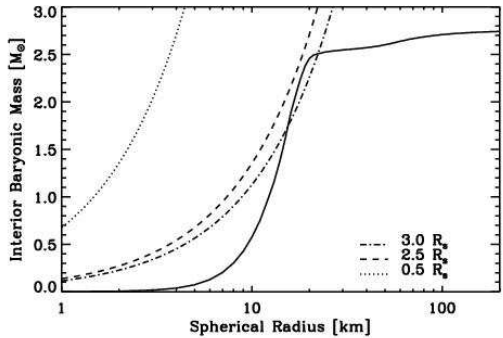}
\caption{
Interior baryonic mass versus spherical radius for the BNS merger model
with no initial spins. The time corresponding to these data is 60\,ms after the start
of the VULCAN/2D simulation. Following \cite{ruffert:96},
we overplot the Schwarzschild radius $R_{\rm s}(m(r))$, scaled by 
0.5, 2.5, and 3, corresponding to the range of masses plotted on the vertical axis.
Our results differ from those of \cite{ruffert:96} in that only a small mass range
falls within the radius 3$\times R_{\rm s}(m(r))$ of the last stable circular orbit 
for an equal mass binary in harmonic coordinates \citep{kidder:92,wex:93}.
This difference likely results from the smaller mass binary system (2$\times$1.4\,\mo 
here compared to 2$\times$1.65\,\mo in their work; the mass is the baryonic mass 
in both cases here) and the stiffer EOS we employ (Shen EOS versus Lattimer \& Swesty EOS),
with an incompressiblity of 180\,MeV.}
\label{fig_mass_versus_rad}
\end{figure}

   In each model, the restart with VULCAN/2D is followed by a transient phase that lasts a few milliseconds 
and that is characterized by high-frequency oscillations of the highest density material, located within 10-20\,km 
of the center. Given the good match between the Shen EOS used by Rosswog and in this work (pressures differ by 
at most a few percent), we speculate that 
these oscillations are caused in part by the glitch introduced through the azimuthal averaging of the 3D SPH 
simulation snaphsot. However, they may also reflect a fundamental dynamical property of this early phase.
Similar oscillations are seen in 3D simulations performed with MAGMA, and have also been reported 
in the literature, e.g. by \cite{baiotti:08}. 
The associated shocks generate thermal energy (note that the thermal part of the pressure is sub-dominant at 
nuclear densities), but this thermal component is negligible compared, for example, with 
what is needed to power the neutrino luminosities we see. Any initial mismatch is thus 
merely a small transient which has a negligible impact on the long-term evolution
of the BNS mergers we study in this work.

\begin{figure}
\epsscale{1.2}
\plotone{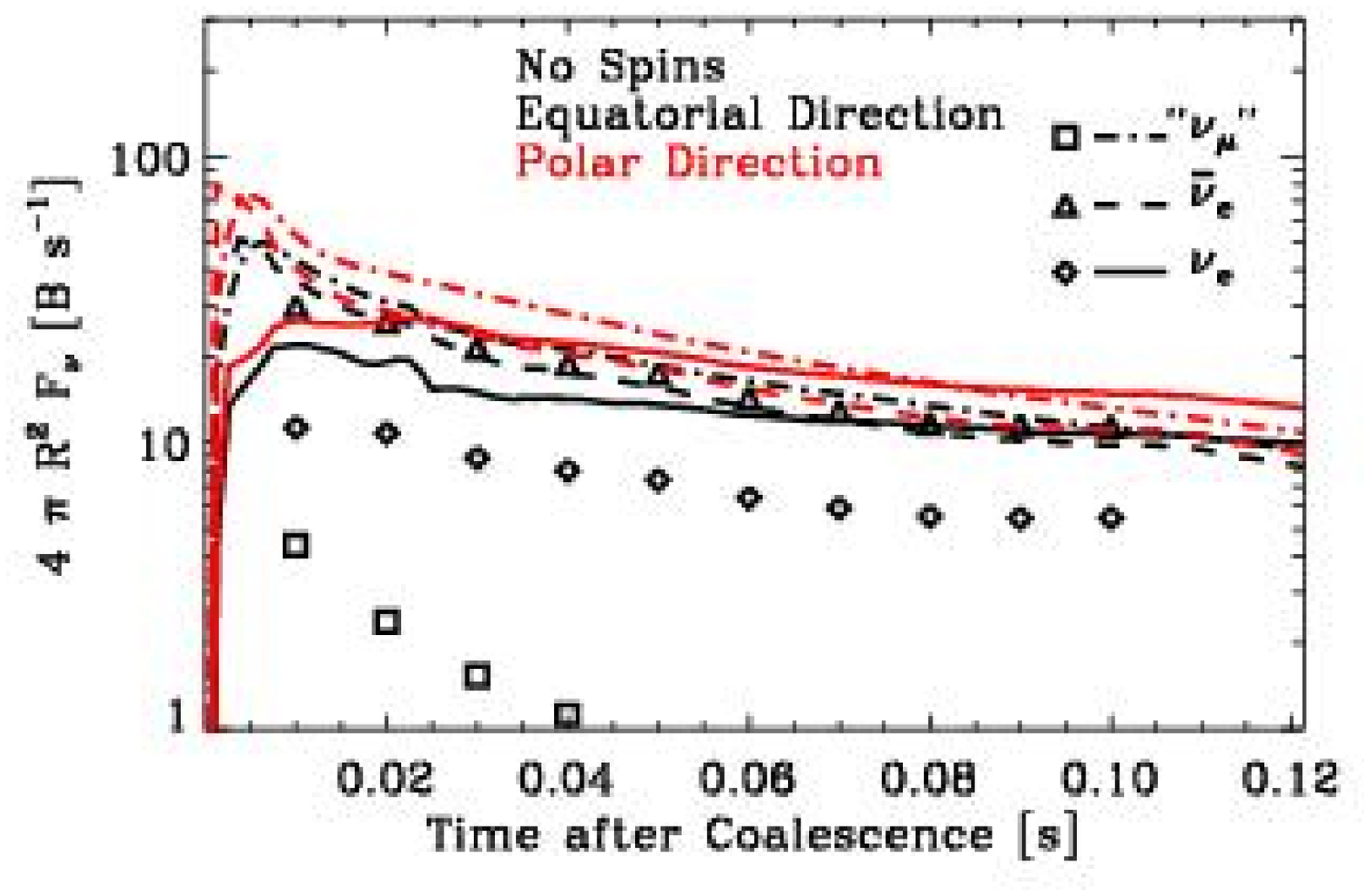}
\plotone{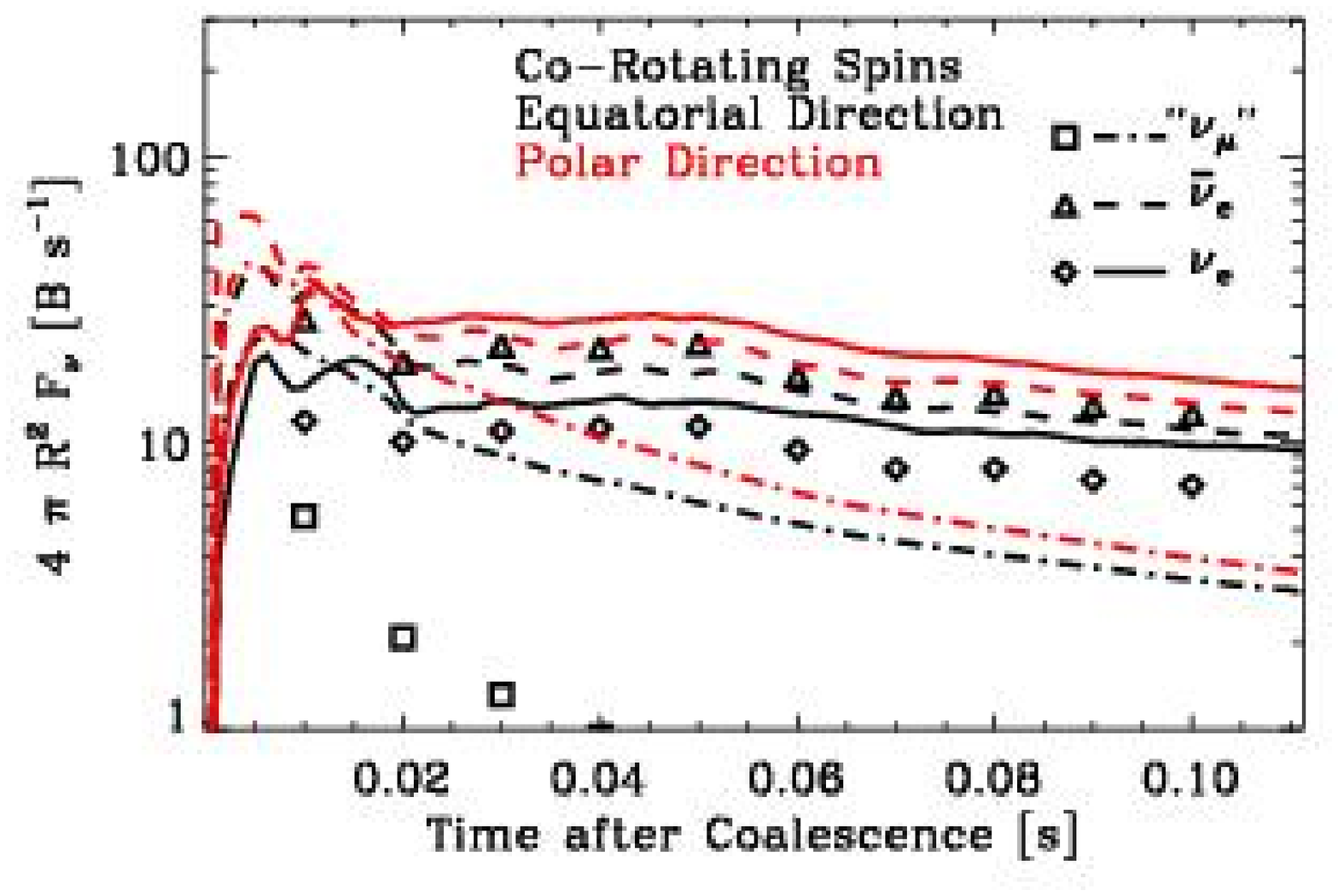}
\plotone{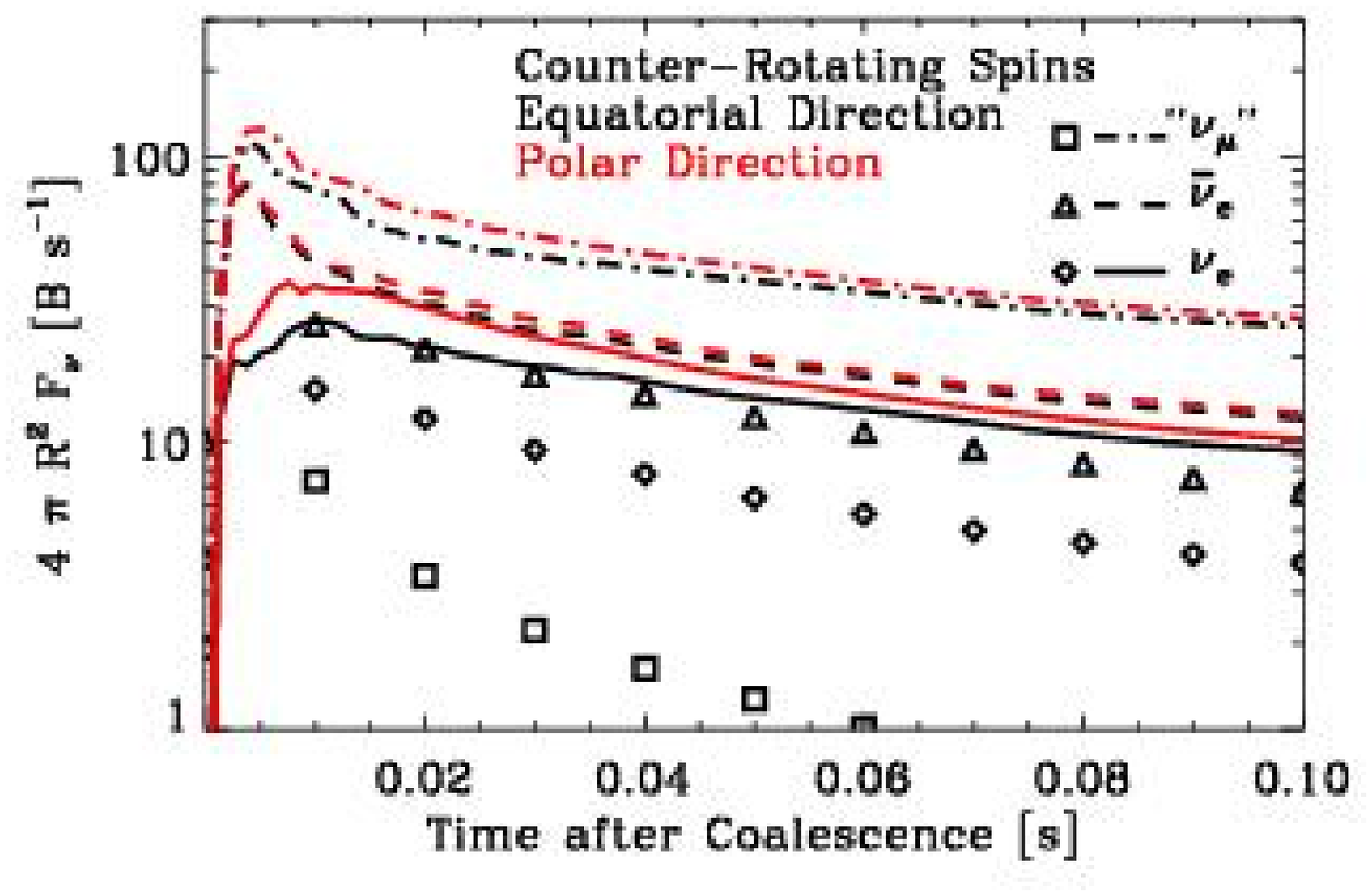}
\caption{
Time evolution of the MGFLD energy-integrated fluxes at $R=$200\,km,
scaled by a factor $4 \pi R^2$ (thus equivalent to a luminosity), for the
$\nu_e$ ({\it solid}), the ${\bar{\nu}}_e$ ({\it dashed}),
and the ``$\nu_{\mu}$'' ({\it dash-dotted}) neutrinos, plotted along the equatorial ({\it black}) 
and polar ({\it red}) directions, 
and for the BNS merger models initially 
with no spins ({\it top}), co-rotating spins ({\it middle}), and counter-rotating spins ({\it bottom}).
We also overplot the neutrino luminosities (diamonds: $\nu_e$; triangles: $\bar{\nu}_e$; 
squares: ``$\nu_\mu$'') computed in \S~\ref{annihil:leakage} using the leakage
scheme of \citet{ruffert:96}, but only for snapshots at 10\,ms intervals. 
Nucleon-nucleon bremsstrahlung processes are not treated in their
scheme and this is the likely cause of the very low ``$\nu_\mu$'' neutrino luminosity
compared to the VULCAN/2D prediction. Note that the fast rise of the VULCAN/2D neutrino 
luminosities reflects in part the light-travel time of $\sim$1\,ms to the 200\,km radius 
where the luminosity is recorded. Note that the luminosity unit used is the 
Bethe, i.e. 10$^{51}$\,erg $\equiv$ 1\,Bethe [1\,B].
}
\label{fig_flux_time}
\end{figure}

\begin{figure}
\epsscale{1.2}
\plotone{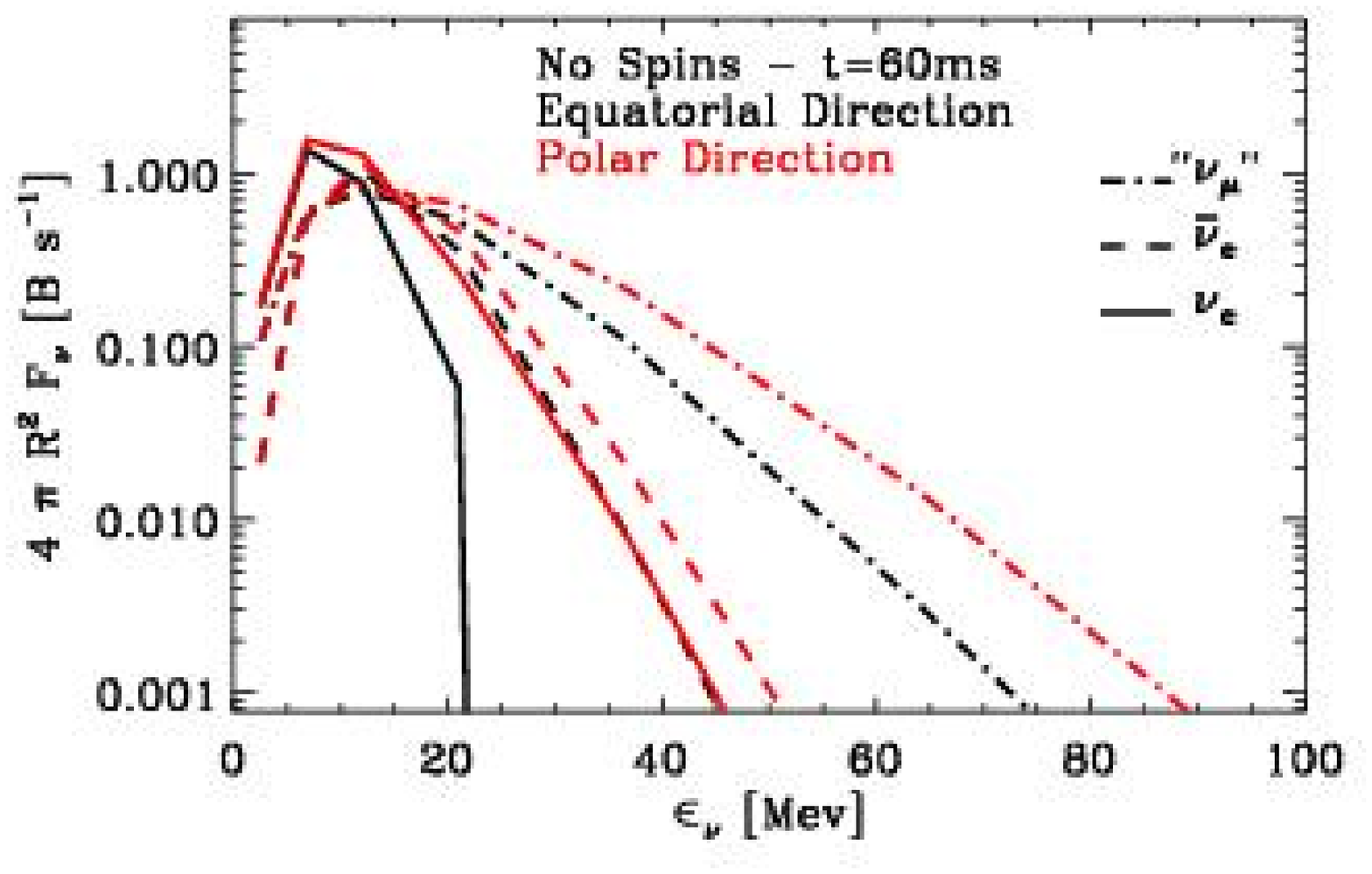}
\plotone{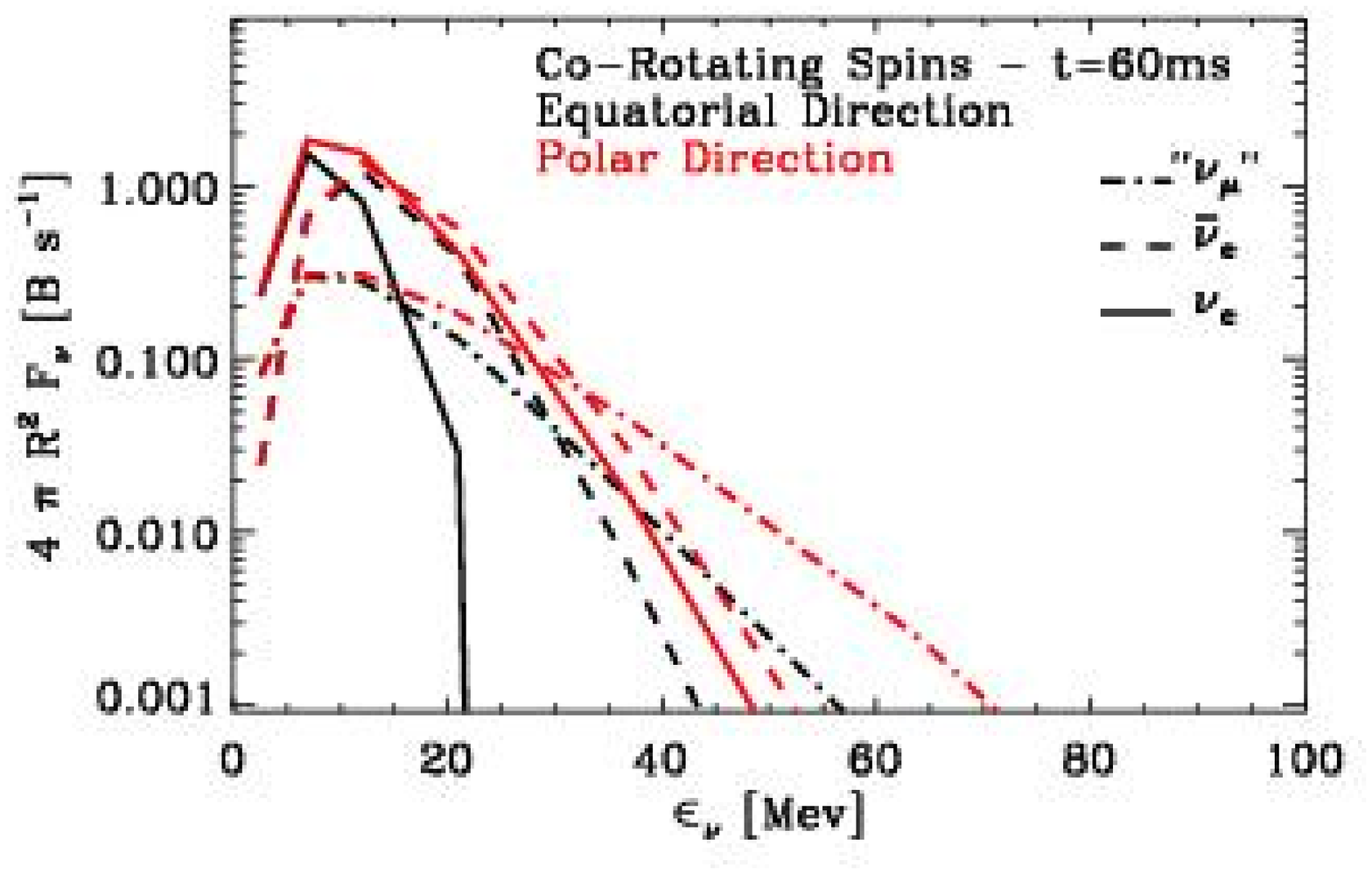}
\plotone{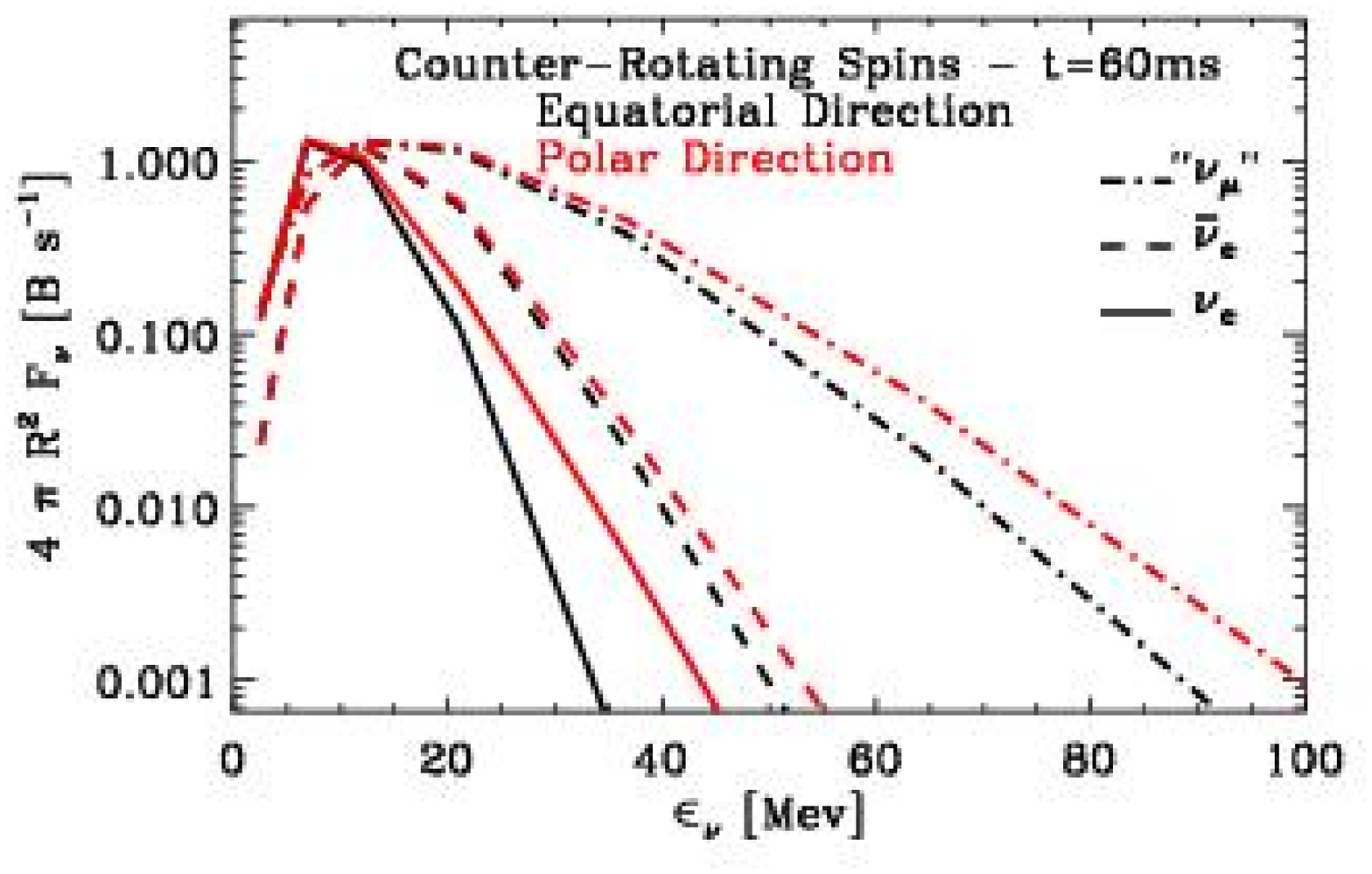}
\caption{
MGFLD flux spectra at $R=$200\,km, scaled by a factor $4 \pi R^2$ (thus equivalent 
to a luminosity spectrum), for $\nu_e$ ({\it solid}), $\bar{\nu}_e$ ({\it dashed}), and ``$\nu_{\mu}$'' 
({\it dash-dotted}) neutrinos, 
for the BNS merger models initially 
with no spins ({\it left}), co-rotating spins ({\it middle}), and counter-rotating spins ({\it right}),
and plotted along the equatorial ({\it black}) and polar
({\it red}) directions. For all models, the time corresponds to 60\,ms after the start of the
simulation.
}
\label{fig_flux_spectrum}
\end{figure}

   The evolutions of the BNS merger models with initially no spins, co-rotating spins, 
and counter-rotating spins, are qualitatively similar. As discussed above, at the start of the VULCAN simulations,
the matter distribution is far from stationary, with a significant amount of material at sub-nuclear densities
and at large distances from the center. Throughout the $\sim$100\,ms of evolution we follow,
matter in the inner few hundred kilometers settles in and comes 
to rest at the neutron star surface, while material beyond a few hundred kilometers and located
at low, near equatorial latitudes, migrates outward due to the large thermal pressure gradient
and the strong centrifugal effects. This expansion also takes place in the vertical direction, 
leading to the formation of a low-density cocoon surrounding an inner and denser disk.
In the intermediate region, i.e., at radii of 50--100\,km and along the equatorial direction,
material evolves towards quasi-Keplerian motion, with little inward or outward migration.
In Figs.~\ref{fig_temp_evol}-\ref{fig_ye_evol}, we show the $T$-$Y_{\rm e}$ distributions
at the start of the VULCAN/2D simulations ({\it top row}), and 100\,ms later ({\it bottom row}),
together with iso-density contours (shown in white) for every decade starting at a maximum value of 
10$^{14}$\,g\,cm$^{-3}$. 

  At 100\,ms after the start of the VULCAN/2D simulations, each merger has reached a 
quasi-steady equilibrium configuration, characterized by a large equator to pole 
radius ratio of $5/1$, visible from the latitudinal variation of the extent of the 
10$^{10}$\,g\,cm$^{-3}$ density contour (bottom row
panels in Fig.~\ref{fig_temp_evol}-\ref{fig_ye_evol}) or from radial slices of the density 
in the polar and equatorial directions (Fig.~\ref{fig_rho_slice}).
This is a general result for fast-rotating neutron stars (or other degenerate objects like white dwarfs) 
which has been documented in various contexts in the past
\citep{ostriker:68,hachisu:86,liu:01,yoon:05,rosswog:02a,kiuchi:08}.
At the start of the VULCAN simulations, high-density material was located 
in a thin ($\sles$20\,km) structure extending $\sles$300\,km in the equatorial direction.
After 100\,ms, 90\% of the total mass is contained within 30\,km of the center
(corresponding to $\sim$2.5\,\mo) and at densities greater than 10$^{13}$\,g\,cm$^{-3}$. 
At the same time, the remaining $\sim$10\% (corresponding to $\sim$0.2\,\mo) is located 
in a quasi-Keplerian disk with an outer edge at $\sles$100\,km 
(Fig.~\ref{fig_rho_slice}--\ref{fig_omega_slice}; see also Fig.~\ref{fig_tauz}).
This is also illustrated in Fig.~\ref{fig_mass_versus_rad} for the no-spin model 
and at 60\,ms after the start of the VULCAN/2D simulation, a figure in which 
we show the interior baryonic mass as a function of spherical radius, 
and how it compares with various multiples of the Schwarschild radius.

\subsection{Neutrino signatures}
\label{sect:neutrinos}

  Neutrino processes of emission and absorption/scattering in BNS merger 
simulations have been treated using leakage schemes by \cite{ruffert:96,ruffert:97}, 
\cite{rosswog:03a}, \cite{setiawan:04,setiawan:06} and \cite{birkl:07}. 
In this approach, the difficult solution 
of the multi-angle multi-group Bolzmann transport equation is avoided by 
introducing a neutrino loss timescale, for both energy and number, and 
applied to optically-thick and optically-thin conditions.
In addition, an estimated rate of electron-type neutrino loss allows
the electron fraction to be updated. Overall, benchmarking of these leakage schemes using more 
sophisticated 1D core-collapse simulations that solve the Boltzmann equation, 
suggests that the resulting neutrino luminosities are within a factor of a few at most 
of what would be obtained using a more accurate treatment.

  The work presented here offers a direct test of this. Moreover, because the neutrino source terms are 
included in the momentum and energy equations solved by VULCAN/2D, we can model the birth and subsequent evolution
of the resulting neutrino-driven wind. Our simulations being 2D and less CPU intensive than 3D SPH simulations,
can also be extended to $\sgreat$100\,ms, and, thus, we can investigate the long-term evolution of the merger.
In particular, we directly address the evolution of the neutrino luminosity and confront this with
the results using the expedient of a steady-state often employed in work 
that is also limited to a short time span of 10-20\,ms after the merger.  

  In the leakage schemes of \cite{ruffert:96,ruffert:97} and \cite{rosswog:03a}, the neutrino emission 
processes treated are electron and positron capture on protons (yielding $\nu_e$) and neutrons 
(yielding ${\bar{\nu}}_e$), respectively, and electron-positron pair annihilation and plasmon decay
(each yielding electron-, $\mu$-, and $\tau$-type neutrinos). For electron-type neutrinos, the 
dominant emission processes are the charged-current $\beta$-processes. For neutrino opacity,
Ruffert et al. include neutrino scattering on nucleons. \cite{rosswog:03a} also treat
electron-type neutrino absorption on nucleons.

\begin{figure}
\epsscale{1.2}
\plotone{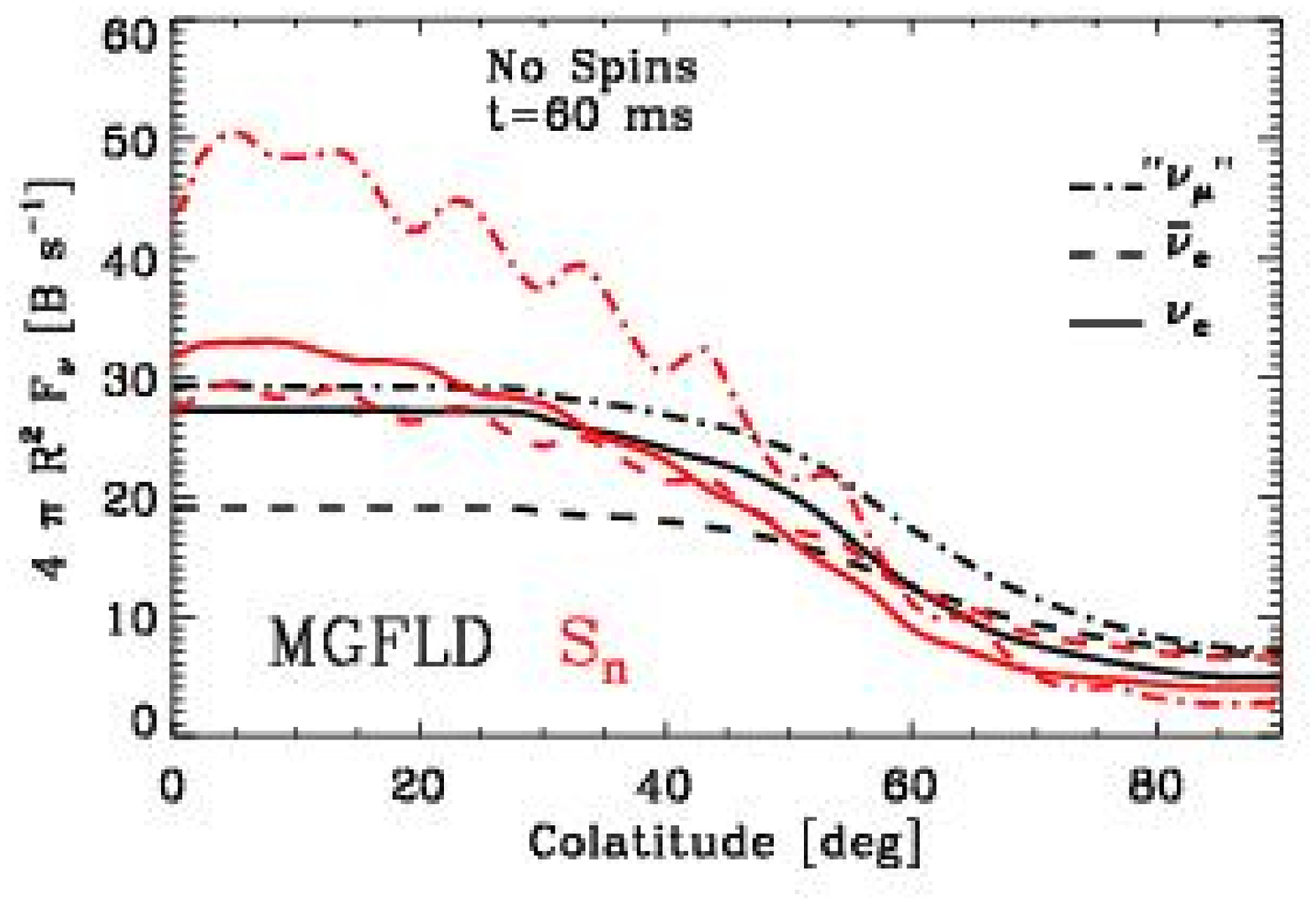}
\plotone{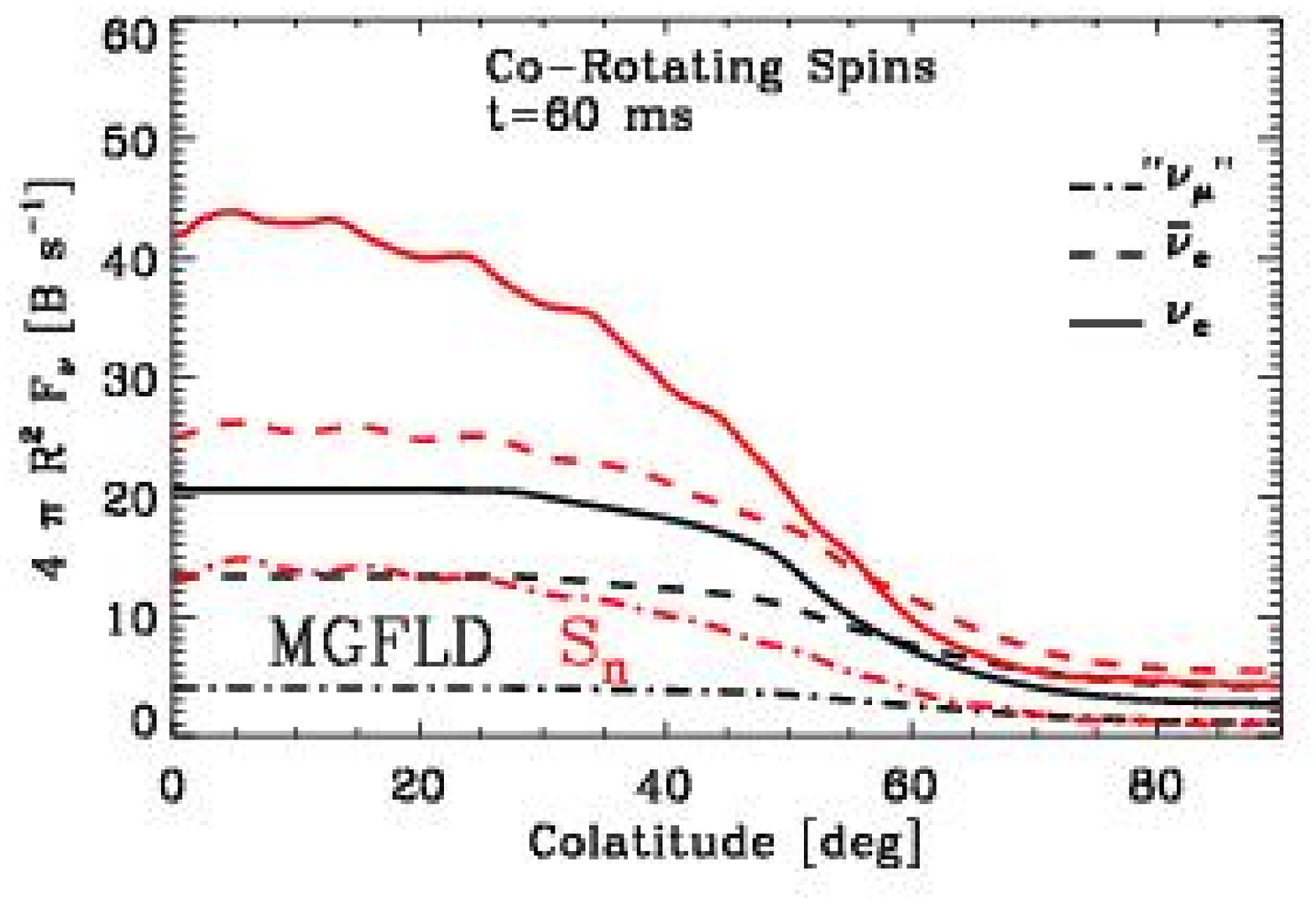}
\plotone{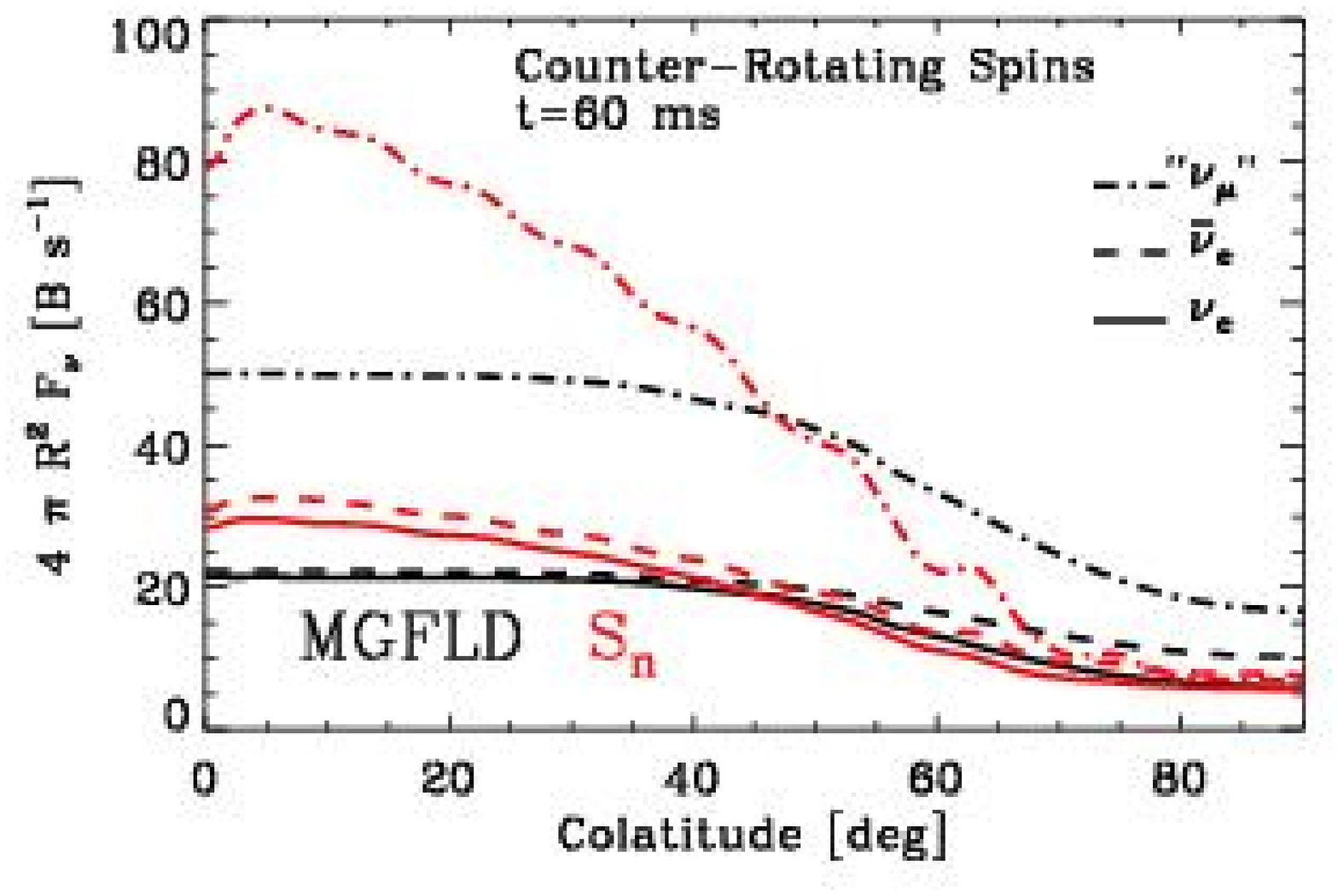}
\caption{
Angular variation from pole to equator of the energy-integrated neutrino 
fluxes at 100\,km (this radius is chosen smaller than for Fig.~\ref{fig_flux_time} 
to better reveal the flux anisotropy), scaled 
by a factor $4 \pi R^2$ (thus equivalent to a luminosity), for the
$\nu_e$ ({\it solid}), the $\bar{\nu}_e$ ({\it dashed}),
and the ``$\nu_{\mu}$'' ({\it dash-dotted}) neutrinos, 
for the BNS merger models initially 
with no spins ({\it top}), co-rotating spins ({\it middle}), and counter-rotating spins ({\it bottom}).
For each model, we show the MGFLD ({\it black}) and $S_n$ ({\it red}; 
using the 16 $\vartheta$-angle calculations) 
predictions at 60\,ms after the start of each simulation.
Note the $S_n$ oscillatory features in the neutrino flux, more prominant 
along the polar direction, which is an artifact of the $S_n$ scheme.
The range of the ordinate axis varies between frames.
}
\label{fig_flux_lat}
\end{figure}

 The neutrino emission, absorption, and scattering processes used in VULCAN/2D simulations are
those summarized in \cite{brt:06}, and include all the above, plus electron neutrino absorption on nuclei
and nucleon-nucleon bremsstrahlung. The latter is the dominant emission process of $\nu_\mu$ and $\nu_\tau$ neutrinos
in PNSs \citep{thompson:00a}. 
In VULCAN/2D, the $\nu_\mu$ and $\nu_\tau$ neutrinos are grouped together and referred to 
as ``$\nu_{\mu}$'' neutrinos. We present in Fig.~\ref{fig_flux_time} the evolution of the neutrino luminosity
for the BNS merger models with initially no spins ({\it top}), 
co-rotating spins ({\it center}), and counter-rotating spins ({\it bottom})
over a typical timespan of $\sgreat$100\,ms, and for an adopted radius of 200\,km along the  
equatorial (polar) directions in black (red).
In each case, we plot the $\nu_e$ ({\it solid line}), the ${\bar{\nu}}_e$ ({\it dashed line}), and
the ``$\nu_{\mu}$'' ({\it dash-dotted line}) neutrinos. For comparison with the predictions of the 
leakage scheme of \cite{ruffert:96}, we have implemented their formalism, following exactly the description 
given in their paper in their Appendix A \& B. We plot the corresponding results at 10\,ms intervals 
for each case (diamonds: $\nu_e$; triangles: $\bar{\nu}_e$; squares: ``$\nu_\mu$''; 
see also \S~\ref{annihil:leakage}).

\begin{figure}[t]
\centering
\plotone{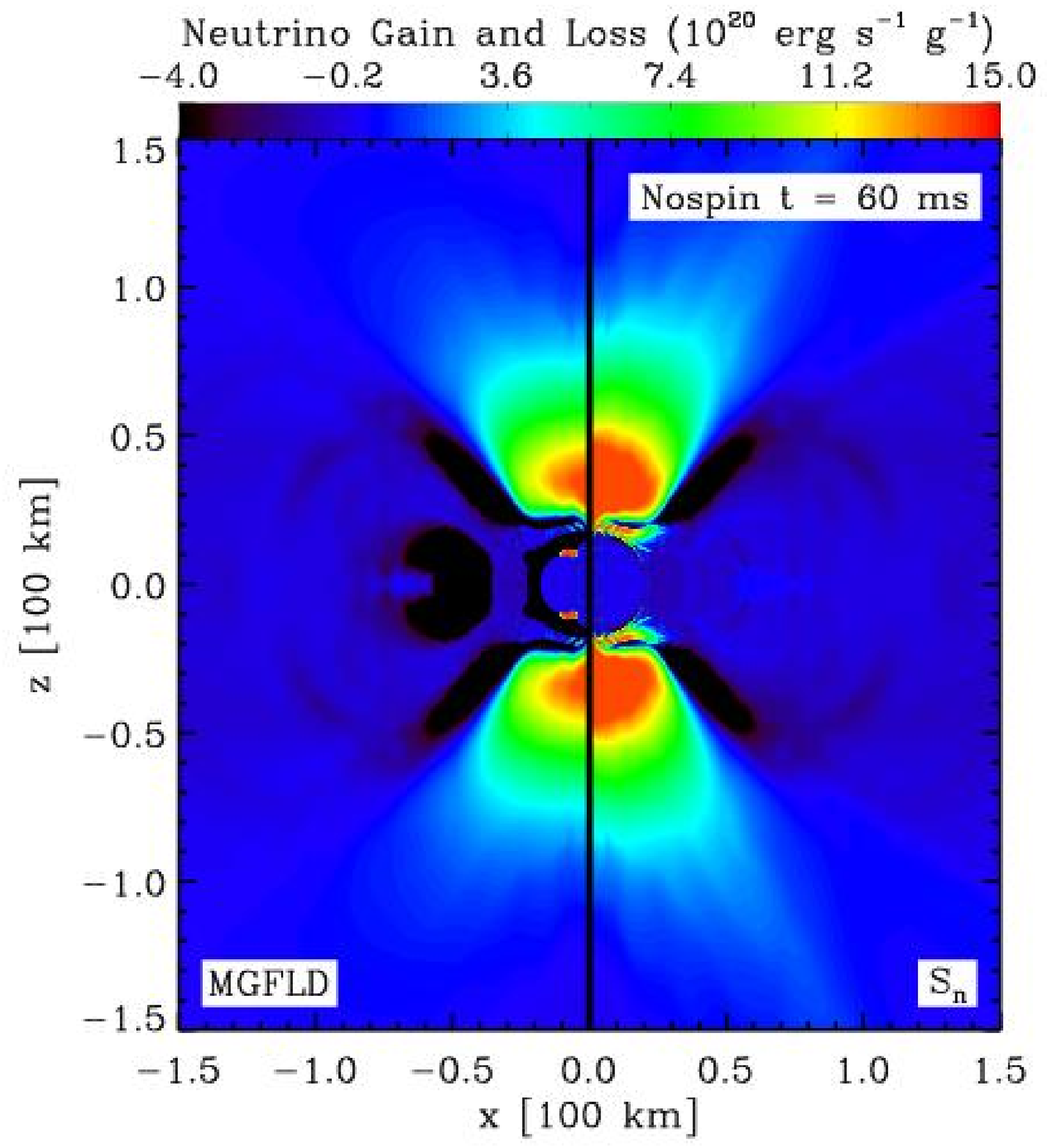}
\caption{Colormaps of energy- and species-integrated specific
neutrino energy deposition (whose volume-integrated values is refered to as $Q(\mathrm{cc})$; 
see Fig.~\ref{fig_edep_tot}) and loss rates in the BNS merger model with no initial 
spin and shown here in units of 10$^{20}$\,erg\,s$^{-1}$~g$^{-1}$).  
This snapshot corresponds to a time of 60\,ms after the start of the simulation.
The left section of the plot depicts the
MGFLD result and the right shows the result of the $S_n$
calculation. Note the distinctively enlarged polar gain regions and
greater specific gain of the S$_n$ result compared to the MGFLD
calculation. This is in part a consequence of the larger polar
neutrino fluxes and overall greater flux asymmetry in the S$_n$ model
(see Fig.~\ref{fig_flux_lat}).  
\label{fig_netgain}}
\end{figure}

  For the BNS merger models with initially no spins, co-rotating spins, and counter-rotating spins,
the neutrino signal predicted by VULCAN/2D 
follows a qualitatively similar evolution. The initial fast rise represents
the time it takes the neutrino emission processes to ramp up to their equilibrium rates, modulated
by the diffusion time out of the opaque core (and out of the not-so-opaque surface layers) 
and the light travel time of $\sim$1\,ms to the radius
of 200\,km where the luminosity is recorded. The peak luminosity occurs at about 5\,ms after the
start of the VULCAN/2D simulations, and is dominated by ${\bar{\nu}}_e$ and ``$\nu_{\mu}$'' neutrinos.
The $\nu_e$ neutrino luminosity is systematically sub-dominant initially compared with the other 
two neutrino species. This is primarily a result of the high neutron richness of the merged object.
Among the models, the stronger neutrino signal, in particular for the ``$\nu_{\mu}$'' neutrinos,
is produced in the counter-rotating model, which achieves the largest central temperature ($\sim$17\,MeV),
followed by the no-spin model ($\sim$15\,MeV). The co-rotating spin model has the weakest neutrino signal 
of all three, mainly because the tidal locking leads to a very smooth merger and correspondingly 
low temperatures (the central temperature is about 10\,MeV). 
Specifically, the angle-integrated, species-integrated, peak neutrino luminosity
for the BNS model with no initial spin is 1.5$\times$10$^{53}$ (2.2$\times$10$^{52}$, 
7.0$\times$10$^{52}$, and 6.7$\times$10$^{52}$), for the co-rotating spin model 1.2$\times$10$^{53}$
(2.2$\times$10$^{52}$, 5.7$\times$10$^{52}$, and 3.9$\times$10$^{52}$), and for the counter-rotating
spin model 2.2$\times$10$^{53}$\,erg\,s$^{-1}$ (2.9$\times$10$^{52}$, 8.2$\times$10$^{52}$, and 1.2$\times$10$^{53}$).
The values given in parentheses correspond in each case to the angle-integrated peak luminosity for the 
$\nu_e$, the $\bar{\nu}_e$, and the ``$\nu_\mu$'' neutrinos, respectively. 
Apart from a significant discrepancy with the ``$\nu_\mu$''-neutrino luminosities, 
the peak neutrino luminosities using our MGFLD approach compare well with those published 
in the literature based on a leakage scheme \citep{ruffert:97,rosswog:03a}. However,
in all cases, cooling of the BNS merger causes a decrease of all neutrino luminosities after the peak 
(i.e., rather than a long-term plateau), with a faster decline rate in the no-spin and counter-rotating 
spin cases. The luminosity decay rate is faster for the ``$\nu_{\mu}$'' neutrinos than for the electron-type
neutrinos. This translates into a strong decrease of the neutrino-antineutrino annihilation rate 
due to its scaling with the square of the neutrino luminosity (see \S~\ref{sect:annihilate}).
In agreement with past work that focused on the initial 10-20\,ms after the onset of coalescence, 
we find that the electron antineutrino luminosity dominates over the electron neutrino luminosity, 
but in contrast with the work of \cite{ruffert:96} and \cite{rosswog:03a}, our ``$\nu_{\mu}$'' neutrino
luminosities are at least one order of magnitude higher. 
We associate this discrepancy with their neglect of the nucleon-nucleon bremsstrahlung 
opacity and emission processes.
Moreover, in the no-spin and co-rotating spin cases, the relationships between the fluxes of
the various neutrino flavors changes significantly with time, partly because they are differently 
affected by neutron star cooling and changes in neutron richness.
While the decrease of the luminosity predicted by VULCAN/2D makes physical sense,
it contrasts with the findings of \cite{setiawan:04,setiawan:06} in their study of neutrino
emission from torus-disks around $\sim$4\,\mo black holes. In their models, they incorporate
physical viscosity in the disk, together with the associated heat release, which keeps
the gas temperature high. In our models, neutrino emission is a huge energy sink, 
not compensated by $\alpha$-disk viscosity heating.

The MGFLD treatment we use permits the calculation of the 
energy-dependent neutrino spectrum and its variation with angle. 
For the BNS merger models with initially no spins ({\it left}), 
co-rotating spins ({\it middle}), and counter-rotating spins ({\it right}),
we show such spectra along the equatorial (black) and polar (red) directions in Fig.~\ref{fig_flux_spectrum},
using a reference radius of 200\,km.
We can also explicitly compute average neutrino energies, here defined as

\begin{eqnarray}
 \sqrt{\langle\varepsilon_{\nu}^2\rangle} \equiv
\left[
    \frac{\int d\varepsilon_{\nu} \varepsilon_{\nu}^2 F_{\nu}(\varepsilon_{\nu},R)}
         {\int d\varepsilon_{\nu} F_{\nu}(\varepsilon_{\nu},R)}
\right]^{\frac{1}{2}}\,.
\end{eqnarray}

At 50\,ms after the start of the VULCAN/2D simulations, such average neutrino energies in the no-spin
BNS model and along the pole are 13.4 ($\nu_e$), 16.7 ($\bar{\nu}_e$), and 25.2\,MeV (``$\nu_\mu$''), 
while they are 10.7, 15.6, and 21.2\,MeV along the equatorial direction (in the same order).
In the co-rotating model and along the polar direction, we have
14.0 ($\nu_e$), 17.0 ($\bar{\nu}_e$), and 22.0\,MeV (``$\nu_\mu$''), and along the equatorial
direction, we have in the same order 9.9, 15.2, and 17.5\,MeV.
In the counter-rotating model and along the polar direction, we have
12.9 ($\nu_e$), 17.1 ($\bar{\nu}_e$), and 27.5\,MeV (``$\nu_\mu$''), and along the equatorial
direction, we have in the same order 11.5, 16.7, and 25.1\,MeV. The oblateness of the compact 
massive BNS merger, with its lower temperature and its more gradual density fall-off along the equator,
systematically places the neutrinospheres in lower temperature regions,
translating into softer neutrino spectra.

\begin{figure*}
\epsscale{0.35}
\plotone{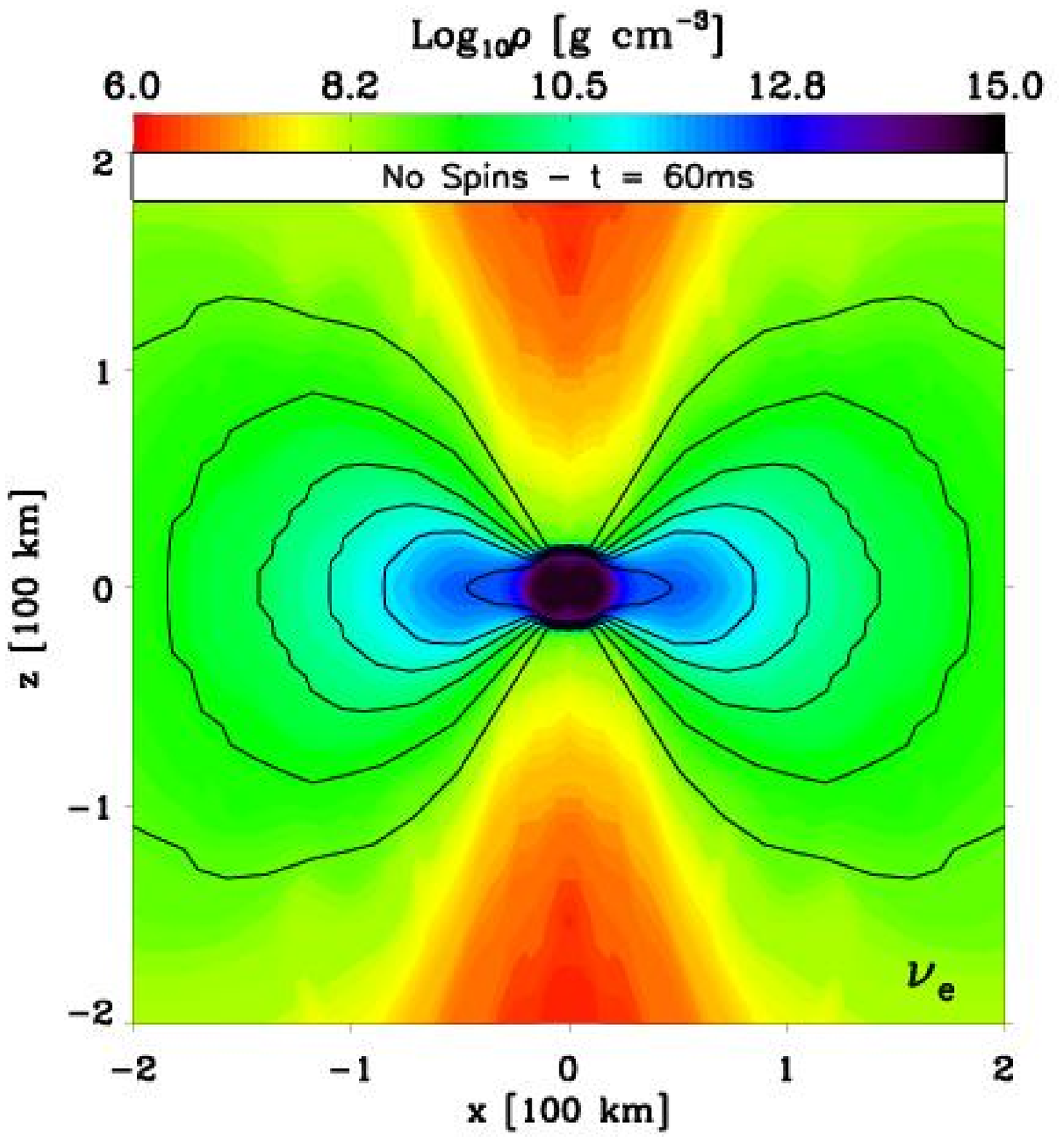}
\plotone{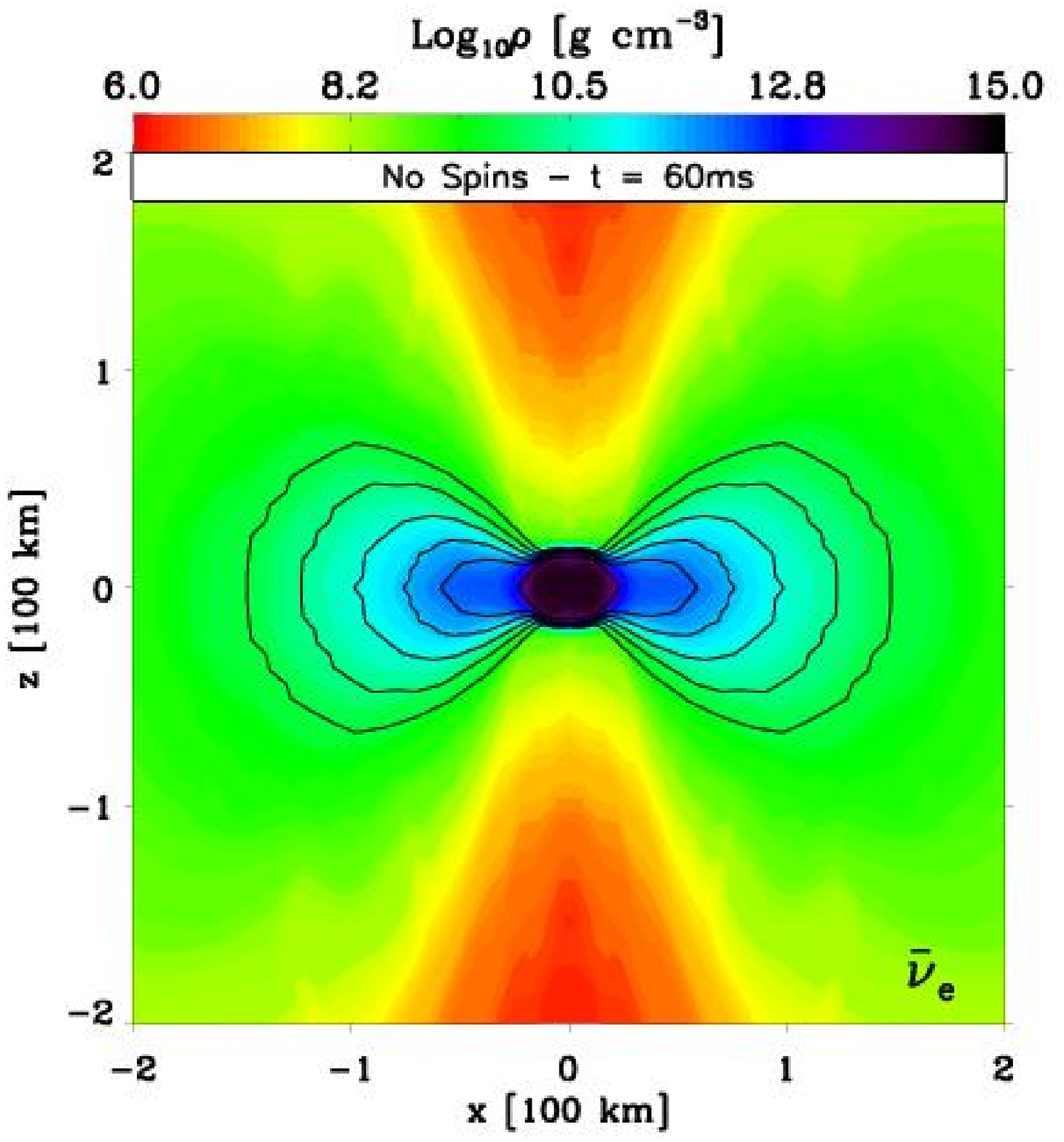}
\plotone{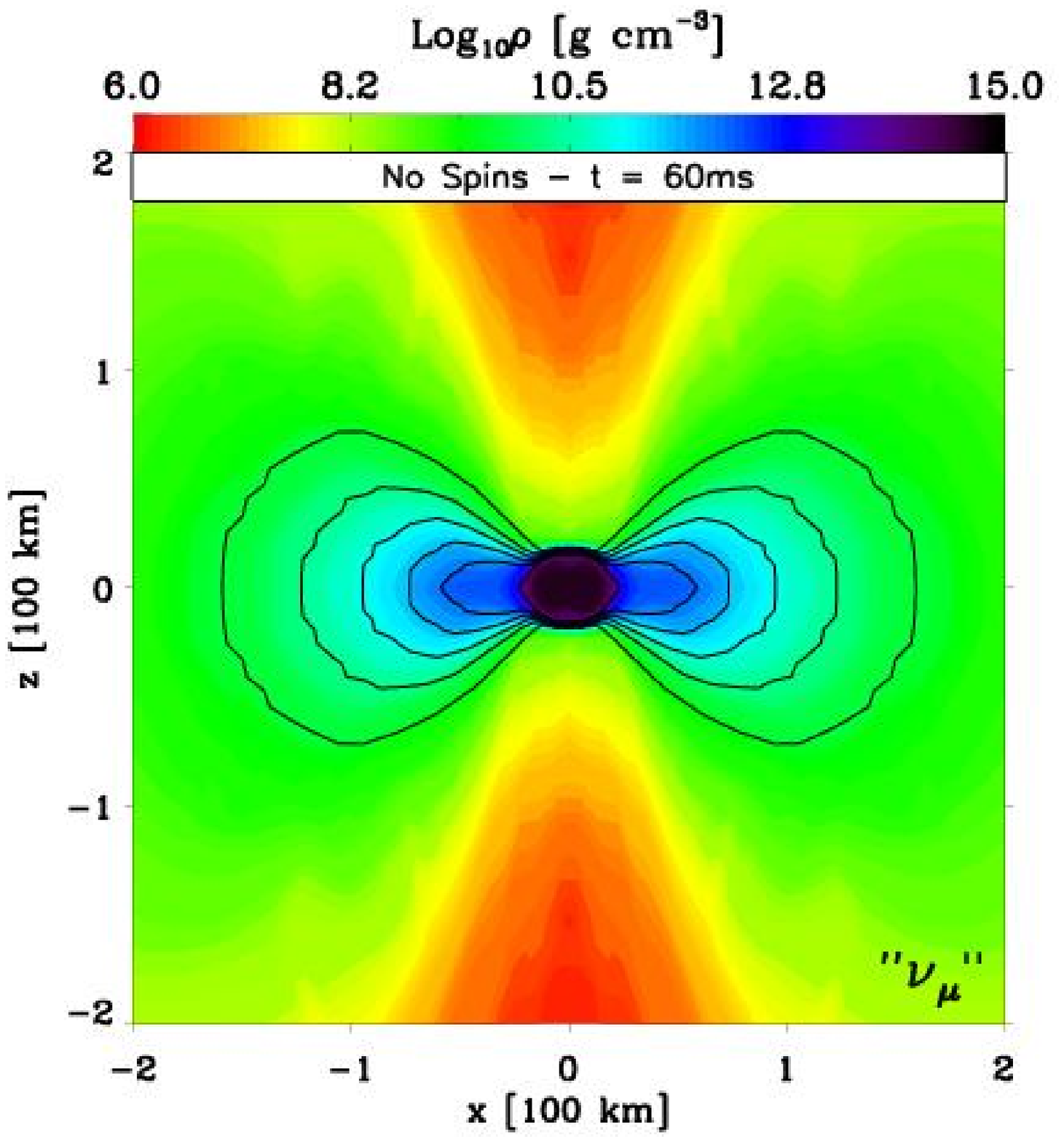}
\plotone{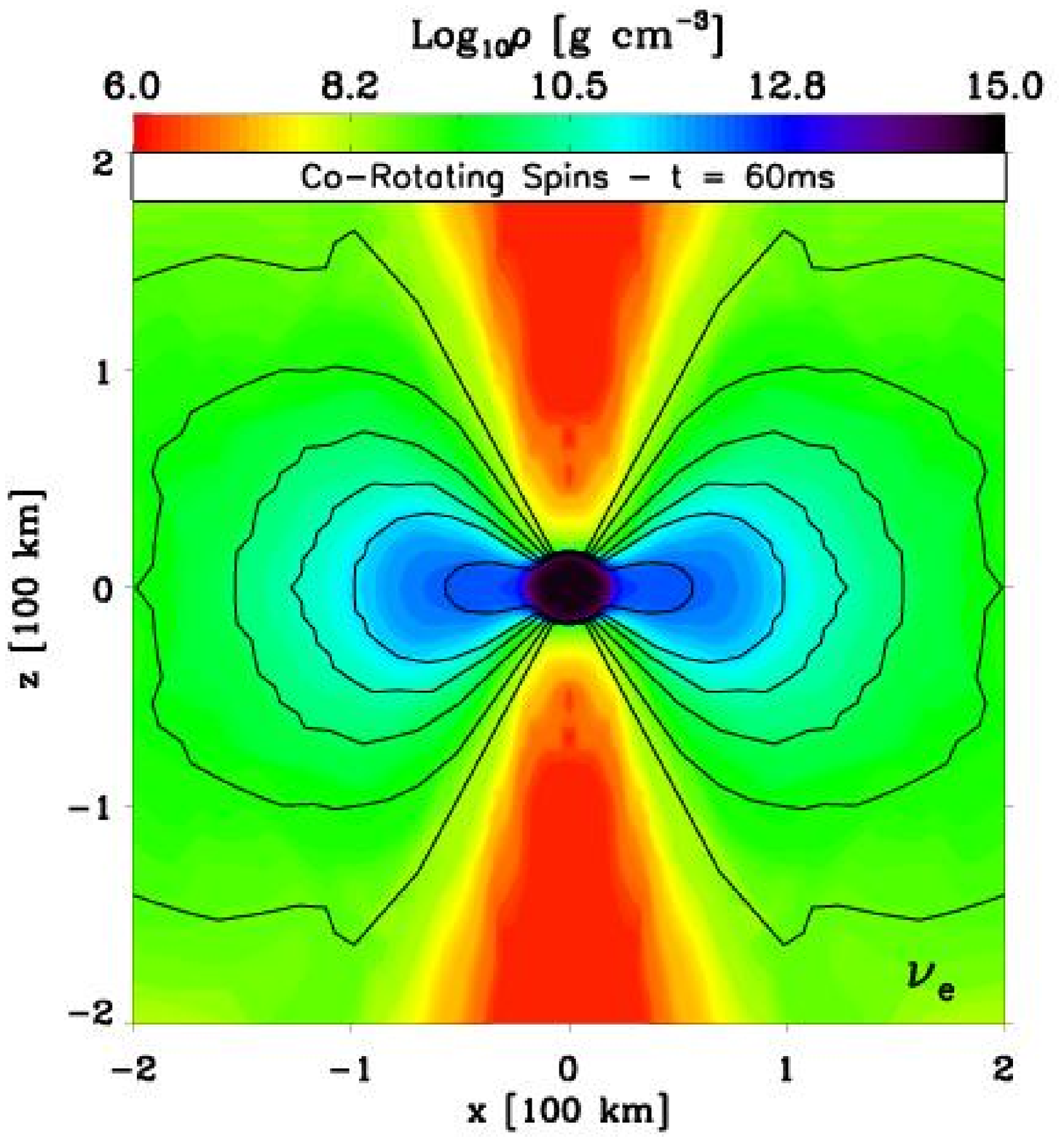}
\plotone{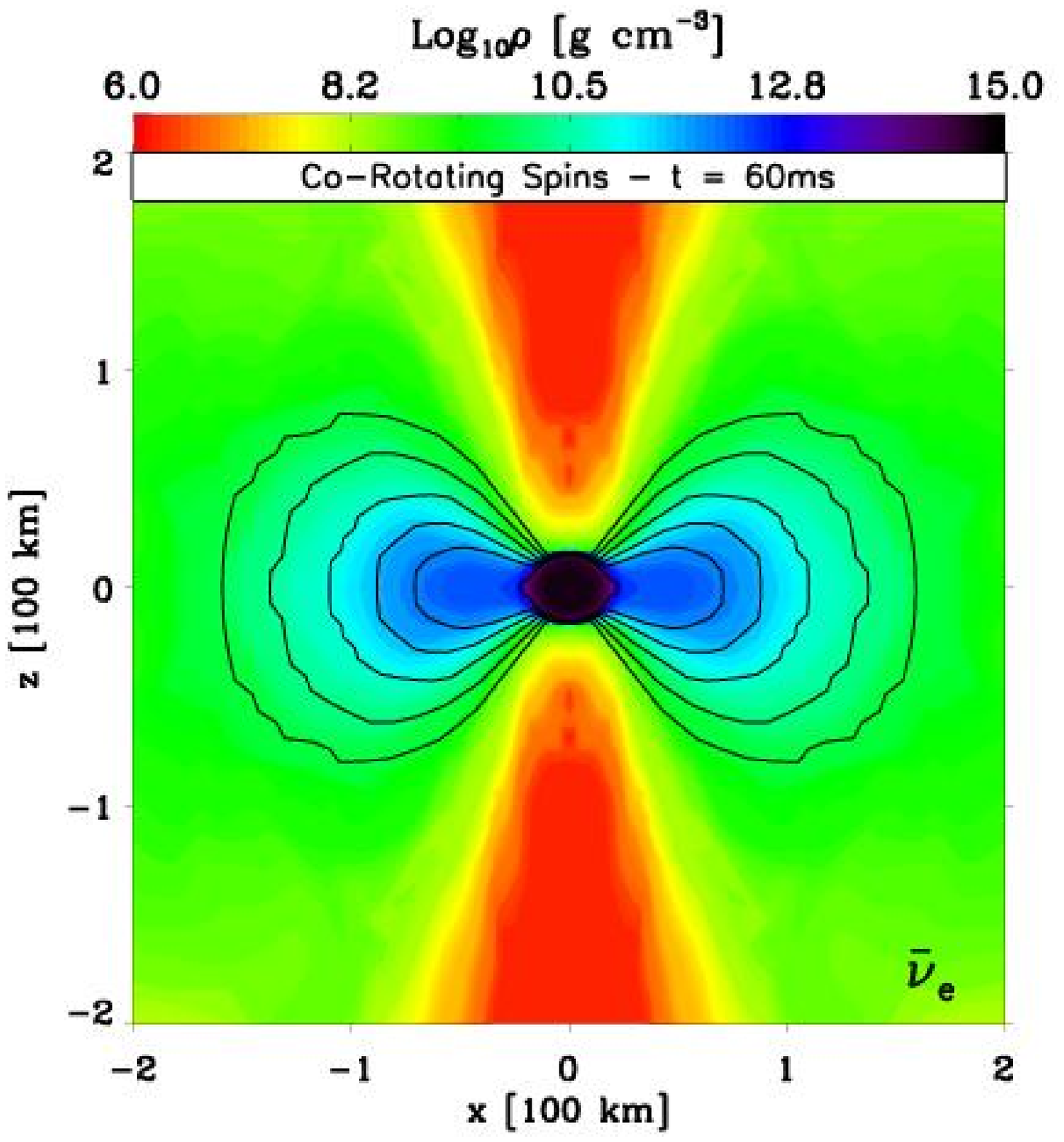}
\plotone{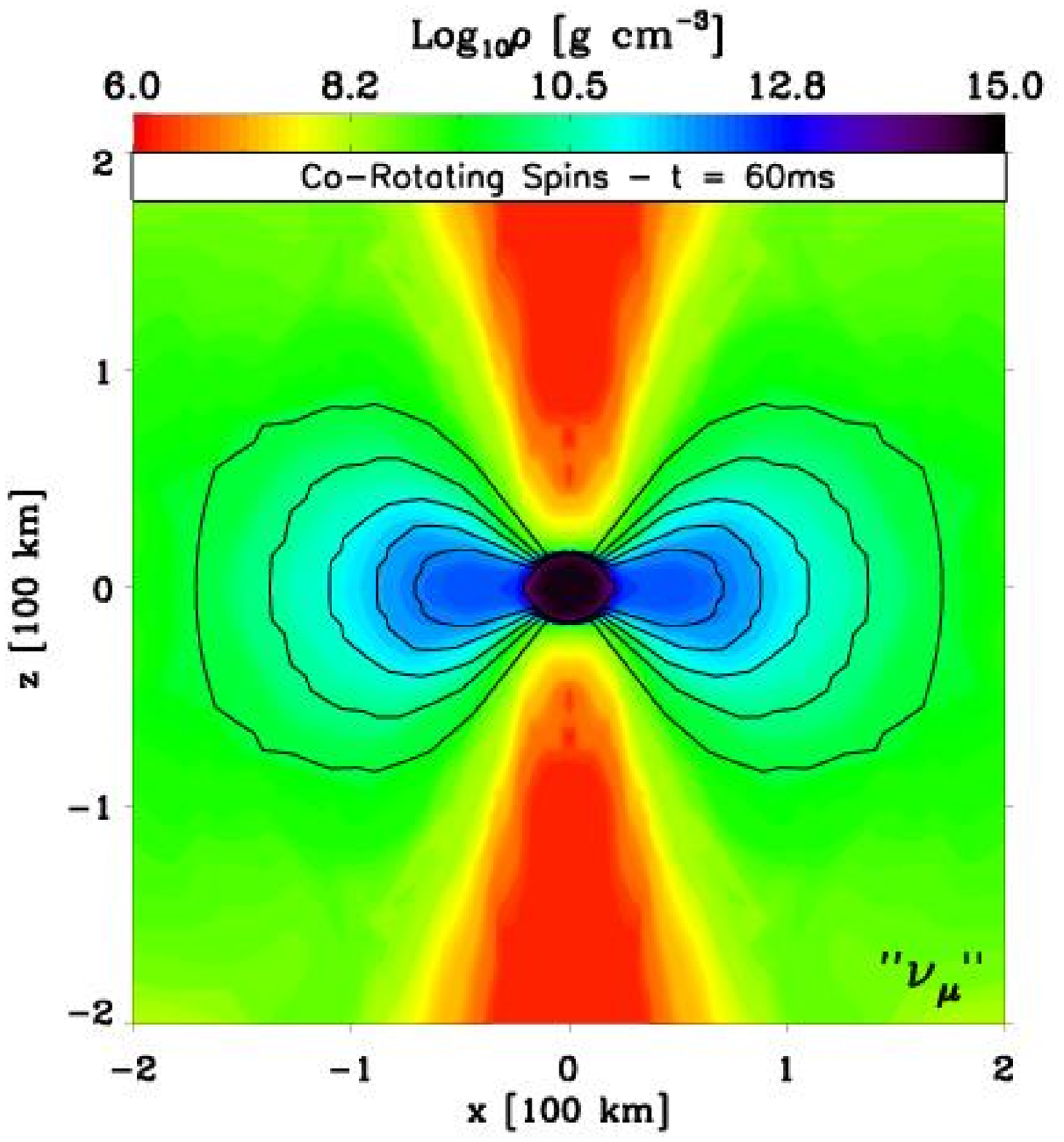}
\plotone{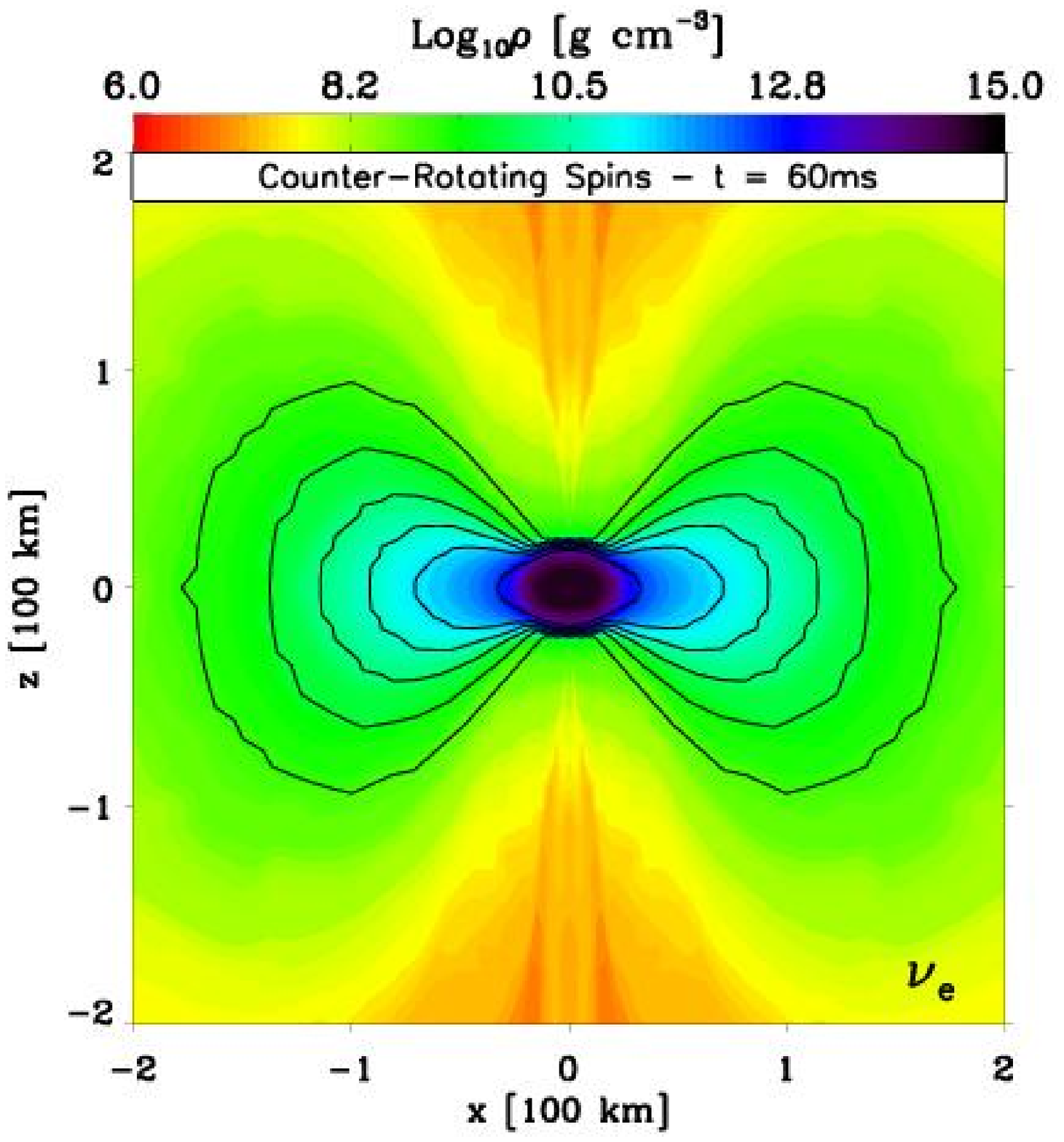}
\plotone{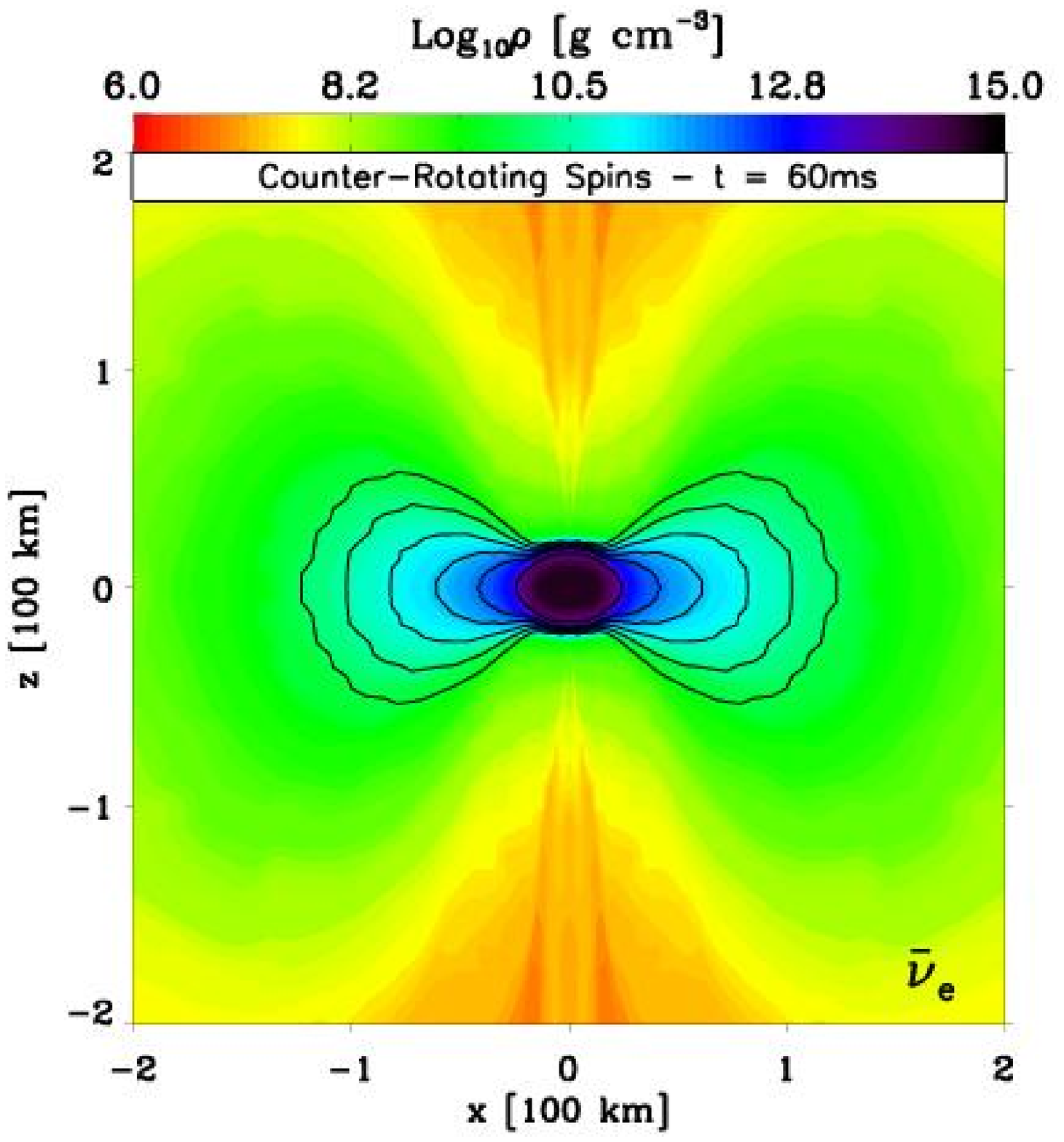}
\plotone{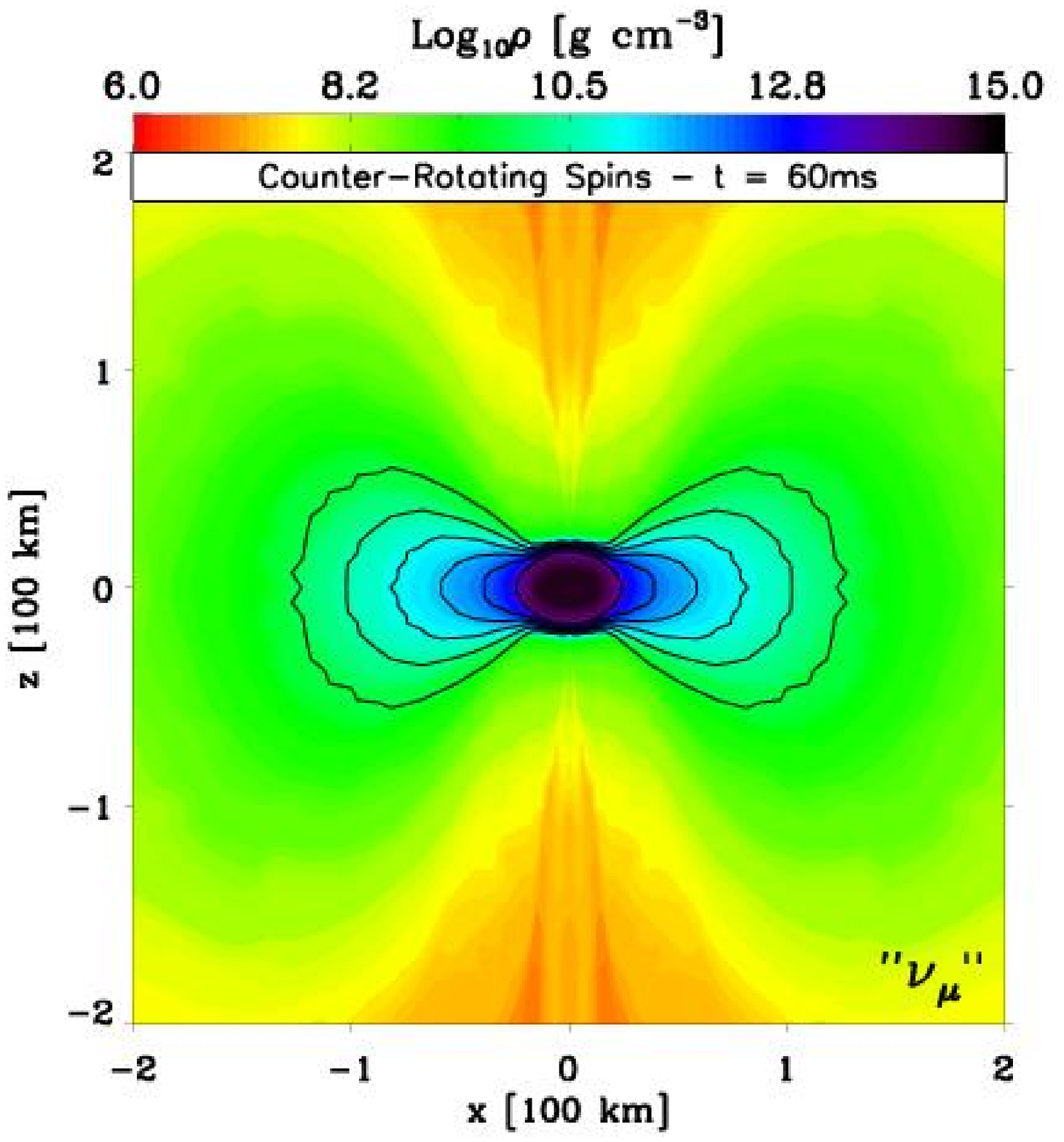}
\caption{
Using a colormap of the density distribution (shown in log scale) 
at 60\,ms after the start of the VULCAN/2D simulations 
for the BNS merger models with initially no spins ({\it top row}), co-rotating spins 
({\it middle row}), and counter-rotating spins ({\it bottom row}), we overplot in each case
the contours corresponding to the energy-dependent and latitude-dependent neutrinosphere 
radii at neutrino energies $\varepsilon_\nu$ of 
2.50, 6.87, 12.02, 21.01, 36.74, 64.25\,MeV, for the $\nu_{\rm e}$ ({\it left column}), 
$\bar{\nu}_{\rm e}$ ({\it middle column}), and ``$\nu_{\mu}$'' ({\it right column}) neutrinos.
Corresponding radii (along a given latitude) increase monotonically with energy 
(matter opacity to neutrinos scales as $\sim\varepsilon_\nu^2$). For these calculations, 
the optical depth is integrated inwards along a fixed latitude starting at the maximum spherical 
radius of 3000\,km. 
}
\label{fig_nu_sphere}
\end{figure*}

  The aspherical density/temperature distributions naturally lead to a strong anisotropy of the radiation
field, whose latitudinal dependence is shown in Fig.~\ref{fig_flux_lat} for the three models (we use the same left
to right ordering), and at a time of 60\,ms after the start of the VULCAN/2D simulations. 
Note the stronger neutrino flux along the polar direction (see also \citealt{rosswog:03a}, 
their Fig.~12), resulting from the 
larger radiating surface seen from higher latitudes, as well as the systematically larger matter temperatures
at the decoupling region along the local vertical (also yielding a harder neutrino spectrum).
In Fig.~\ref{fig_flux_lat}, we include in red the corresponding neutrino fluxes computed 
with the 16 $\vartheta$-angle $S_n$ scheme,
which shows comparable fluxes along the equator, but considerably larger ones at higher latitudes.
We also show in Fig.~\ref{fig_netgain} the energy- and species-integrated specific
neutrino energy deposition and loss rates in the no-spin BNS
model at 60\,ms after the start of the simulation, for both the MGFLD and the $S_n$ calculations.
Notice how much more emphasized the anisotropy is with the multi-angle, $S_n$, solver, and how
enhanced is the magnitude of the deposition along the polar direction.
This is a typical property seen for fast-rotating PNSs simulated with such 
a multi-angle solver \citep{ott:08}.

\begin{figure}
\epsscale{1.2}
\plotone{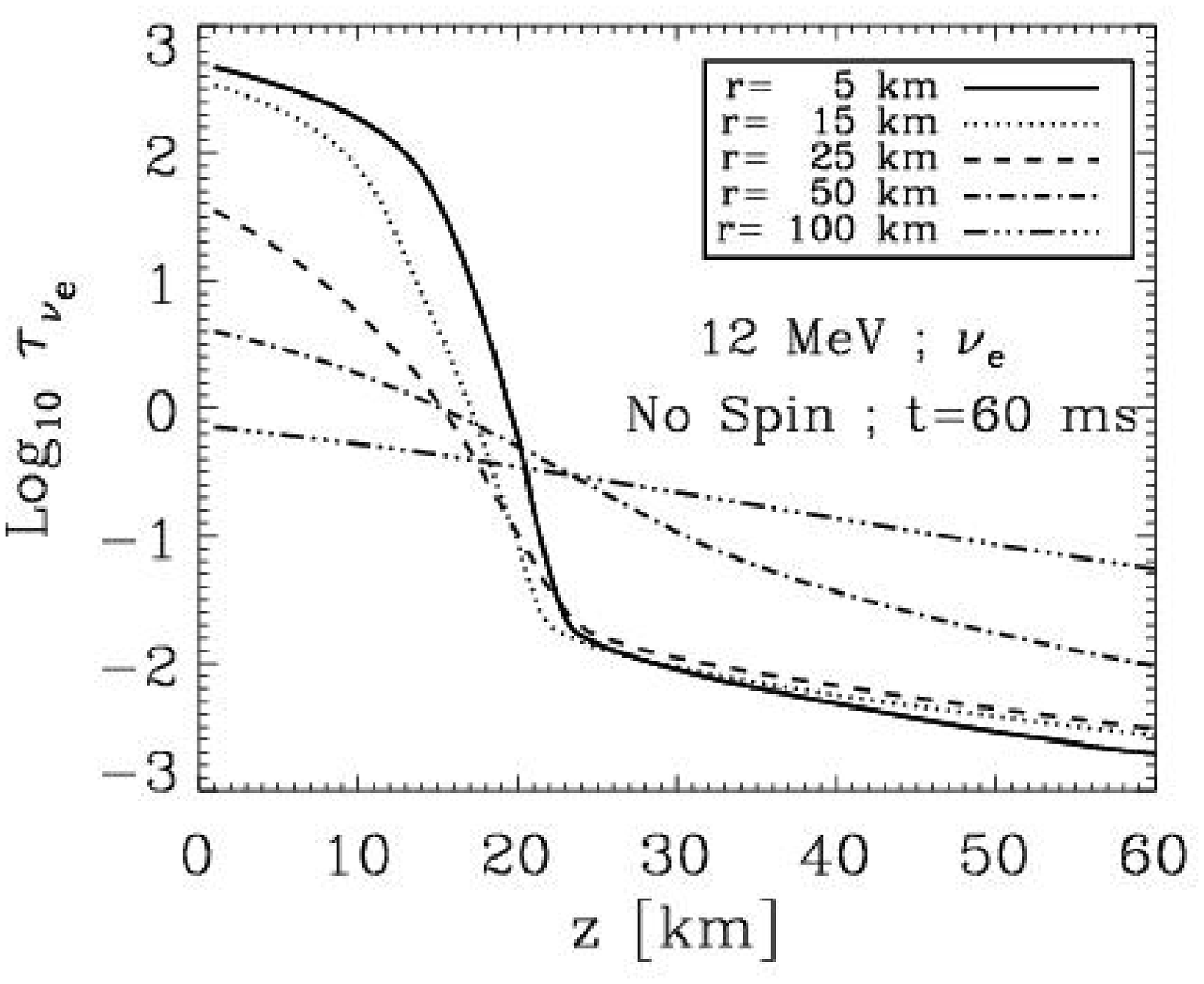}
\caption{Variation of the log of the optical depth along the $z$ direction for the $\nu_{\rm e}$ 
neutrino at 12.02\,MeV in the BNS merger model with no initial spins,
plotted for a selection of cylindrical radii, 5, 15, 25, 50, and 100\,km (see legend 
for the corresponding linestyles).
The time corresponds to 60\,ms after the start of the VULCAN/2D simulations.
Notice that the disk transitions from optically-thick to optically-thin conditions at 
$\sim$100\,km, and that optical depth increases to a few hundreds only inside the SMNS.
For most of the disk, the diffusion time (eq.~\ref{eq:tdiff}) for 12.02\,MeV $\nu_{\rm e}$ 
neutrinos is on the order  of a few milliseconds. The disk is therefore only moderately
optically-thick for neutrinos with energies at the peak of the spectral energy distribution.
}
\label{fig_tauz}
\end{figure}

In Fig.~\ref{fig_nu_sphere}, we show the energy-dependent neutrinosphere radii 
for the $\nu_{\rm e}$ ({\it left}), $\bar{\nu}_{\rm e}$ ({\it middle}), and ``$\nu_{\mu}$'' ({\it right}) neutrinos, 
and for the BNS merger models with initially no spins ({\it top row}), 
co-rotating spins ({\it middle row}), and counter-rotating spins ({\it bottom row}).
These decoupling radii manifest a strong variation with latitude, but also with energy due to the approximate 
$\varepsilon_\nu^2$ dependence of the material opacity to neutrinos.
The diffusion of neutrinos out of the opaque core, which leads to global cooling of the neutron star,
is also energy dependent.
The location-dependent diffusion timescale $t^{\rm diff}_{\nu_i}(r,z)$ is given by \citep{mihalas:84,ruffert:96}:

\begin{eqnarray}
\label{eq:tdiff}
 t^{\rm diff}_{\nu_i}(r,z) \approx 3 \tau_{\nu_i} \Delta R / c \,, 
\end{eqnarray}

where $c$ is the speed of light, $\Delta R$ is the local density scale height, $\tau_{\nu_i}$ is the optical depth 
(species and energy dependent), and $\nu_i$ is the neutrino species (i.e., $\nu_e$, $\bar{\nu}_e$, or ``$\nu_{\mu}$'').
For the $\nu_e$ neutrinos, we find an optical depth in the core that varies from 
a few $\times$10$^2$ for 10\,MeV neutrinos to a few $\times$10$^4$ for 100\,MeV neutrinos. 
These optical depths 
translate into diffusion times that are on the order of 10\,ms to 1\,s.
The quasi-Keplerian disk that extends from 20--30 to 100\,km 
is moderately optically-thick to neutrinos at the peak of the energy distribution (i.e., at 10-20\,MeV).
Heat can thus leak out in the vertical direction over a typical diffusion time of a few tens 
of milliseconds (see Fig.~\ref{fig_tauz}). This is comparable to the accretion time scale of the disk itself 
\citep{setiawan:04,setiawan:06}, which suggests that this neutrino energy is available
to power relativistic ejecta, and is not advected inwards with the gas, as proposed by \cite{dimatteo:02}. 

  Compared to PNSs formed in the ``standard'' core-collapse of a massive star, SMNSs 
formed through coalescence have a number of contrasting properties: 1) most of the material is 
at nuclear density, unlike in PNSs where newly accreted material deposited at the PNS surface 
radiates its thermal energy as it contracts and cools; 2) most of the material is neutron rich, 
thereby diminishing the radiation of electron-type neutrinos in favor of antielectron-type neutrinos. 
In PNSs, electron-type neutrinos are an important cooling agent, but have large opacities. In SMNSs,  
the material is strongly deleptonized and electron-type neutrinos are non-degenerate. They
suffer relatively lower opacity and can therefore diffuse out more easily.
The opacity of other neutrino species is even lower and therefore more prone to energy leakage through
radiation; 3) energy gain through accretion is modest since there is merely a few 0.1\,\mo of material 
outside of the SMNS. In our three merger models, the internal energy budget of such shear-heated 
SMNSs is $\sgreat$10$^{53}$\,erg, half of which is thermal energy. With a total neutrino
luminosity of $\sim$5$\times$10$^{52}$\,erg\,s$^{-1}$, the cooling time scale for these
SMNS should be on the order of a second. This is significantly, but not dramatically, shorter
than the 10-30\,s cooling time scale of hot PNSs, i.e the time it takes these objects to radiate
a total of $\sim$3$\times$10$^{53}$erg of internal energy. 

A fraction of the diffusing core energy is absorbed through charge-current reactions (with associated 
volume-integrated energy $\dot{E}(\mathrm{cc})$) and turned into internal energy in a 
``gain'' layer at the surface of the SMNS, primarily on the pole-facing side and at high-latitudes 
(see Fig.~\ref{fig_netgain}). This neutrino energy deposition is the origin of a thermally-driven 
wind whose properties we now describe.

\subsection{Neutrino-driven wind}
\label{sect:dynamics}

\begin{figure*}
\epsscale{0.5}
\plotone{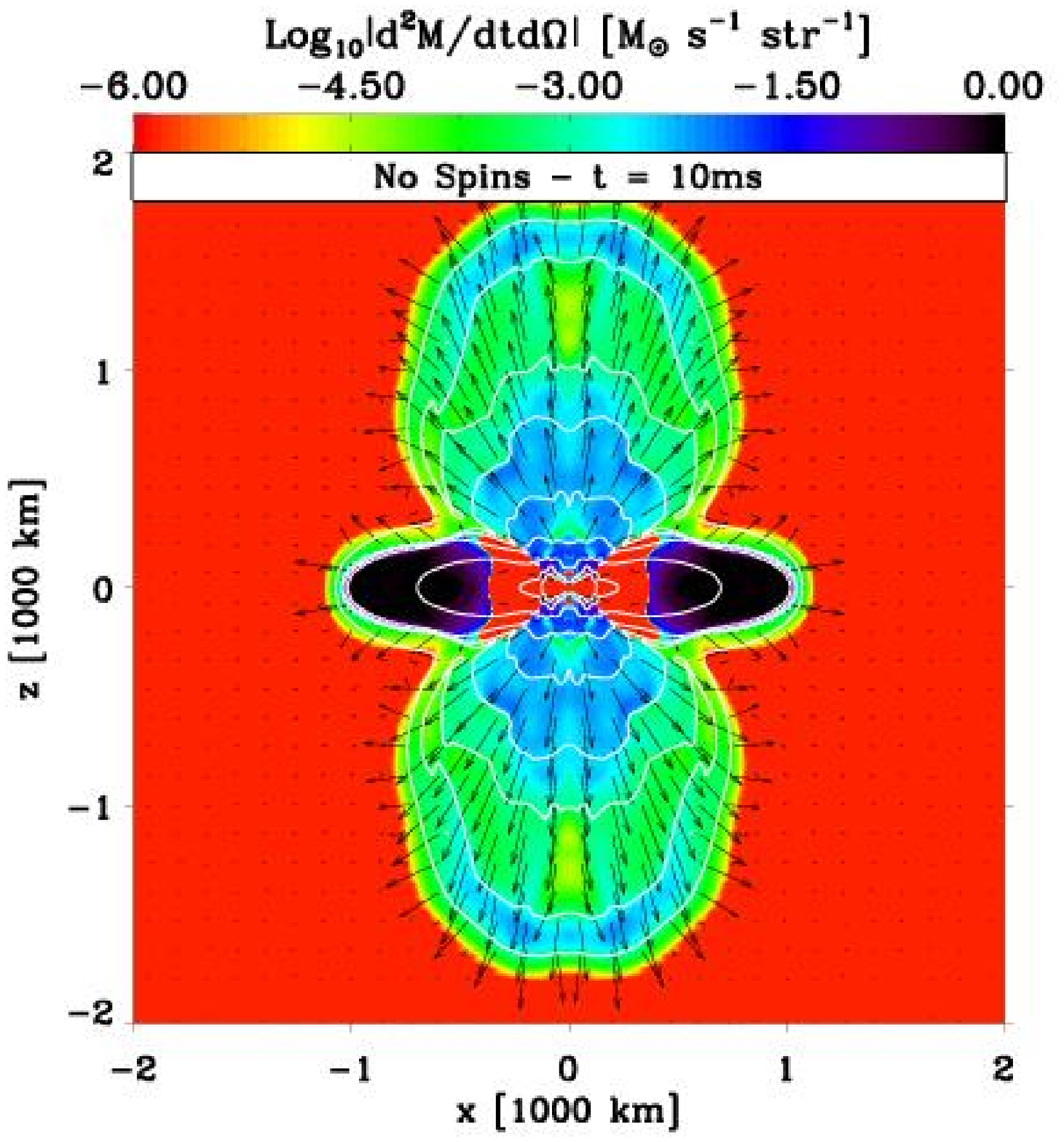}
\plotone{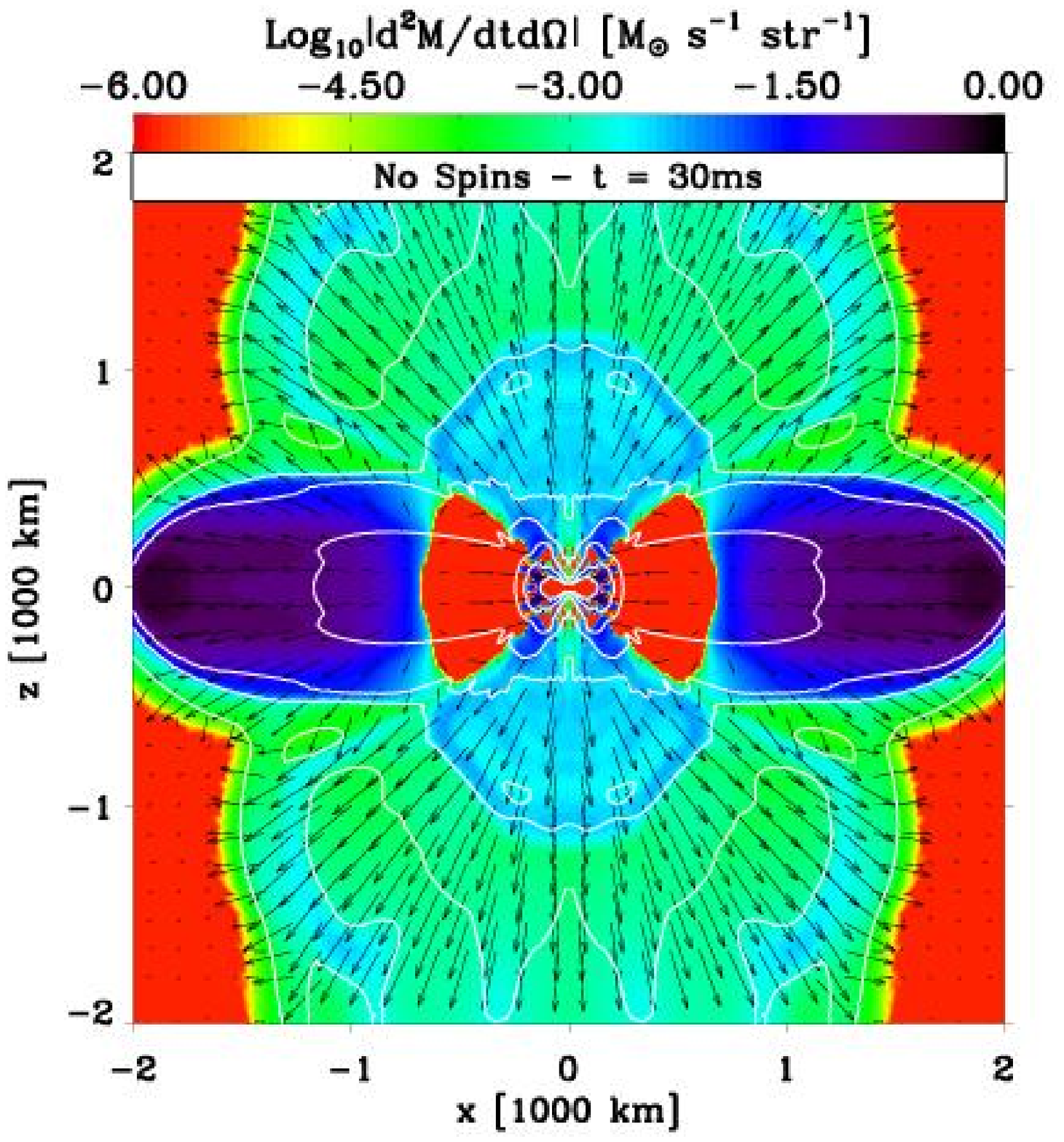}
\plotone{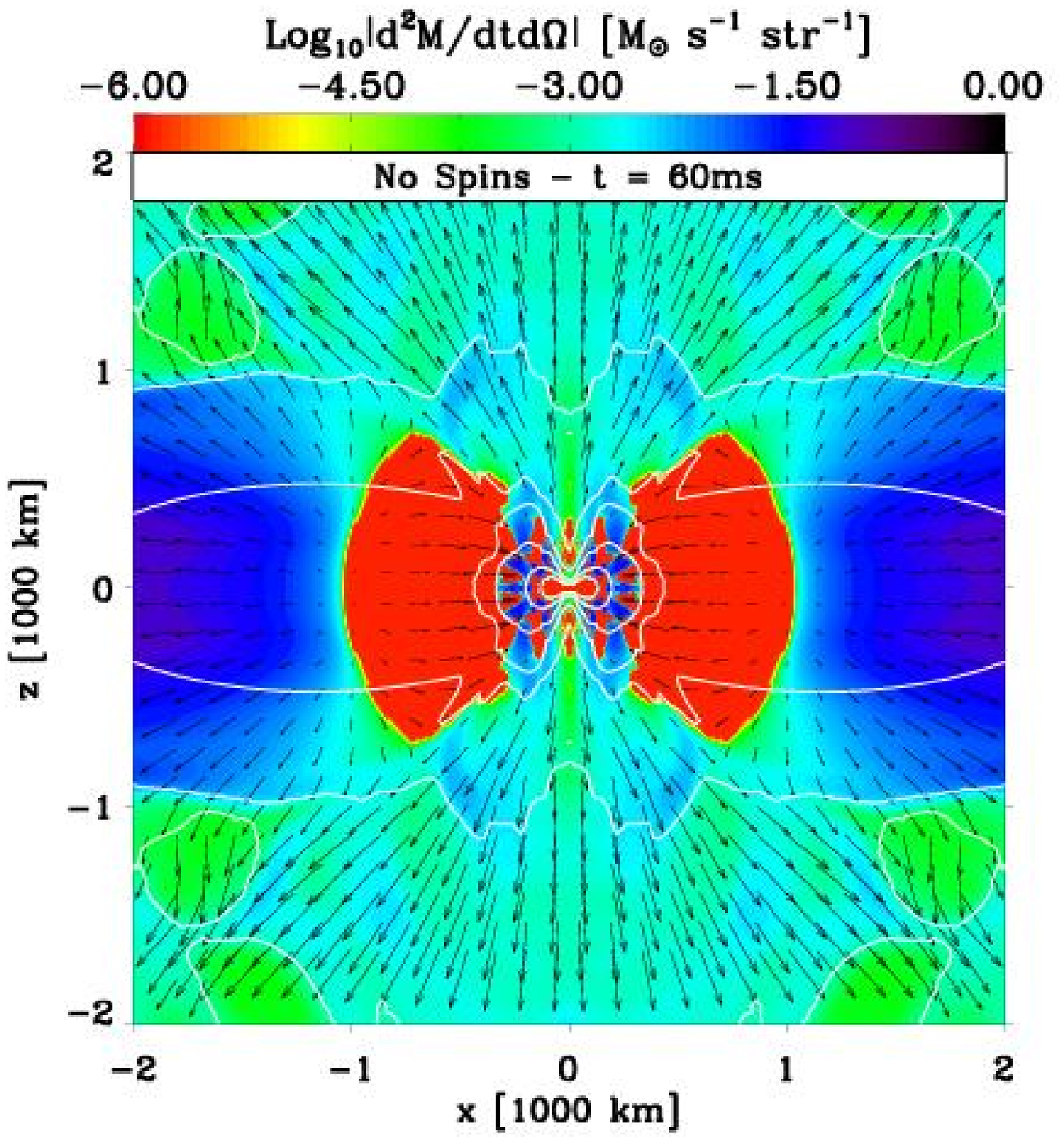}
\plotone{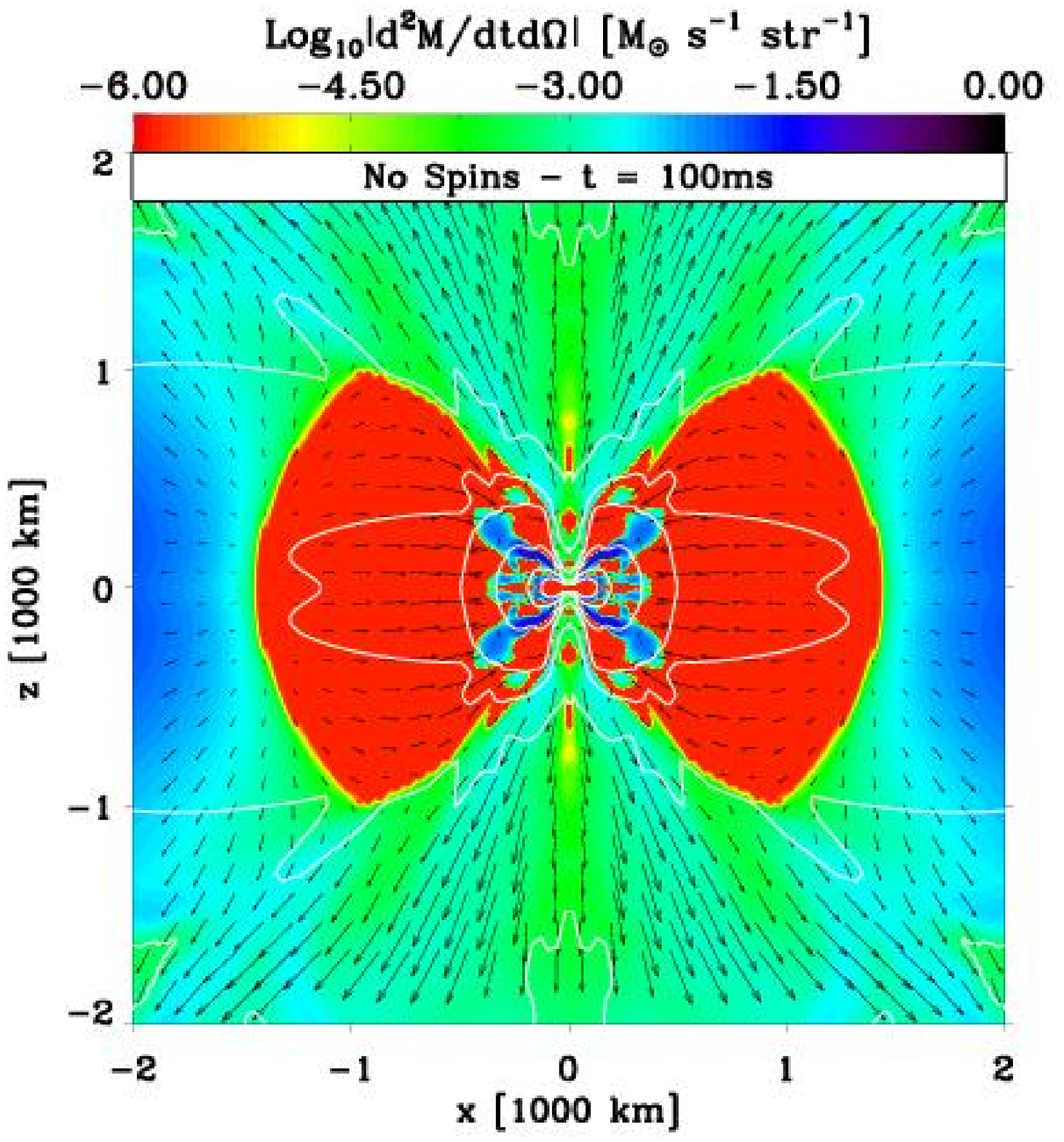}
\caption{Colormaps of the log of the mass loss rate per steradian ($d^2M/dtd\Omega$, in units 
of \mo\,s$^{-1}$\,str$^{-1}$)
for the no-spin BNS merger model at 10\,ms ({\it top left}), 30\,ms ({\it top right}), 
60\,ms ({\it bottom left}), 
and 100\,ms ({\it bottom right}) after the start of the VULCAN/2D simulation, and depicting the 
mass loss associated with the initial transient, followed by the neutrino-driven wind.
The displayed region covers 2000$\times$2000\,km$^2$. Regions that are infalling or 
denser than 10$^{10}$\,g\,cm$^{-3}$ are shown in red,
and velocity vectors, overplotted in black, have a length saturated at 7\% of the width of the display
for a magnitude of 30,000\,\kms. Notice the concomitant mass loss from the poles down to mid-latitudes 
(the wind) and the expansion of BNS merger material at near-equatorial latitudes.
}
\label{fig_wind}
\end{figure*}

A few milliseconds after the start of our simulations and onset of the burst of neutrinos from the 
BNS merger, a neutrino-driven wind develops. It relaxes
into a quasi-steady state by the end of the simulations at 100\,ms. Note that in this quasi 
steady-state configuration, and as shown in Fig.~\ref{fig_netgain}, the energy gain/losses
due to neutrino in the polar direction is mostly net heating, with no presence of a sizable net cooling region.
In Fig.~\ref{fig_wind}, we show this evolution for the no-spin BNS merger model for
four selected times: 10, 30, 60, and 100\,ms.
One can see the initial transient feature, which advects out and eventually leaves the grid. 
This transient phase is not caused by neutrino energy deposition, but is rather due to infall
at small radii and in the polar regions of the ambient medium we placed around the merged object,
as well as due to the sound waves generated by the core during the initial quakes. One can
then see the strengthening neutrino-driven wind developing in the polar funnel.
The side lobes of the SMNS confine the ejecta at radii $\sles$500\,km to a small angle 
of $\sim$20$^{\circ}$ about the poles, but this opening angle grows at largers distances
to $\sim$90$^{\circ}$. The confinement at small radii also leads to the entrainment of material
from the side lobes, overloading the wind and making it choke.
The radial density/velocity profile of this wind flow is thus kinked, with variations 
in velocity that can be as large as the average asymptotic velocity value of $\sim$30000\,\kms 
(Fig.~\ref{fig_rho_slice}). 
As shown in Fig.~\ref{fig_wind}, the mass loss associated with the neutrino-driven wind is
on the order of a few 10$^{-3}$\,\mo\,s$^{-1}$\,str$^{-1}$ at a few tens of milliseconds, but decreases
to 10$^{-3}$--10$^{-4}$\,\mo\,s$^{-1}$\,str$^{-1}$ at 100\,ms. The wind is also weaker along the poles
than along the 70$^{\circ}$-80$^{\circ}$ latitudes. The associated angle-integrated mass loss summed over
100\,ms approaches 10$^{-4}$\,\mo and is made up in part of high $Y_{\rm e}$ material ($\sim$0.5) along
the pole, but mostly of low $Y_{\rm e}$ material ($\sim$0.1-0.2) along mid-latitudes.
Thus, ``r-process material'' will feed the interstellar medium through this neutrino-driven wind.
This latitudinal variation of the electron fraction at large distances 
is controlled by the relative strength of the angle- and time-dependent $\nu_e$ and $\bar{\nu}_e$ 
neutrino luminosities, the expansion timescale of the ``wind'' parcels, and the
neutron richness at their launching site \citep{thompson:00b}.
Along the pole and at the SMNS surface, the low-density, high neutron-richness, and relatively 
stronger $\nu_e$ neutrino luminosities at late times in the no-spin and co-rorotating spin models
lead to a high asymptotic $Y_{\rm e}$ value (high proton-richness).
Despite the relatively high resolution employed in our simulations ($\sim$300\,m in the radial direction
at $\sim$20\,km), higher resolution would be needed to resolve this region accurately. 
Although we expect this trend would hold
at higher resolution, it would likely yield lower asymptotic values of the electron fraction
along the pole. 

\begin{figure}
\plotone{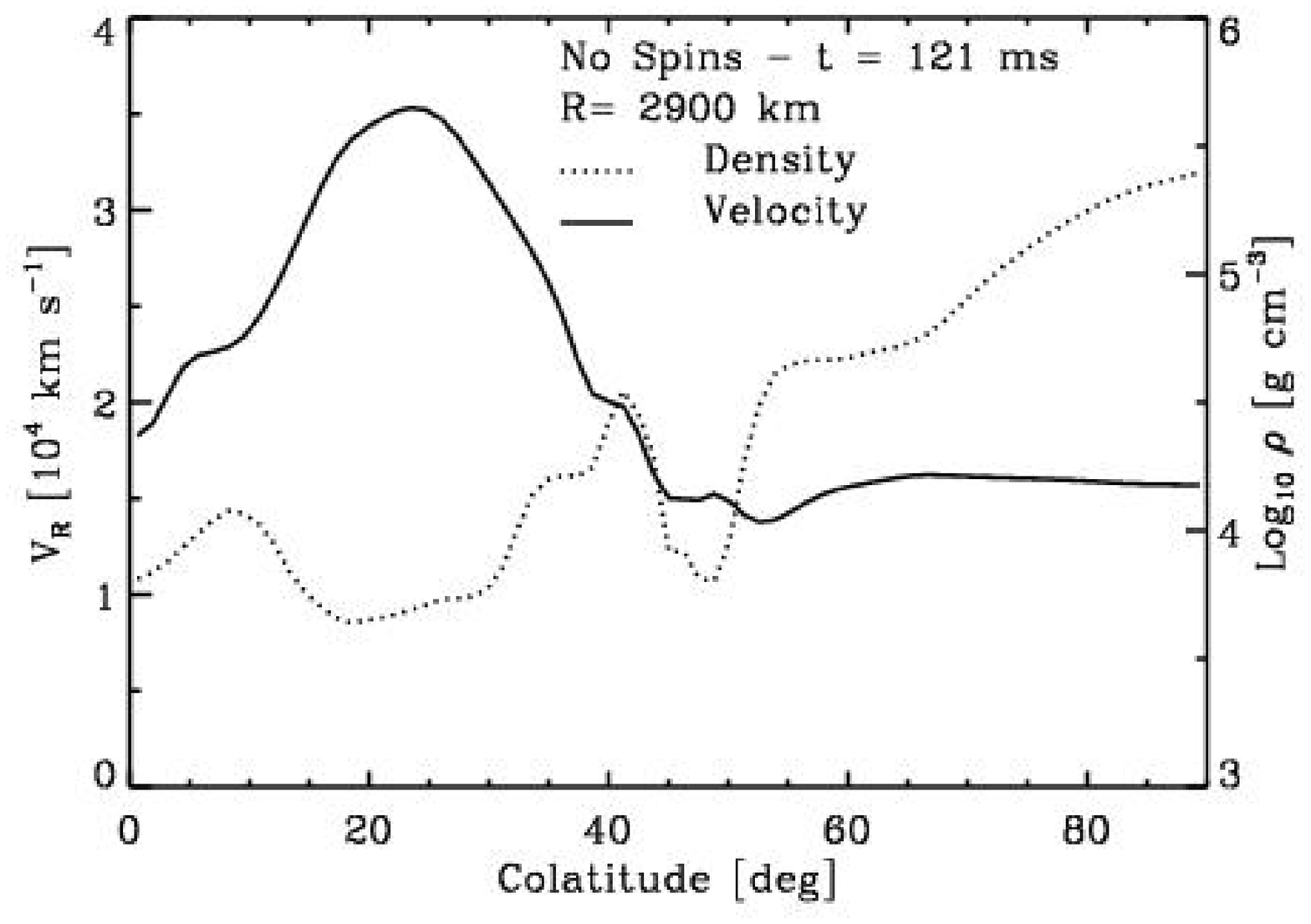}
\caption{Variation of the radial velocity ({\it solid}) and density ({\it dotted}; Log scale) 
versus colatitude
and at a spherical radius of 2900\,km for the BNS merger model with no initial spin. 
The time plotted is 121\,ms after the start of the MGFLD VULCAN/2D simulation.
Note that the escape speed at this radius and for a $\sim$2.8\,\mo central object is $\sim$16000\,\kms,
so most of the mass near equatorial latitudes will remain trapped in the gravitational potential of the
SMNS.}
\label{fig_rhov_lat}
\end{figure}

Along the equatorial direction, low $Y_{\rm e}$ material ($\sim$0.1-0.2) migrates outward,
but its velocity is below the local escape speed and it is unclear how much will eventually
escape to infinity.\footnote{Note that in the counter-rotating model, an axis problem  
in the form of a low-density, high-velocity, narrow region starts at the onset of the neutrino-driven
wind and persists throughout the simulation. This spurious feature is, however, localized and therefore
does not influence the global properties of the simulation.}
This is further illustrated in Fig.~\ref{fig_rhov_lat} where we show
the angular variation of the density and velocity at 2900\,km in the no-spin BNS model,
at 121\,ms. The BNS merger is thus cloaked along the poles by material with
a density in excess of 10$^4$\,g\,cm$^{-3}$, while along lower latitudes even denser material 
from the side lobes
obstructs the view from the center of the SMNS. Importantly, wind material will feed the polar regions
for as long as the merger remnant remains gravitationally stable. Being so heavily baryon-loaded, the 
outflow can in no way be accelerated to relativistic speeds with high-Lorentz factors.
In this context, the powering of a short-hard GRBs is impossible before black hole formation.

As seen in Fig.~\ref{fig_omega_slice}, the rotational profile in the inner 100\,km is
strongly differential in the no-spin and counter-rotating spin cases, while it is quasi-uniform
in the co-rotating spin case (see also \citealt{rosswog:02a}, their Fig.~17).
The good conservation of specific angular momentum in VULCAN/2D
is in part responsible for the preservation of the initial rotation
profile throughout the simulation. In reality, such a differential rotation should not survive.
Our 2D axisymmetric setup inhibits the development of tri-axial instabilities that arise at modest and 
large ratios of rotational and gravitational energies \citep{rampp:98,centrella:01,saijo:03,ott:05,ott:07b}, 
i.e., under the conditions that prevail here.
Moreover, our good, but not excellent, spatial resolution prevents the modeling 
of the magneto-rotational instability, 
whose effect is to efficiently redistribute angular momentum \citep{balbus:91,pessah:06,pessah:07}, and 
dissipate energy, and, in the present context, lead to mass accretion onto the SMNS.
The magneto-rotational instability, operating on an rotational timescale, could lead to solid-body rotation
within a few milliseconds in regions around $\sim$10\,km, and within a few tens of milliseconds in regions
around $\sim$100\,km.
Note that this is a more relevant timescale than the typical $\sim$100\,s that characterizes 
the angular momentum loss through magnetic dipole radiation \citep{rosswog:02a}.
Importantly, in the three BNS merger models we study, we find that the free energy of rotation
(the energy difference between the given differentially-rotating object and that
of the equivalent solid-body rotating object with the same cumulative angular momentum)
is very large. Modulo the differences of a factor of a few between models, it reaches $\sim$5$\times$10$^{51}$\,erg
inside the supermassive neutron star (regions with densities greater than 10$^{14}$\,g\,cm$^{-3}$),
but is on the order of 2$\times$10$^{52}$\,erg in the torus disk, regions with densities 
between 10$^{11}$ and 10$^{14}$\,g\,cm$^{-3}$. Similar conditions in the core-collapse
context yield powerful, magnetically- (and thermally-) driven explosions 
\citep{leblanc:70,bisnovatyi:76,akiyama:03,ardeljan:05,moiseenko:06,obergaulinger:06,burrows:07b,dessart:07}. 
Rotation dramatically enhances the rate of mass ejection
by increasing the density rather than the velocity of the flow, even possibly halting 
accretion and inhibiting the formation of a black hole \citep{dessart:08}. In the present context,
the magneto-rotational effects, which we do not include here, would considerably enhance the mass flux
of the neutrino-driven wind. Importantly, the loss of differential rotational energy needed to 
facilitate the gravitational instability is at the same time delaying it through
the enhanced mass loss it induces.
Work is needed to understand the systematics of this interplay, and how much rotational energy 
the back hole is eventually endowed with.

\cite{oechslin:07b}, using a conformally-flat approximation to GR and an SPH code, 
find that BNS mergers of the type 
discussed here and modeled with the Shen EOS, avoid the general-relativistic gravitational instability 
for many tens of milliseconds after the neutron stars first come into contact. 
\cite{baumgarte:00}, and more recently \cite{morrison:04}, \cite{duez:04,duez:06}, and \cite{shibata:06}, 
using GR (and for some using a polytropic EOS), find that imposing even modest levels of differential 
rotation yields a significant increase by up to 50\% in the maximum mass that can be supported stably,
in particular pushing this value beyond that of the merger remnant mass after coalescence.
Surprisingly, \cite{baiotti:08}, using a full GR treatment but with a simplifed (and soft) EOS, 
find prompt black hole formation in such high mass progenitors. 
Despite this lack of consensus, the existence of neutron stars with a gravitational
mass around 2\,\mo favors a high incompressibility of nuclear matter, such as is in the Shen EOS,
and suggests that SMNSs formed through BNS merger events may survive for tens of milliseconds 
before experiencing the general-relativistic gravitational instability. 
In particular, the presence of a significant amount of material (a few tenths of a solar mass) 
located on wide orbits in a Keplerian disk, reduces the amount of mass that resides initially
in the core, i.e., prior to outward transport of angular momentum.

\begin{figure*}
\epsscale{.35}
\plotone{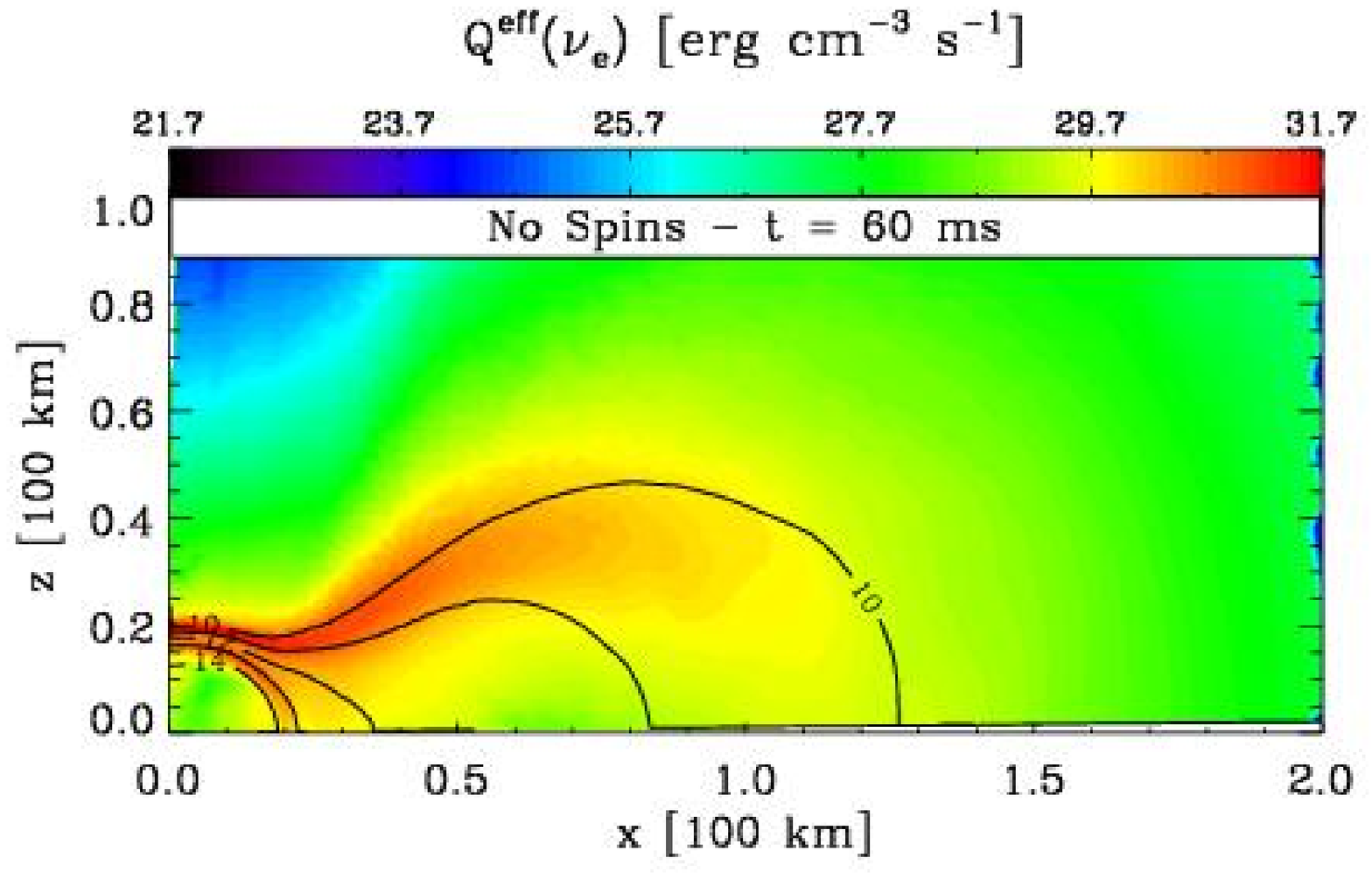}
\plotone{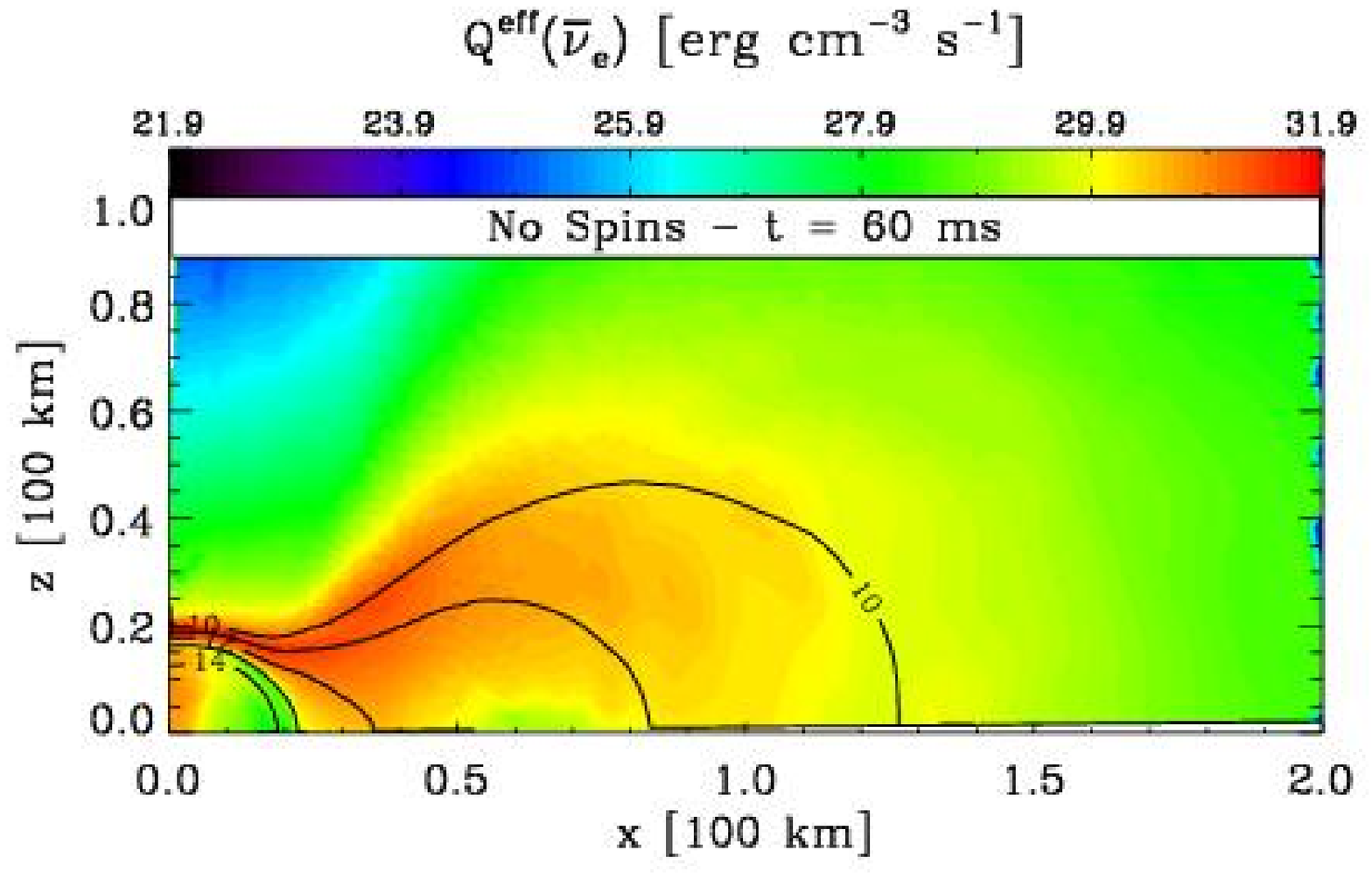}
\plotone{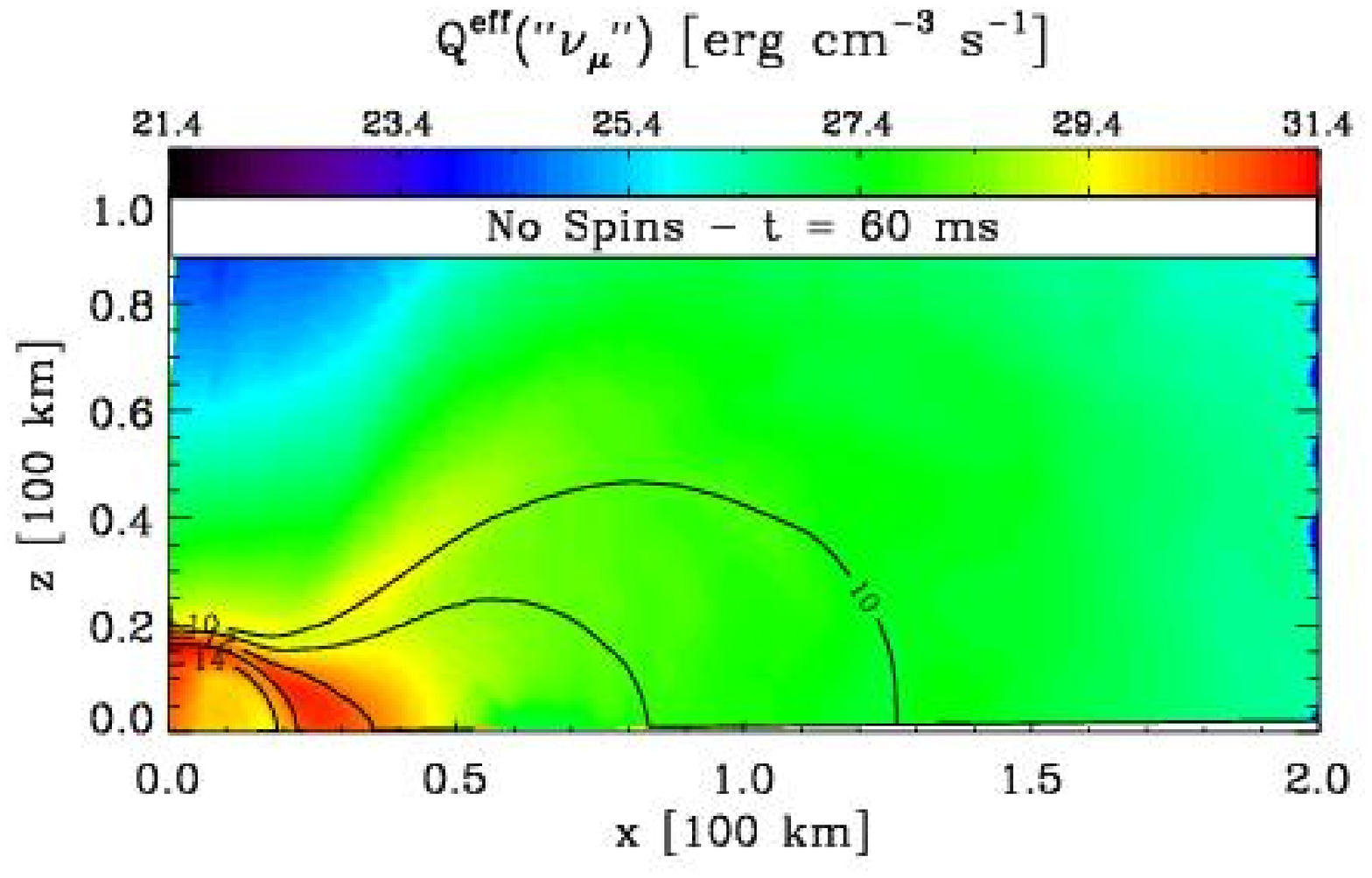}
\plotone{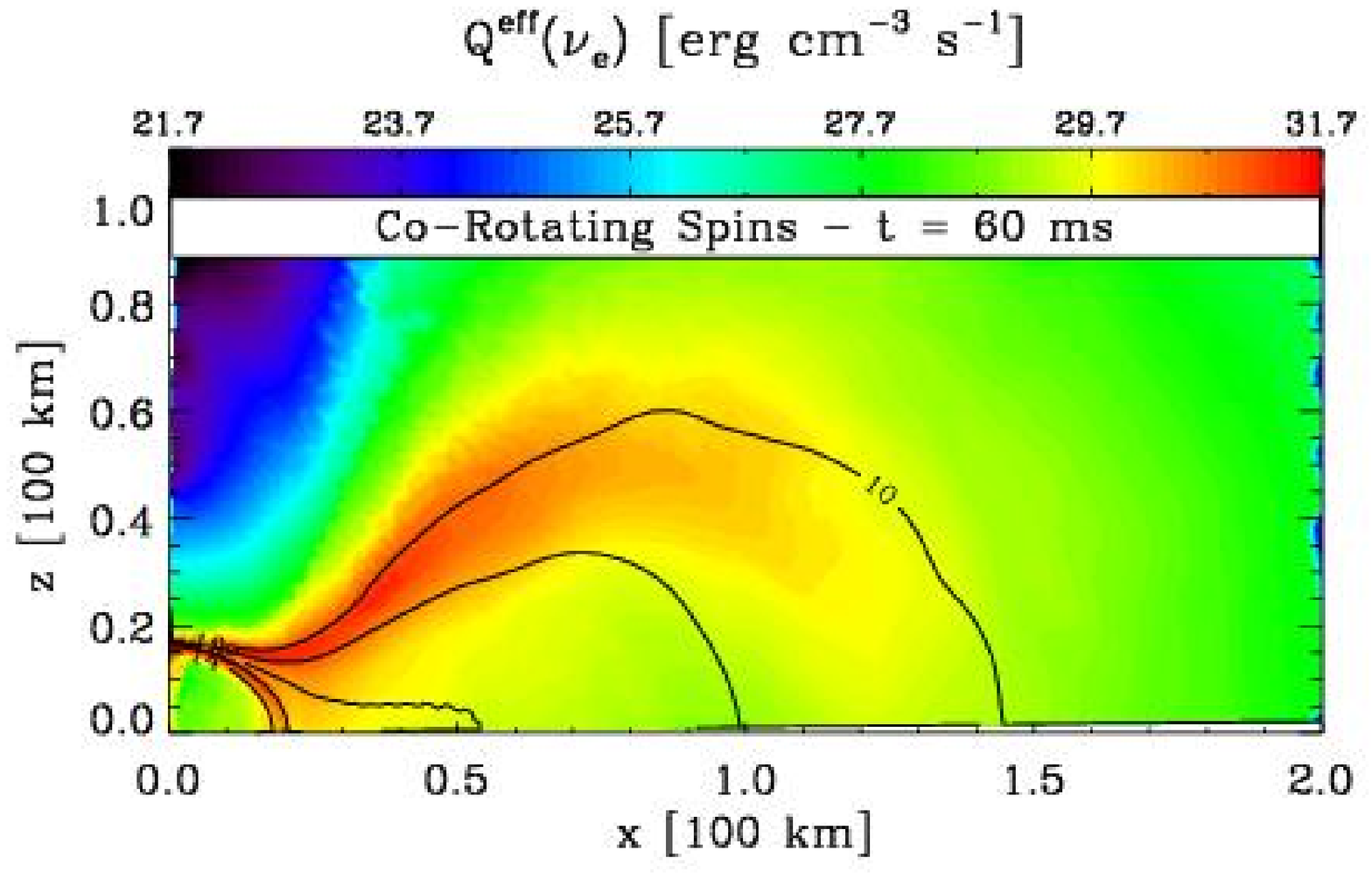}
\plotone{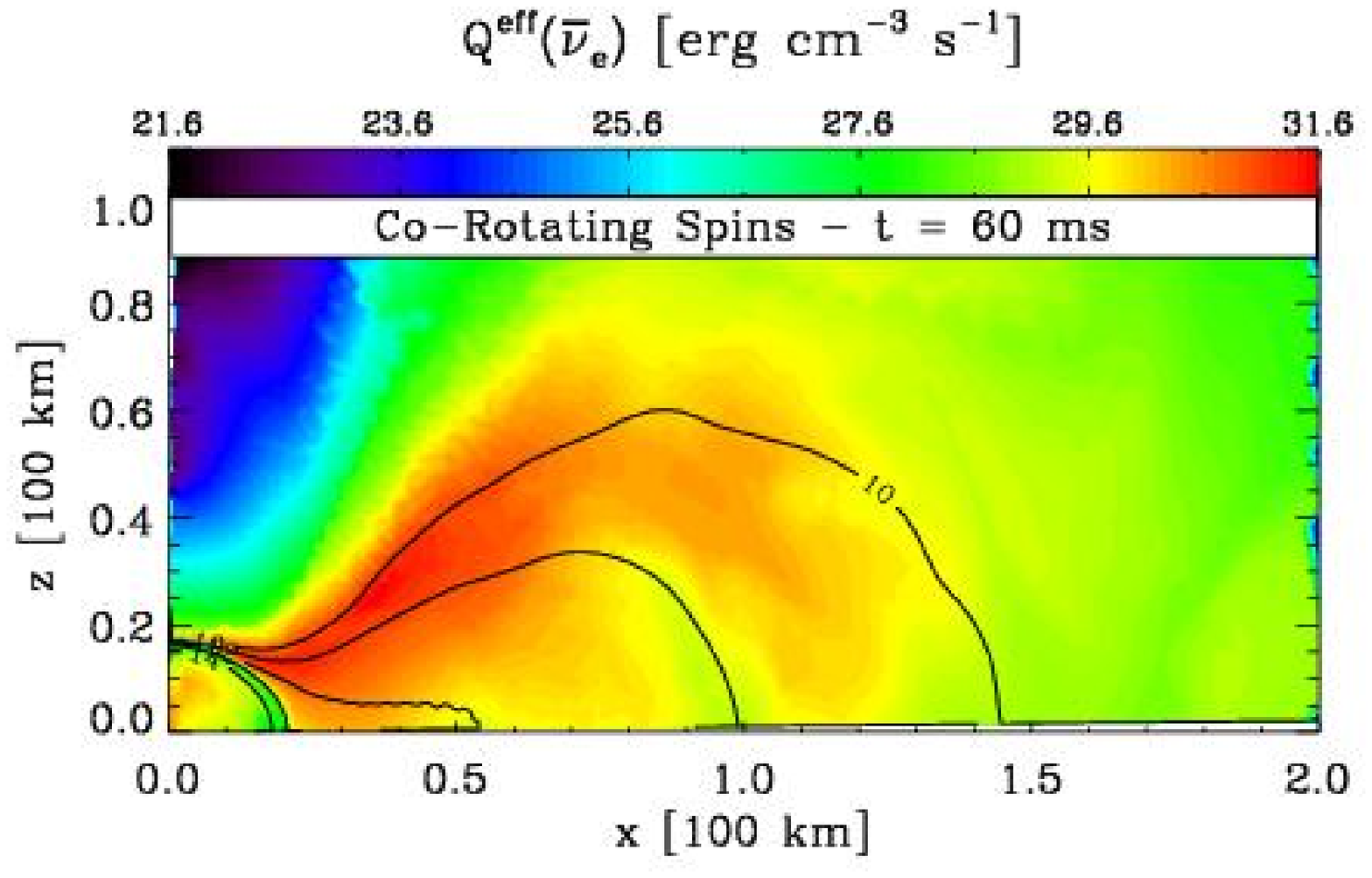}
\plotone{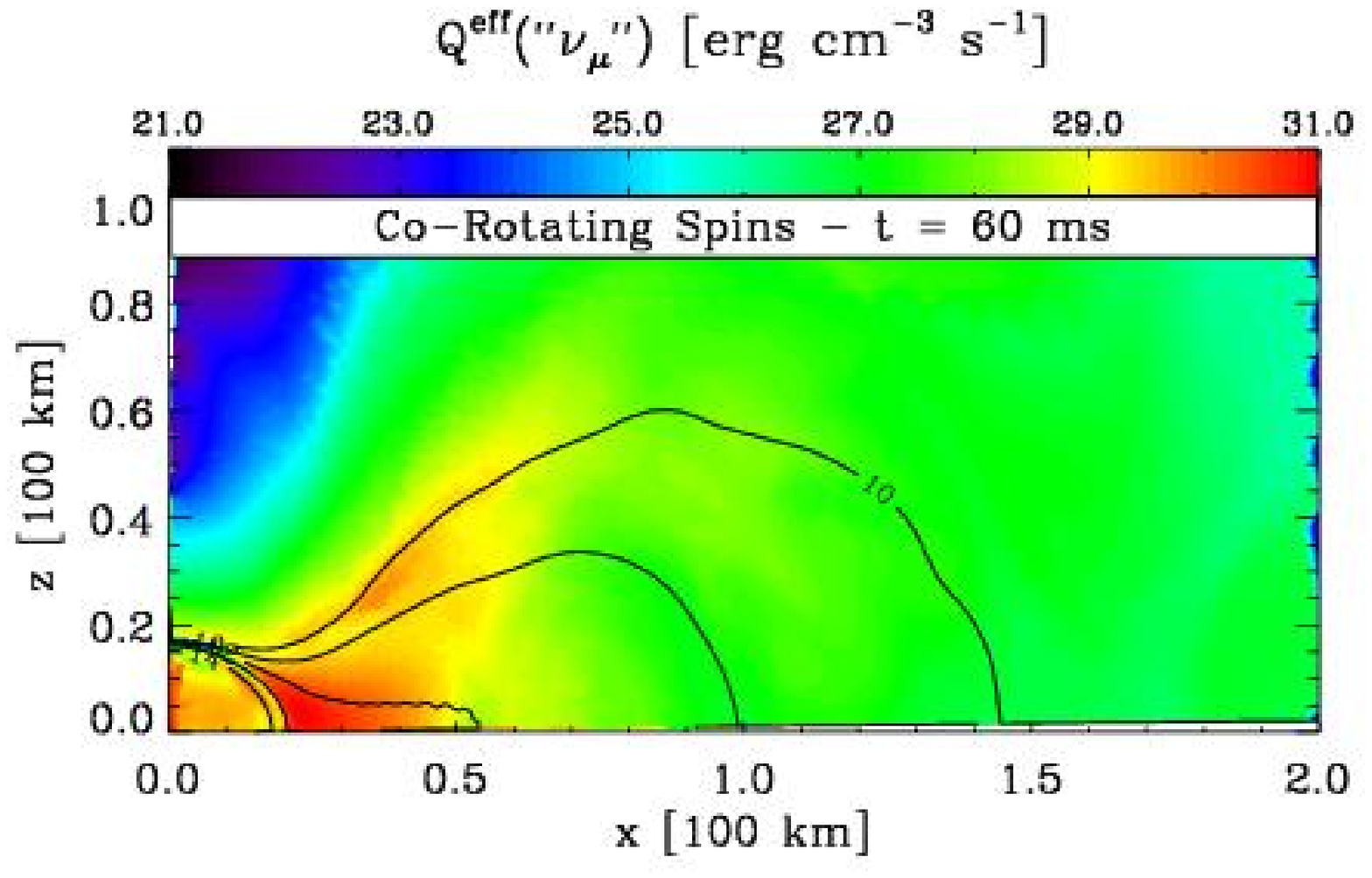}
\plotone{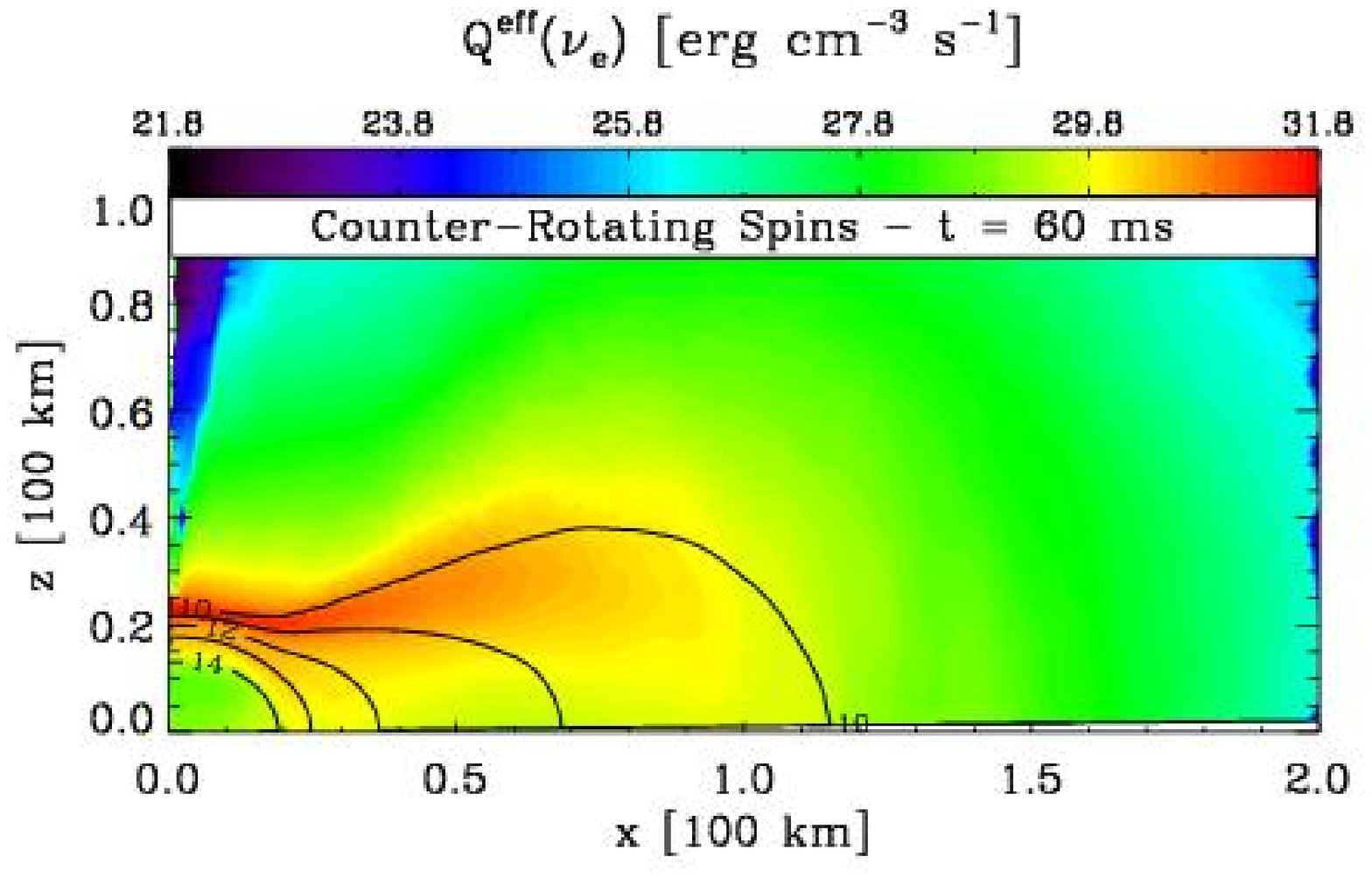}
\plotone{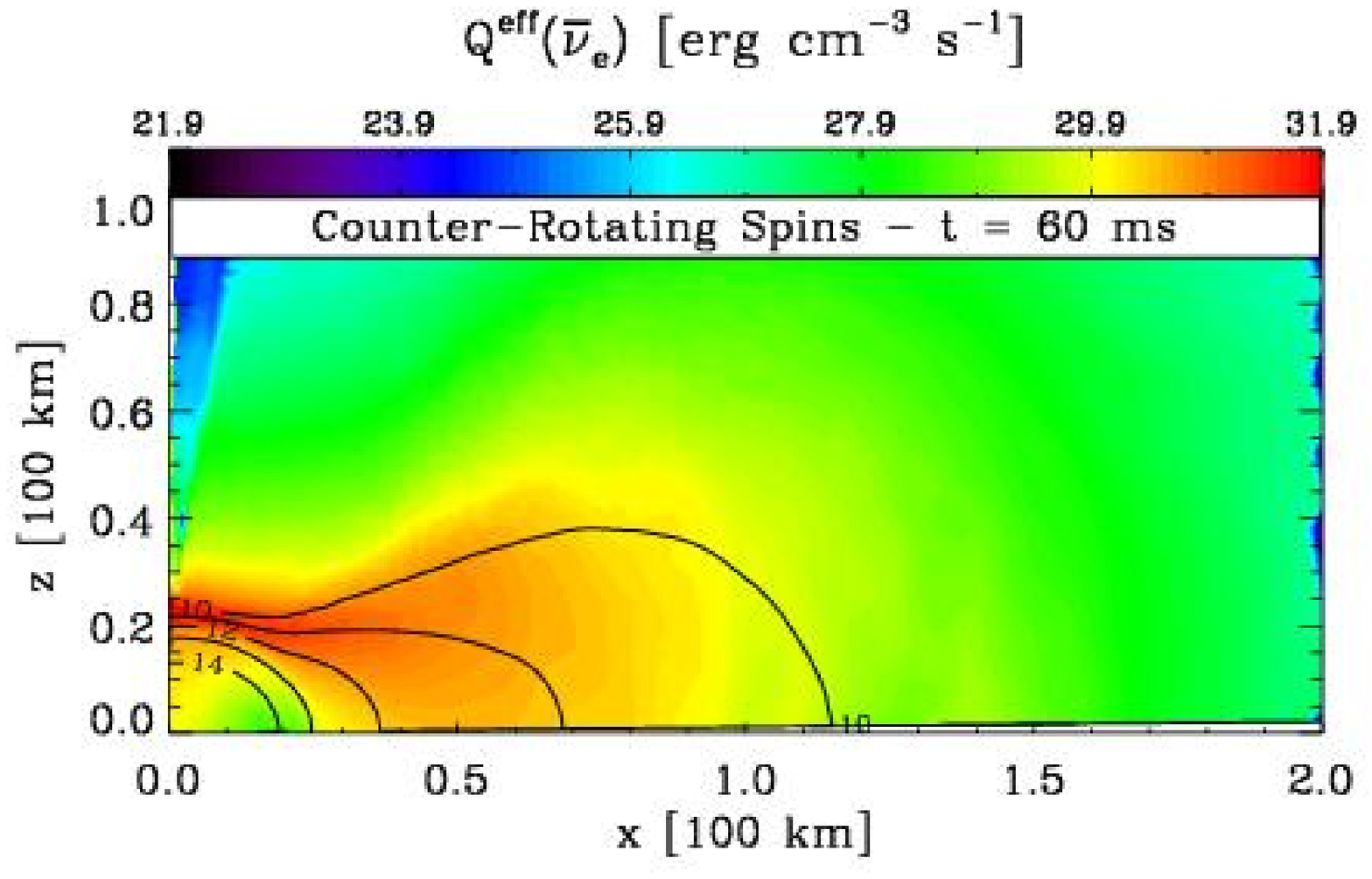}
\plotone{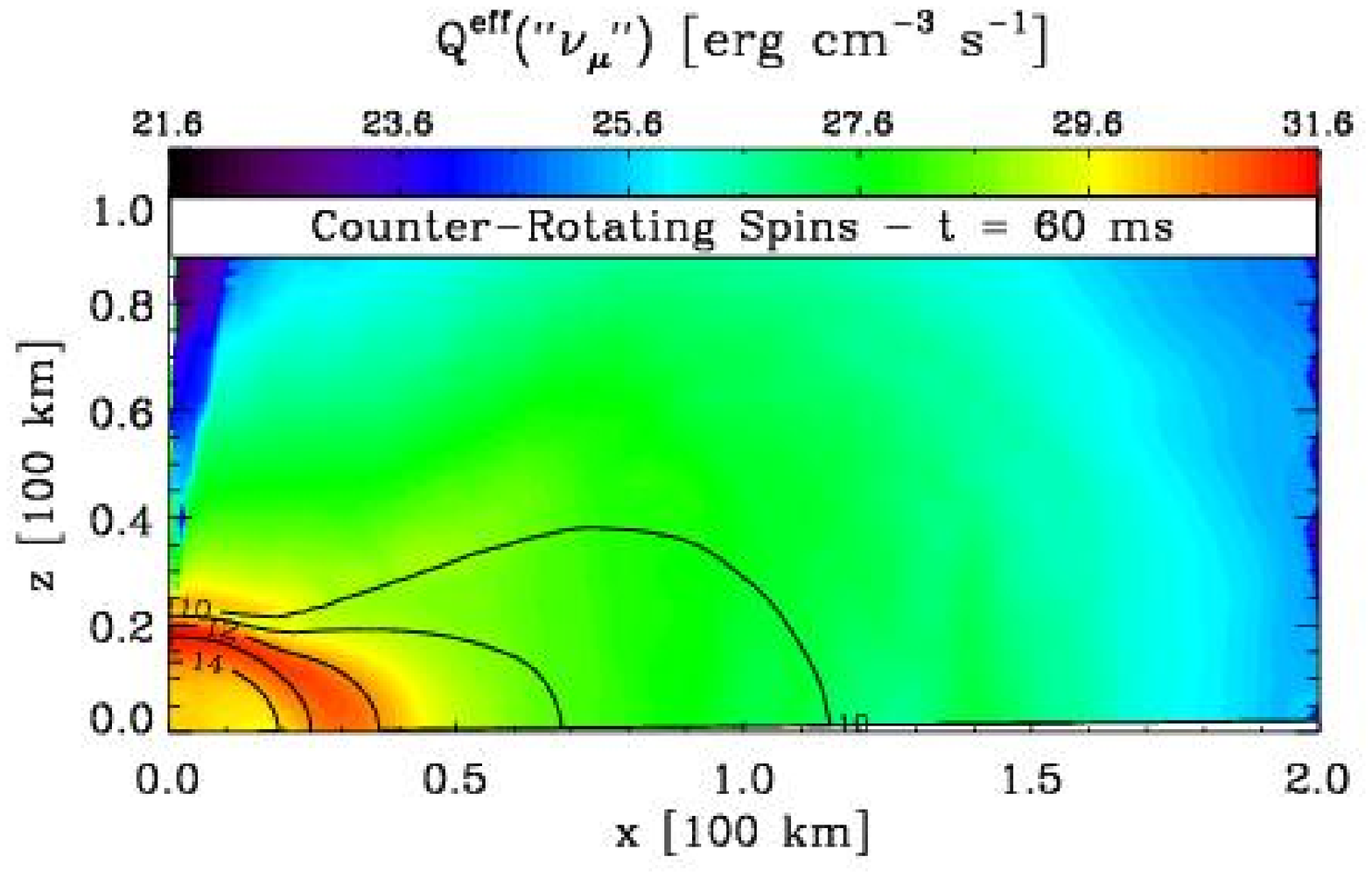}
\caption{
Colormap of the log of the ``effective'' emissivity $Q^{\rm eff}(\nu_i)$ for $\nu_{\rm e}$ ({\it left}), 
$\bar{\nu}_{\rm e}$ ({\it middle}), and ``$\nu_{\mu}$'' ({\it right}) neutrinos
at 60\,ms after the start of the simulation for the BNS merger configuration with 
no initial spins ({\it top row}), co-rotating spins ({\it middle row}), and 
counter-rotating spins ({\it bottom row}).
We also overplot iso-density contours at every decade between 10$^{10}$ and 10$^{14}$\,g\,cm$^{-3}$,
to allow a visual assessment of the effect of the adopted density cut (at 10$^{11}$\,g\,cm$^{-3}$)
on the computation of the annihilation rate. Note how severely this cut truncates the emission of 
``$\nu_{\mu}$'' neutrinos. The calculation is based on the formalism developed and presented 
by \cite{ruffert:96}.
}
\label{fig_emis_leakage}
\end{figure*}

\section{$\nu$--$\bar{\nu}$ Annihilation}
\label{sect:annihilate}

  Besides being prime candidates for gravitational wave emission, binary neutron
star mergers may lead to short-hard GRBs (for an overview, see, e.g., \citealt{piran:05,nakar:07}).
In this context, the high energy radiation arises from non-thermal radiation associated with
relativistic ejecta. The annihilation of neutrino pairs, by the process

\begin{eqnarray}
\nu_i + \bar{\nu}_i \rightarrow e^+ \, +\, e^-\,\,\,;\,\, i \in
\{e,\mu,\tau\}\,\,,
\end{eqnarray}

represents one possible powering source for high-Lorentz factor baryon-free ejecta beamed into a
small solid angle about the rotation axis of the BNS merger
(\citealt{ruffert:97,ruffert:99,asano:00,asano:01,
wmiller:03,kneller:06,birkl:07,rosswog:03c,rosswog:05}, and references therein).

Here, we revisit this proposition by studying quantitatively the energetics of neutrino-antineutrino 
annihilation in the three BNS mergers we simulated with VULCAN/2D.
First, in \S~\ref{annihil:leakage}, we 
apply the approach used in previous work and based on the leakage scheme of \cite{ruffert:96,ruffert:97}.
This approach was also used by \cite{rosswog:03a}, but for merger events whose dynamical evolution
was computed independently.\footnote{SPH versus grid-based code; different EOS; different resolution.} 
Note that similar $\nu_i\bar{\nu}_i$ annihilation rate calculations, i.e., without the full 
momentum-space angular dependence, have been carried out in the core-collapse and PNS
contexts, both with Newtonian gravity \citep{goodman:87,cooperstein:86,cooperstein:87,janka:91} and in
GR \citep{salmonson:99,bhatta:07}.
Then, in \S~\ref{annihil:sn} we present a new formalism based on moments of the neutrino specific 
intensity. The angle and energy dependence of the neutrino radiation field is obtained by post-processing
individual MGFLD VULCAN/2D snapshots (see \S~\ref{sect:results}) with the multi-angle, $S_n$, variant.
In both cases, because the power associated with neutrino-antineutrino annihilation is sub-dominant 
compared to that associated with the charge-current reactions prior to black hole formation,
it can be estimated only through a post-processing step.

\begin{deluxetable*}{lcccccc}
\tablewidth{18cm}
\tabletypesize{\scriptsize}
\tablecaption{$\nu_i\bar{\nu}_i$ annihilation rates using the leakage and the $S_n$ schemes}
\tablehead{
 & \multicolumn{2}{c}{No Spins} & \multicolumn{2}{c}{Co-Rotating Spins} & \multicolumn{2}{c}{Counter-Rotating Spins} \\
\hline
\colhead{Time}&  
\colhead{$\int dV Q^+$(${\nu_e}\bar{\nu}_e$)}&   
\colhead{$\int dV Q^+$(``${\nu_\mu}\bar{\nu}_\mu$'')}& 
\colhead{$\int dV Q^+({\nu_e}\bar{\nu}_e$)}&   
\colhead{$\int dV Q^+$(``${\nu_\mu}\bar{\nu}_\mu$'')}& 
\colhead{$\int dV Q^+({\nu_e}\bar{\nu}_e$)}&   
\colhead{$\int dV Q^+$(``${\nu_\mu}\bar{\nu}_\mu$'')} \\ 
\hline
\colhead{(ms)} & \multicolumn{2}{c}{B\,s$^{-1}$} & \multicolumn{2}{c}{B\,s$^{-1}$} & \multicolumn{2}{c}{B\,s$^{-1}$} 
}
\startdata
& \multicolumn{6}{c}{Leakage Scheme} \\
\hline
10    &   1.56(-1)   &    3.28(-5)   &    2.05(-1)   &    1.01(-4)   &    2.18(-1)   &    1.40(-4)   \\  
20    &   9.28(-2)   &    5.99(-6)   &    4.79(-2)   &    2.38(-6)   &    1.05(-1)   &    1.01(-5)   \\  
30    &   3.80(-2)   &    1.15(-6)   &    3.70(-2)   &    1.43(-6)   &    5.95(-2)   &    3.35(-6)   \\  
40    &   3.11(-2)   &    8.54(-7)   &    4.67(-2)   &    1.08(-6)   &    3.65(-2)   &    1.36(-6)   \\  
50    &   2.82(-2)   &    6.23(-7)   &    5.67(-2)   &    1.35(-6)   &    2.16(-2)   &    5.81(-7)   \\  
60    &   1.78(-2)   &    2.57(-7)   &    4.43(-2)   &    8.74(-7)   &    1.50(-2)   &    3.00(-7)   \\  
70    &   1.32(-2)   &    1.63(-7)   &    2.69(-2)   &    3.78(-7)   &    1.18(-2)   &    1.79(-7)   \\  
80    &   1.22(-2)   &    1.67(-7)   &    3.55(-2)   &    8.16(-7)   &    1.01(-2)   &    1.21(-7)   \\  
90    &   1.47(-2)   &    2.94(-7)   &    2.96(-2)   &    6.64(-7)   &    7.64(-3)   &    7.64(-8)   \\  
100   &   1.60(-2)   &    3.73(-7)   &    2.54(-2)   &    5.41(-7)   &    6.48(-3)   &    5.47(-7)   \\
\hline
 & \multicolumn{6}{c}{$S_n$ Scheme} \\
\hline
10    &   1.81(-1)    &     1.44(-2)   &  5.97(-1)     &    9.19(-3)  &   3.63(-1)      &   4.76(-2)   \\
20    &   6.82(-2)    &     6.41(-3)   &  9.22(-2)     &    1.15(-3)  &   8.92(-2)      &   1.70(-2)   \\
30    &   3.95(-2)    &     3.96(-3)   &  4.59(-2)     &    4.59(-4)  &   4.08(-2)      &   1.01(-2)   \\
40    &   2.71(-2)    &     2.58(-3)   &  4.13(-2)     &    2.82(-4)  &   2.77(-2)      &   7.83(-3)   \\
50    &   2.18(-2)    &     1.78(-3)   &  4.19(-2)     &    1.89(-4)  &   1.92(-2)      &   5.84(-3)   \\
60    &   1.82(-2)    &     1.39(-3)   &  3.59(-2)     &    1.34(-4)  &   1.28(-2)      &   4.27(-3)   \\
70    &   1.47(-2)    &     1.02(-3)   &  2.31(-2)     &    8.70(-5)  &   1.04(-2)      &   3.76(-3)   \\
80    &   1.11(-2)    &     6.94(-4)   &  2.30(-2)     &    6.22(-5)  &   8.49(-3)      &   3.28(-3)   \\
90    &   1.01(-2)    &     5.25(-4)   &  1.84(-2)     &    4.28(-5)  &   7.13(-3)      &   2.83(-3)   \\
100   &   9.67(-3)    &     3.93(-4)   &  1.59(-2)     &    3.02(-5)  &   6.21(-3)      &   2.46(-3)   
\enddata
\tablecomments{{\it Upper half:} Listing of the volume-integrated rates for the $\nu_e\bar{\nu}_e$ and 
``$\nu_\mu\bar{\nu}_\mu$'' annihilation rates using the approach of \cite{ruffert:96,ruffert:97}, 
as summarized in \S \ref{annihil:leakage}. The time for each calculation refers to the time since 
the start of the VULCAN/2D simulation. Numbers in parenthesis correspond to powers of ten.
The strong decrease of the energy deposition rates reflects 
the fading of the neutrino luminosity due to the cooling of the SMNS.
A density cut above 10$^{11}$\,g\,cm$^{-3}$ is applied for both emission and deposition sites.
{\it Lower half:} Same as upper half, but this time using the 8 $\vartheta$-angle $S_n$ calculations,
performed through a post-processing of the MGFLD radiation-hydrodynamics snapshots 
computed with VULCAN/2D. The same hydrodynamics background ($\rho$, $T$, $Y_{\rm e}$ distribution)
is used for the leakage and the $S_n$-schemes. 
}
\label{tab:edep}
\end{deluxetable*}

\subsection{Formalism based on a leakage scheme and the approach of \citet{ruffert:96}}
\label{annihil:leakage}

  Estimation of the neutrino-antineutrino annihilation rate using the \cite{ruffert:96,ruffert:97} approach
is done in two steps. First, using the processes described in \S \ref{sect:neutrinos}, 
the instantaneous rate of neutrino emission $Q(\nu_i)$ is computed for all grid cells. 
It is subsequently weighted by the direction-dependent factor that depends 
on the radiative diffusion timescale $t^{\rm diff}_{\nu_i}$ and neutrino emission 
timescale $t^{\rm loss}_{\nu_i}$ relevant for that cell.
Since we are primarily interested in the energy deposition in the polar regions, we consider only the  
cylindrical-$z$ direction when estimating $t^{\rm diff}_{\nu_i}$. 
The effective emissivity $Q^{\rm eff}(\nu_i)$ then has the form
\begin{eqnarray}
Q^{\rm eff}(\nu_i) = \frac{Q(\nu_i)}{1 + \left(t^{\rm loss}_{\nu_i}\right)^{-1} 
t^{\rm diff}_{\nu_i}}\,\,,
\end{eqnarray}
where expressions for the various components are given explicitly in Appendices A \& B of \cite{ruffert:96}.

In Fig.~\ref{fig_emis_leakage}, we show the effective emissivity resulting from this leakage scheme and 
with the neutrino processes of \cite{ruffert:96} for the  $\nu_e$ ({\it left column}), $\bar{\nu}_e$ 
({\it middle column}), and ``$\nu_{\mu}$'' ({\it right column}) neutrinos,
for the BNS merger models with initially no spins ({\it top row}), 
co-rotating spins ({\it middle row}), and counter-rotating spins ({\it bottom row}).
For each process, emission is conditioned by the competing elements of optical depth,
density, temperature, and electron fraction. In practice, it peaks in regions that are dense, hot, but not too
optically thick, i.e., at the surface of the SMNS (see density contours in Fig.~\ref{fig_emis_leakage}, 
overplotted in black). Optical depths in excess of 100 for all neutrino energies 
make the inner region (the inner $\sim$15\,km, where densities have nuclear values) 
a weak ``effective'' emitter, their contribution operating on a 0.1--1\,s timescale.
The dominant emission associated with the ``$\nu_{\mu}$'' neutrinos originates from a considerably higher-density
region than that associated with the electron-type neutrinos, i.e.
10$^{12}$-10$^{13}$\,g\,cm$^{-3}$ compared to 10$^{9}$-10$^{11}$\,g\,cm$^{-3}$, 
with a corresponding ``leakage'' luminosity in 
all three merger models that is typically an order of magnitude smaller than predicted by 
VULCAN/2D (see Fig.~\ref{fig_flux_time}).
This low ``$\nu_{\mu}$'' emissivity is likely caused by the neglect of 
nucleon-nucleon bremsstrahlung processes in the approach of Ruffert et al., 
which leads to a smaller decoupling radius for ``$\nu_{\mu}$'' neutrinos, 
and, therefore, an underestimate of the size of the radiating surface from which they emerge 
(shown in Fig.~\ref{fig_nu_sphere}, {\it right column}). In practice, the decoupling radius
is energy-dependent, but here we stress the systematic reduction of the decoupling radius 
for all ``$\nu_{\mu}$'' neutrino energies because of the neglect of this extra opacity source. 
The importance of the bremsstrahlung process for ``$\nu_{\mu}$'' emissivity and 
opacity has been emphasized by \citet{thompson:00a} for ``hot'' PNSs.
The SMNS that results from the merger is considerably heated by shocks and shear, 
and is thus also in a configuration where such bremsstrahlung processes are important and should be 
included.

From the effective emissivity distribution computed for each neutrino flavor with the leakage scheme, 
\cite{ruffert:97} then sum the contributions from all pairings between grid cells. 
For completeness, we briefly reproduce here the presentation of \cite{ruffert:97}. 
The integral to be computed for the energy-integrated $\nu_i\bar{\nu}_i$  
(representing equivalently $\nu_{\rm e}\bar{\nu}_{\rm e}$ or ``$\nu_{\mu}\bar{\nu}_{\mu}$'') annihilation rate at position
$\vec{r}$  is 

\begin{eqnarray}\label{eq:leakage1}
\lefteqn{Q^+(\nu_i\bar\nu_i) =  {1\over 4}\,
{\sigma_0\over c\,\rund{m_e c^2}^2}
 \, \Biggl\lbrace \,
{\rund{C_1 + C_2}_{\nu_i\bar\nu_i}\over 3}\,\cdot } \nonumber\\
 & & \oint_{4\pi}{\rm d}
\Omega\,I_{\nu_i}\,\oint_{4\pi}{\rm d}\Omega'\,
I_{\bar\nu_i}\,
\eck{\ave{\epsilon}_{\nu_i} + \ave{\epsilon}_{\bar\nu_i}}\,
\rund{1 - \cos\Phi}^2\,+
\nonumber\\
& + & C_{3,\nu_i\bar\nu_i}\rund{m_e c^2}^2\cdot \nonumber \\
& & \oint_{4\pi}{\rm d}\Omega\,I_{\nu_i}
\,\oint_{4\pi}{\rm d}\Omega'\,I_{\bar\nu_i}\,
{\ave{\epsilon}_{\nu_i} + \ave{\epsilon}_{\bar\nu_i}\over
\ave{\epsilon}_{\nu_i}\,\ave{\epsilon}_{\bar\nu_i}}\,
\rund{1 - \cos\Phi} \,\, \Biggr\rbrace \,\, .
\end{eqnarray}

$I_{\nu_i}$ is the neutrino specific intensity. $\Omega$ and $\Omega'$ are the solid angles
subtended by the cells producing the neutrino and antineutrino radiation incident
from all directions. 
$C_1$, $C_2$, and $C_3$ are related to the weak coupling constants $C_{\rm A}$ and $C_{\rm V}$ 
and depend on the neutrino species (see \citealt{ruffert:97} as well as \S~\ref{annihil:sn}). 
$\sigma_0$ is the baseline weak interaction cross section, 1.705 $\times 10^{-44}$\,cm$^{2}$,
$c$ is the speed of light, $m_e$ is the electron mass. 
$\Phi$ is the angle between the neutrino and antineutrino beams, entering
the annihilation rate formulation through the term $\rund{1 - \cos\Phi}$ (squared or not)
which thus gives a stronger weighting to larger angle collisions. This is what makes the dumbell-like 
morphology of BNS mergers such a prime candidate for $\nu_i\bar{\nu}_i$ annihilation
over spherical configurations (see also \S~\ref{sect:appendix}).
$\ave{\epsilon}_{\nu_i}$ and $\ave{\epsilon}_{\bar\nu_i}$ are the 
$\nu_i$ and $\bar\nu_i$ mean neutrino energies, whose values we adopt for consistency 
from the simulations of \cite{ruffert:97}, i.e., 12, 20, and 27\,MeV for $\nu_{\rm e}$,
$\bar{\nu}_{\rm e}$, and ``$\nu_{\mu}$'', respectively (our values are within 
a few times 10\% of these, so this has little impact on our discussion).
Note also that the average neutrino energies for all three species 
are much larger than $m_e c^2$, and therefore make the second term in the above 
equation negligible (it is about a factor of 1000 smaller than the first one, 
and also has a much weaker large-angle weighting). The total annihilation rate 
is then the sum $Q^+(\nu_{\rm e}\bar{\nu}_{\rm e}) + Q^+($``$\nu_{\mu}\bar{\nu}_\mu$'').

In practice, we turn the angle integrals into discretized sums through the transformation
\begin{eqnarray}
\oint_{4\pi} {\rm d}\Omega\,I_{\nu}\ \longrightarrow\ \
\sum_k \Delta\Omega_k\cdot I_{\nu,k} \, ,
\label{eq:discret}
\end{eqnarray}  
where $\Delta\Omega_k$ is the solid angle subtended by the cell $k$ as seen from the location $\vec{r}$.
\cite{ruffert:97} applied their method to 3D Cartesian simulations, while our simulations are in
2D and have both axial symmetry, and mirror symmetry about the equatorial plane.
Making use of these symmetry properties, we remap our VULCAN/2D simulation from a 2D (cylindrical) 
meridional slice onto a 3D spherical volume which covers the space with a uniform grid of 
40 zones for the
2$\pi$ azimuthal direction and 20 zones for the $\pi$ polar direction, while the reduced 
radial grid has a contant spacing in the log and uses 20 zones between 12 and 120\,km. 
We compute the annihilation rate at locations in a 2D meridional slice of this new volume, 
with 16 uniformly-spaced angles from the pole to the equator, and 40 zones
with a constant spacing in the log between 15 and 120\,km.\footnote{We also perform higher resolution, 
accuracy-check, calculations, but our results are already converged at this lower resolution.} 
To estimate the annihilation rate at $\vec{r}$, one needs to estimate the flux received from all 
cells $\vec{r}_k$.
\cite{ruffert:97} assume that neutrino radiation is isotropic in the half space around the outward 
direction given by the density gradient $\vec{n}_\rho$ at $\vec{k}$. Moreover, they approximate
 each emitting cell volume as a sphere, whose radius at $(r_k,\theta_k,\phi_k)$ is 
\begin{eqnarray}
D_k = \left( \frac{\Delta\phi_k \Delta\mu_k \Delta r_k^3}{4 \pi}   \right)^{1/3}\, ,
\end{eqnarray}
where $\mu_k = \cos \theta_k$. With our location dependent cell volume, we obtain 
\begin{eqnarray}
\sum_k \Delta\Omega_k \cdot I_{\nu,k} \approx \frac{1}{\pi} \sum_k 
                                        \frac{\Delta \mu_k \Delta \phi_k \Delta r_k^3}
                                             {3 d_k^2}Q^{\rm eff}(\nu_i)\,\, , 
\end{eqnarray}
where $d_k = |\vec{r} - \vec{r}_k|$. When we sum over discrete cells, 
the cosine of the angle between $\vec{k}$ and $\vec{k^\prime}$ is given by 
$\cos \Phi_{kk^\prime} = (\vec{r} - \vec{r}_k)(\vec{r}-\vec{r}_{k^\prime}) / 
                        (|\vec{r} - \vec{r}_k||\vec{r}-\vec{r}_{k^\prime}|) 
$.

\begin{figure*}[t]
\centering
\epsscale{0.35}
\plotone{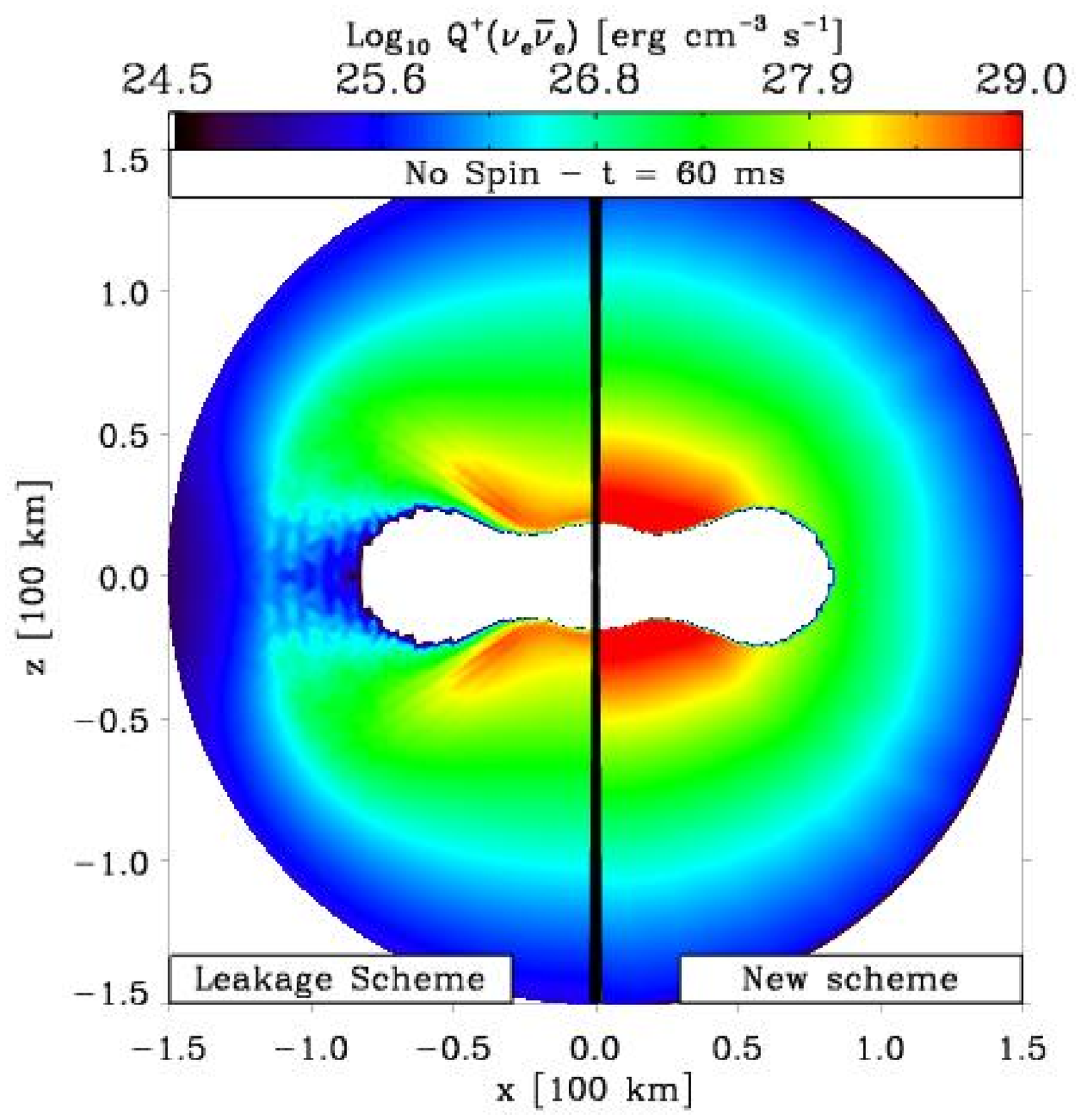}
\plotone{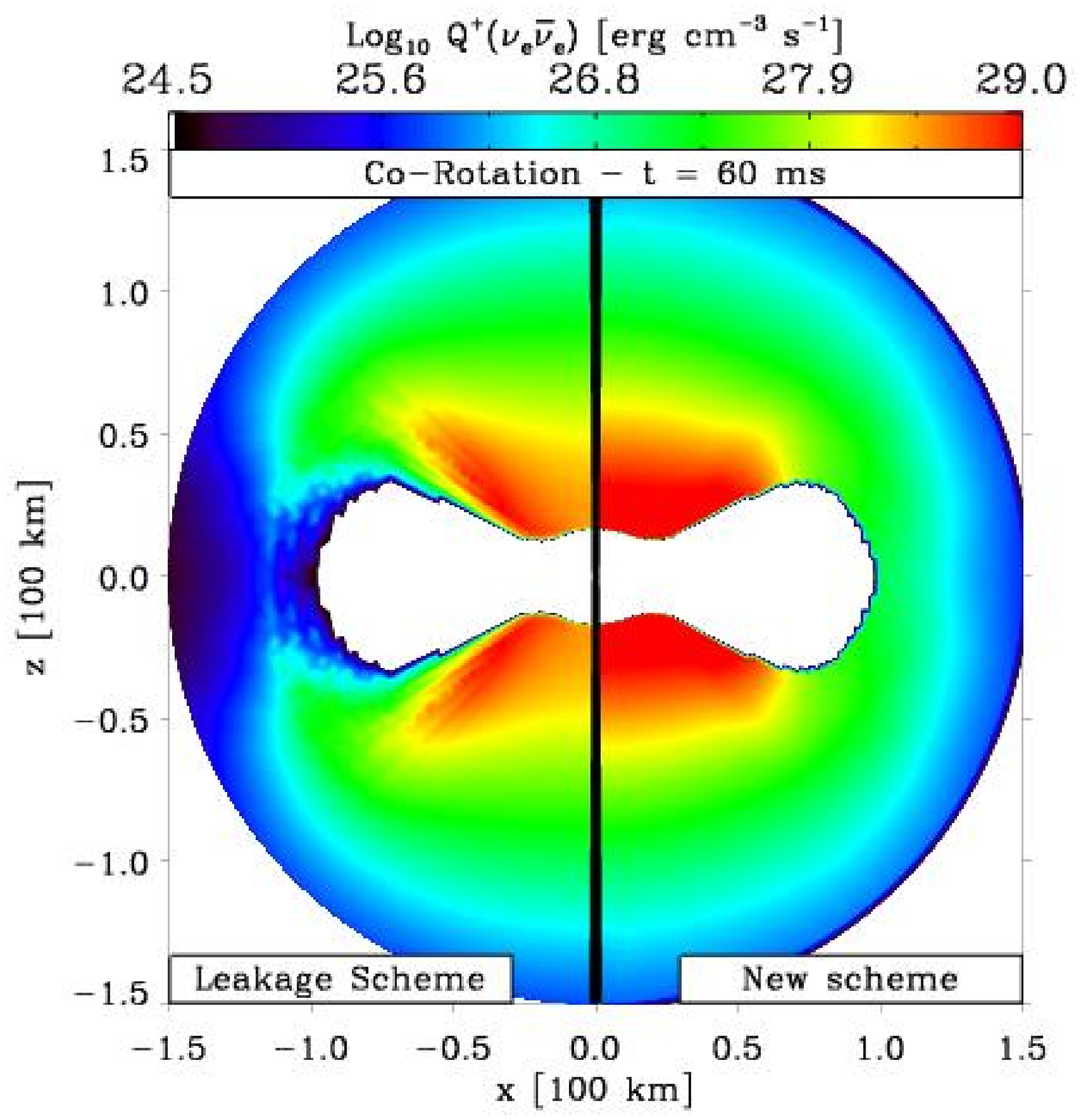}
\plotone{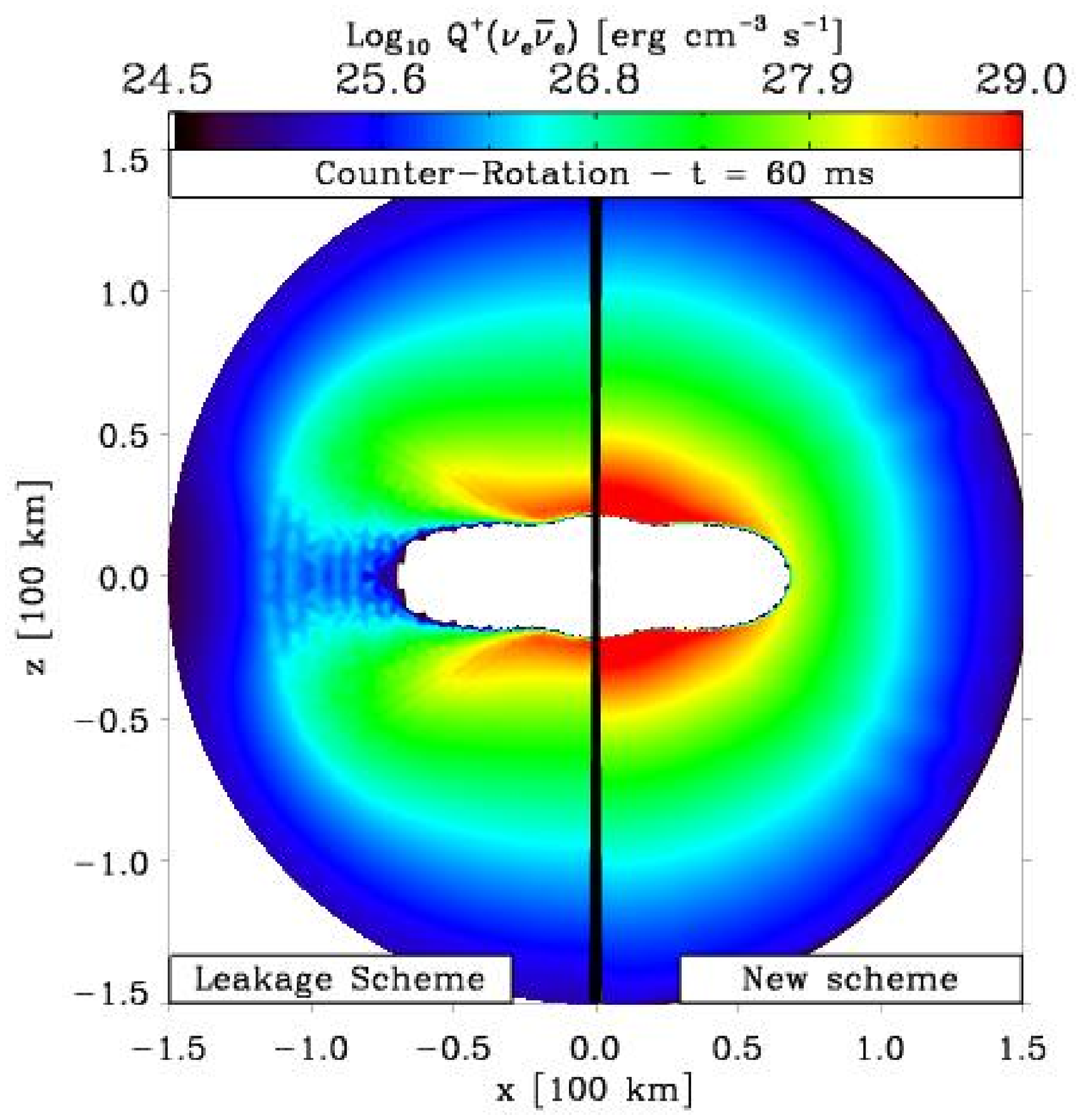}
\caption{Colormaps of the log of the energy deposition $Q^+(\nu_{\rm e}\bar{\nu}_{\rm e})$ 
by electron neutrino pair annihilation 
in the BNS merger models with initially no spin ({\it left}), co-rotating spins 
({\it middle}), and counter-rotating spins ({\it right}), and at 60\,ms after the start of the 
VULCAN/2D simulation. In each case, the evaluation is done with two different methods.
In the left half of each panel, we show the results using the leakage scheme (\citealt{ruffert:96,ruffert:97}; 
see also \S~\ref{annihil:leakage}).
In the right half, we show the corresponding results using
the new formalism presented in \S~\ref{annihil:sn} using 16 $\vartheta$-angles.
We set the minimum of the colorbar to white for all regions with 
densities greater than 10$^{11}$\,g\,cm$^{-3}$.
The maximum of the color bar is set at 10$^{29}$\,erg\,cm$^{-3}$\,s$^{-1}$ to improve the 
visibility. Maximum values differ in fact by large factors between the two methods (in contrast with
the good agreement in the cumulative values). For the model with no initial spins, we obtain maxima
at 1.0$\times$10$^{30}$ ($S_n$), compared with 6.5$\times$10$^{28}$\,erg\,cm$^{-3}$\,s$^{-1}$ (leakage),
and in the same order, we have 5.2$\times$10$^{29}$ and 1.4$\times$10$^{29}$\,erg\,cm$^{-3}$\,s$^{-1}$ 
for the co-rotating model, and 1.4$\times$10$^{30}$ and 2.1$\times$10$^{29}$\,erg\,cm$^{-3}$\,s$^{-1}$
for the counter-rotating model.
We do not show the distribution for the ``$\nu_\mu\bar{\nu}_\mu$'' annihilation, which is qualitatively similar to
that of $\nu_{\rm e}\bar{\nu}_{\rm e}$, being merely weaker everywhere by about an order of magnitude.
With the leakage scheme, the contrast in annihilation rate between these two neutrino types is even greater, primarily
because  ``$\nu_\mu$" neutrino emission occurs at densities in excess of 10$^{12}$\,g\,cm$^{-3}$,
thus, beyond our density cut. 
In the snapshots shown here and with the leakage scheme, we find integral net energy deposition 
rates by neutrino pair annihilation $\dot{E}(\nu_{\rm i}\bar{\nu}_{\rm i})$ (summed over all neutrino species)
of 1.78$\times$10$^{49}$, 4.43$\times$10$^{49}$, 
and 1.50$\times$10$^{49}$\,erg\,s$^{-1}$ for the BNS merger models with initially no
spins, co-rotating spins, and counter-rotating spins, respectively.
In the same order, but with the $S_n$ scheme (and using 16 $\vartheta$-angles), we find
2.13$\times$10$^{49}$, 3.54$\times$10$^{49}$, and 1.44$\times$10$^{49}$\,erg\,s$^{-1}$.
Note that these numbers are at least one to two orders of magnitude smaller than the corresponding rates 
due to charged-current neutrino absorption (see left and right panels of  Fig.~\ref{fig_edep_tot}). 
}
\label{fig:nu_nubar_NS2}
\end{figure*}

  Following \cite{ruffert:97}, we apply selection criteria to determine whether to include a contribution.
Specifically, emission and deposition sites must be in regions with a density lower than 10$^{11}$\,g\,cm$^{-3}$.
Interestingly, the leakage scheme predicts a very small ``effective'' emissivity from high-density regions, due to the very long
diffusion times from those optically-thick regions. Relaxing the density cuts in the
calculations leads only to a 50\% enhancement in volume-integrated annihilation rate
$\dot{E}(\nu_{\rm e}\bar{\nu}_{\rm e})$, but an increase of 
a factor of $\sim$20 for $\dot{E}($``$\nu_{\mu}\bar{\nu}_{\mu}$''). 
As pointed out earlier, the location in very high-density regions of the ``$\nu_{\mu}$'' emitting cells 
means that most of the emitting volume is truncated by the adopted 10$^{11}$\,g\,cm$^{-3}$ density cut,
while results converge when this cut is increased to densities of $\sim$10$^{13}$\,g\,cm$^{-3}$.
Even with the latter, $\dot{E}($``$\nu_{\mu}\bar{\nu}_{\mu}$'') is still three orders
of magnitude smaller than $\dot{E}(\nu_{\rm e}\bar{\nu}_{\rm e})$, likely a result of
the neglect of bremsstrahlung, and the unfavorable $(1-\cos\Phi_{k{k^\prime}})^2$ for this
compact emitting configuration.

  By contrast with the rather large uncertainty introduced through the somewhat arbitrary density cuts,
the annihilation rate calculation is weakly affected by increasing the resolution. In the no-spin BNS
model at 50\,ms, changing the number of radial-angle zones for the deposition sites 
$\{nr_{\rm edep},nt_{\rm edep}\}$ from (40,16) to (60,48)
and for emitting sites $\{nr_{\rm emit},nt_{\rm emit},np_{\rm emit}\}$ from (20,20,40) to (30,30,60)
increases $\dot{E}(\nu_{\rm e}\bar{\nu}_{\rm e})$ from 2.82$\times$10$^{49}$ to
3.14$\times$10$^{49}$\,erg\,s$^{-1}$ (a $\sim$10\% increase) and increases $\dot{E}($``$\nu_{\mu}\bar{\nu}_{\mu}$'')
from 6.23$\times$10$^{44}$ to 7.59$\times$10$^{44}$\,erg\,s$^{-1}$ (a $\sim$20\% increase). 
The same test done for the no-spin BNS model at 60\,ms after the start of the simulations 
yields a $\dot{E}(\nu_{\rm e}\bar{\nu}_{\rm e})$ which is 
identical to within 1\%, while $\dot{E}($``$\nu_{\mu}\bar{\nu}_{\mu}$'') increases by $\sim$10\% 
in the higher resolution model. Hence, higher resolution does not change these values by a significant amount.

  In the left half of each panel in Fig.~\ref{fig:nu_nubar_NS2}, we show the distribution in a 2D meridional slice 
(it is in fact axisymmetric by construction) for the annihilation rate $Q^+(\nu_{\rm e}\bar{\nu}_{\rm e})$ of 
electron-type neutrinos, for the BNS merger models with initially no spins ({\it left}), 
co-rotating spins ({\it middle}), and counter-rotating spins ({\it right}).
Volume-integral values $\dot{E}(\nu_{\rm i}\bar{\nu}_{\rm i})$ as a function of time 
for all three models are shown in Fig.~\ref{fig_edep_tot} 
({\it left panel; dashed line joined by star symbols}) and given in Table~\ref{tab:edep}.
The deposition rate is maximum near the poles, where the large-angle collisions
occur, and close to the peak emission sites (shown in Fig.~\ref{fig_emis_leakage}), 
i.e., near the SMNS surface, with peak values on the order of 
10$^{29}$\,erg\,cm$^{-3}$\,s$^{-1}$ for all three models.
In this formalism, $\dot{E}($``$\nu_{\mu}\bar{\nu}_{\mu}$''$)$ ({\it middle panel} of Fig.~\ref{fig_edep_tot}) 
is at least three orders of magnitude smaller than $\dot{E}(\nu_{\rm e}\bar{\nu}_{\rm e})$. 
The peak volume-integrated energy deposition reaches a few 10$^{50}$\,erg\,s$^{-1}$, typically two orders
of magnitude below that achieved by charge-current reactions, whose associated energy deposition drives
the neutrino-driven wind (\S~\ref{sect:results}; right panel of Fig.~\ref{fig_edep_tot}).

\subsection{Formalism based on multi-angle transport and moments of the specific intensity}
\label{annihil:sn}

\begin{figure}
\plotone{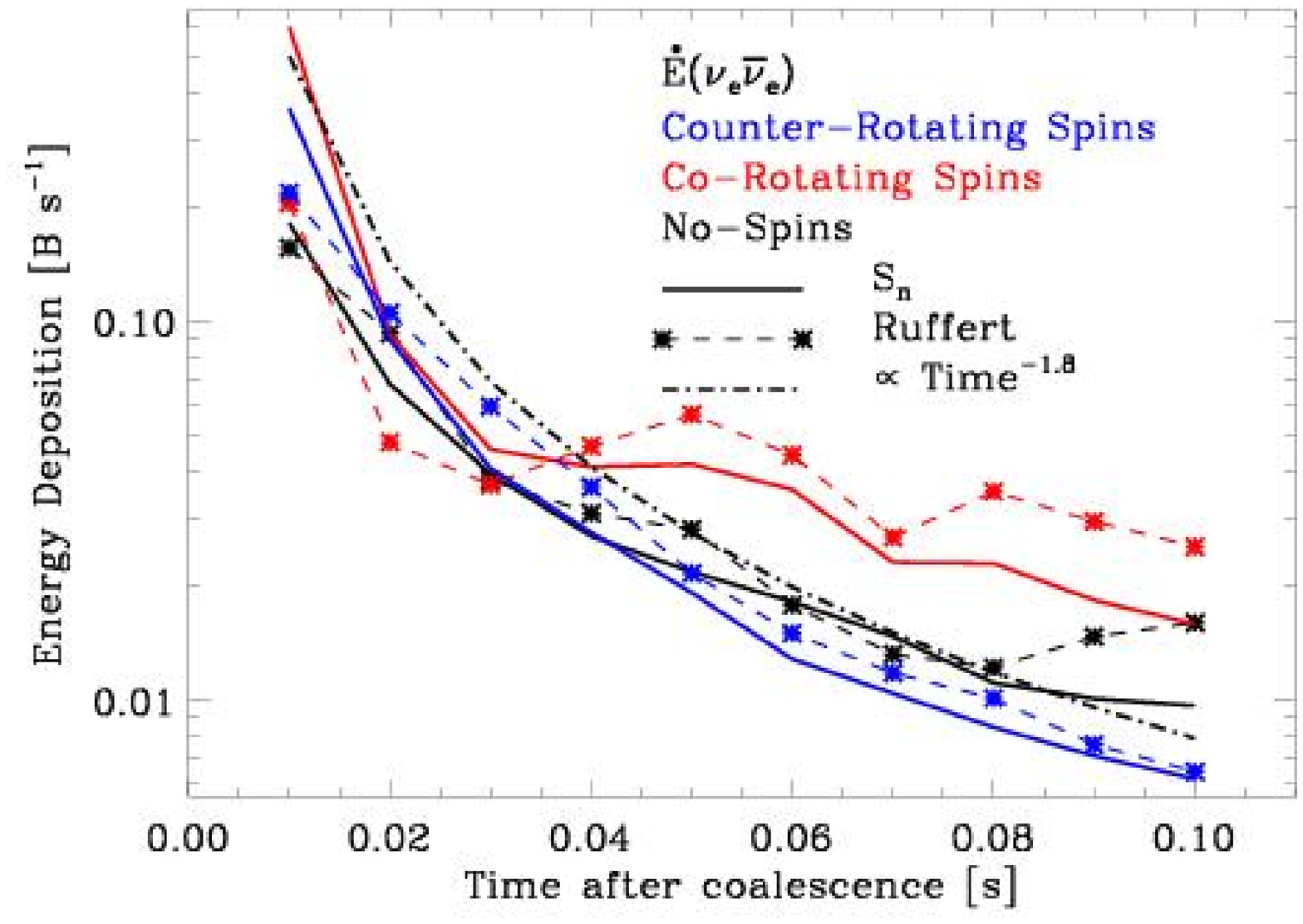}
\plotone{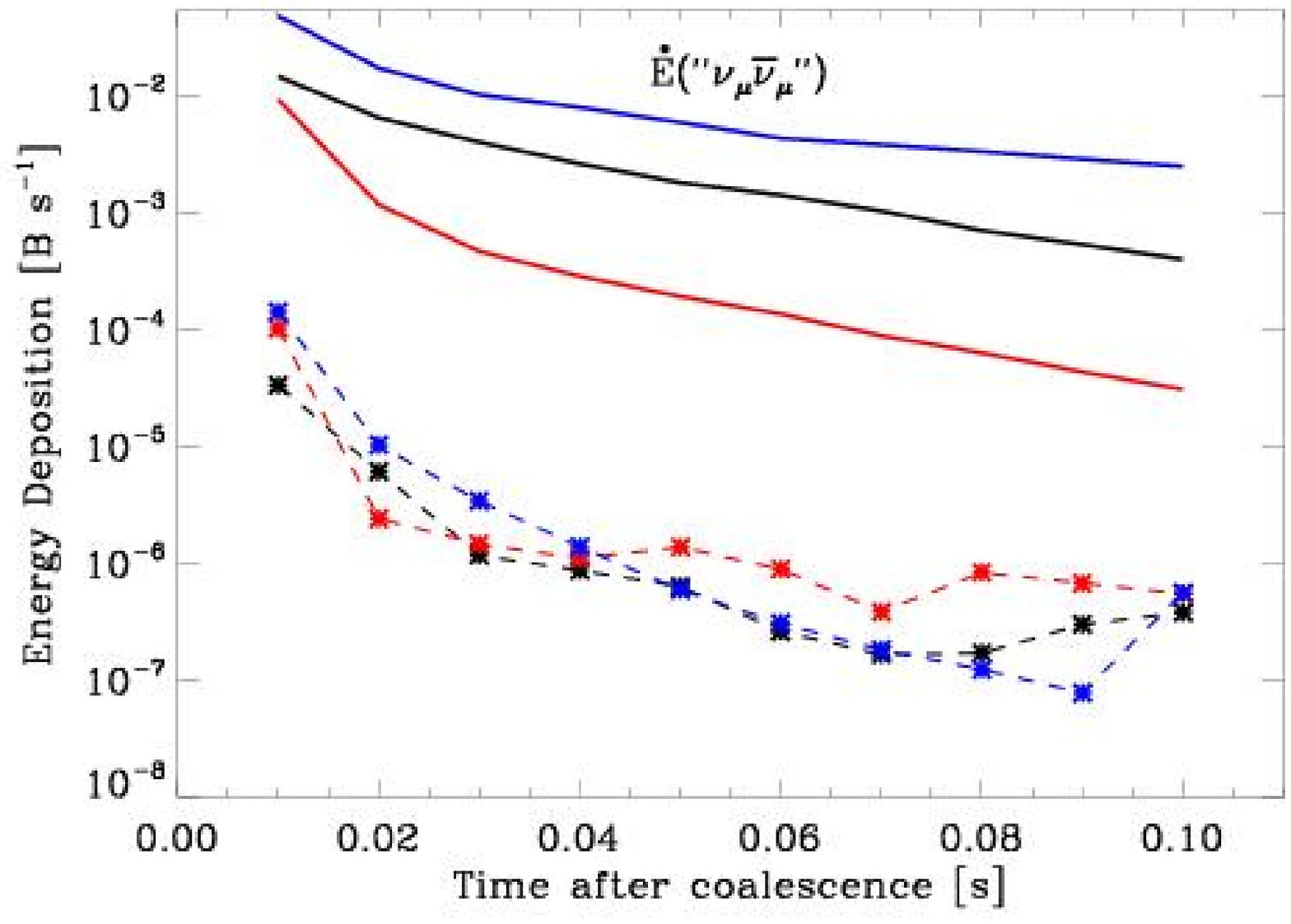}
\plotone{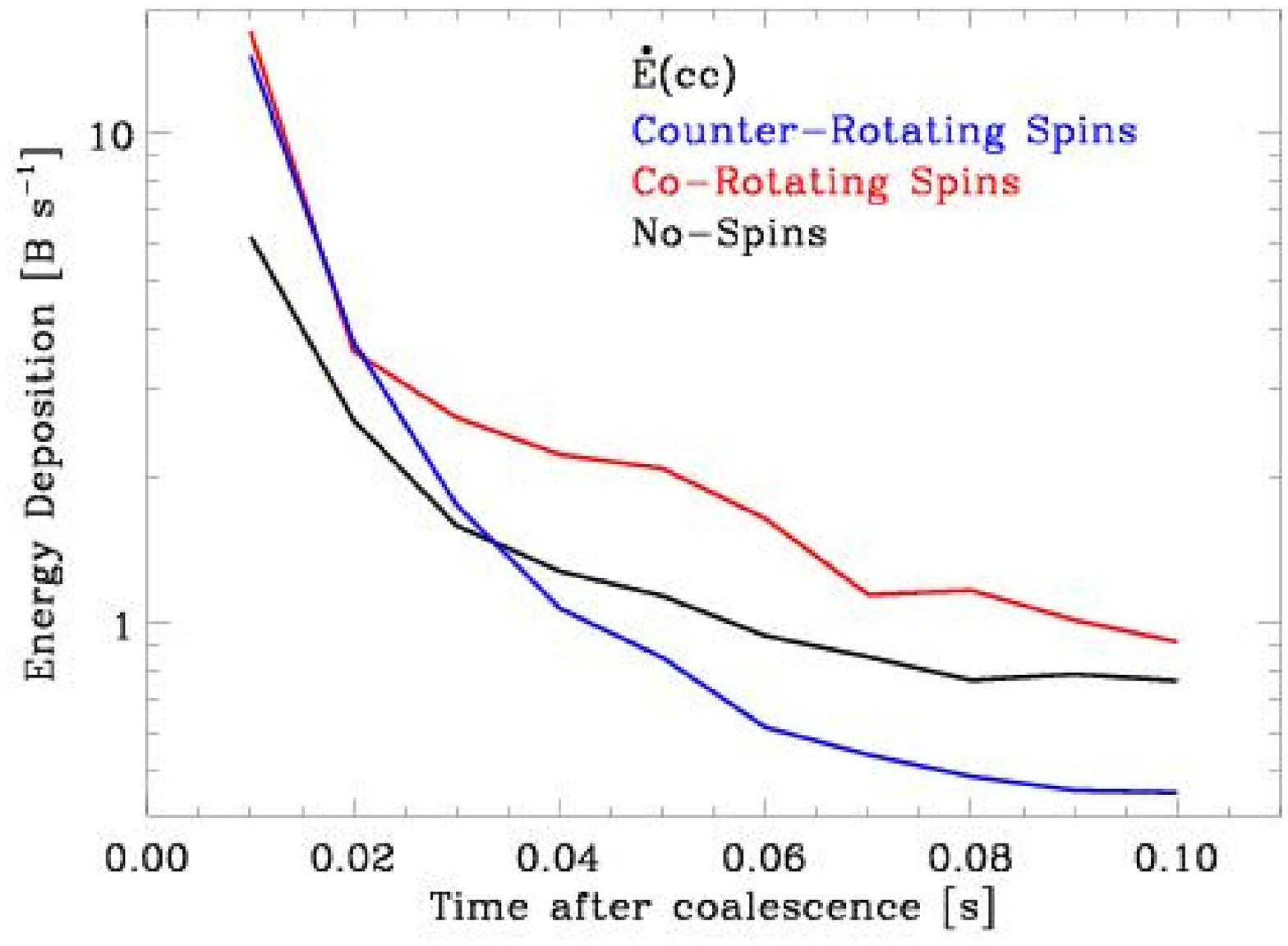}
\caption{
Volume-integrated energy deposition rate $\dot{E}(\nu_{\rm e}\bar{\nu}_{\rm e})$ ({\it top}) 
and the $\dot{E}$(``$\nu_{\mu}\bar{\nu}_{\mu}$'') ({\it middle})
neutrino-antineutrino annihilation processes, as well as that due to charge-current reactions ({\it bottom})  
for the models initially with no spins ({\it black}), co-rotating spins ({\it red}), and counter-rotating
spins ({\it blue}) and computed with the formalism presented in \S~\ref{annihil:sn}.
The dashed lines (symbols highlight the times computed) represent for each model the corresponding result 
based on the leakage scheme (\S~\ref{annihil:leakage}), following the method of \cite{ruffert:96,ruffert:97}.
The computations are performed every 10\,ms, from 10 to 100\,ms after the start of the VULCAN/2D simulations.
Note the fast decrease of the energy deposition rate with time ($\propto t^{-1.8}$; dotted line), 
following the decrease of all neutrino luminosities (see Fig.~\ref{fig_flux_time}) and the contraction 
of the SMNS. Note that \cite{setiawan:04,setiawan:06} obtain a slightly weaker 
dependence ($\propto t^{-1.5}$), but with a black-hole/torus-disk configuration and a treatment 
of the physical viscosity for energy dissipation through shear in the differentially-rotating disk.
Note that the energy unit used is the Bethe, i.e. 10$^{51}$\,erg $\equiv$ 1\,Bethe [1\,B].
}
\label{fig_edep_tot}
\end{figure}

   We now present a formalism for the computation of the $\nu_i\bar{\nu}_i$ annihilation rate 
that directly exploits the neutrino-transport solution computed with 
VULCAN/2D in our BNS mergers, 
rather than using evaluations based on the leakage scheme \citep{ruffert:96,ruffert:97}.

  Our approach is to use the angle-dependent neutrino specific intensity calculated for 
snapshots at 10\,ms intervals for these BNS merger models using the multi-angle, $S_n$, 
scheme (\citealt{livne:04,ott:08}; \S~\ref{sect:vulcan}). 
In the Appendix, and for completeness, we present such annihilation rate calculations, but 
in the context of the post-bounce phase of a 20\,\mo progenitor \citep{ott:08} and of the 
accretion induced collapse (AIC) of a massive and fast rotating white dwarf \citep{dessart:06b}.
The formalism presented here applies to all cases equivalently.

Following \cite{brt:06} and \cite{janka:91}, 
the expression for the local energy deposition rate $Q^+({\nu_i{\bar {\nu}}_i})$ at a position
$\vec{r}$ due to annihilation of $\nu_i{\bar {\nu}}_i$ pairs into
electron-positron pairs is given to leading order by
\begin{eqnarray}
\label{eq:anni}
 Q^+({\nu_i{\bar {\nu}}_i}) &=& \frac{1}{6} \frac{\sigma_0 (C_A^2 +
   C_V^2)_ {\nu_i {\bar {\nu}}_i}}{c(m_e c^2)^2} \int_0^{\infty}
   d\varepsilon_\nu \int_0^{\infty} d\varepsilon_\nu' (\varepsilon_\nu +
   \varepsilon_\nu')\,\,\times \nonumber\\ &&\oint_{4\pi} d\Omega
   \oint_{4\pi} d\Omega' I_{\nu_i} I'_{{\bar {\nu}}_i} (1 -
   \cos\Phi)^2 \,\,,
   \label{eq:nunubar_general}
\end{eqnarray}
where $I_{\nu_i} \equiv I_{\nu_i}(\varepsilon_\nu,\vec{r},\vec{n},t)$ is
the $\nu_i$-neutrino specific intensity at energy $\varepsilon_\nu$, location
$\vec{r}$, along the direction $\vec{n}$, and at time $t$. The primes
denote the anti-particle.  The angle $\Phi$ is the angle between the
directions $\vec{n}$ and $\vec{n}'$ of the neutrino and antineutrino,
i.e., $\cos\Phi = \vec{n}\cdot \vec{n}'$.   

The formulation we present applies equally to all neutrino species, as
long as the values of the weak coupling constants $C_A$ and $C_V$ are
appropriately set.  We have $C_V = 1/2 + 2\sin^2\theta_{\rm W}$ for
the electron types, $C_V = -1/2 + 2\sin^2\theta_{\rm W}$ for the
$\nu_\mu$ and $\nu_\tau$ types, and $C_A = \pm 1/2$, with
$\sin^2\theta_{\rm W}=0.23$.  The other variables have their usual
meanings. Note that this formulation neglects the phase-space blocking
by the final-state electron-positron pair which is relevant only
at high densities and negligible in the semi-optically thin
regime where pair annihilation may contribute to the net heating.

Now, setting the flux factor $\vec{h} = \vec{H}/J$ and the
Eddington-tensor factor ${\textsf k}={\textsf K}/J$ and
by expanding the $\cos\Phi = \vec{n}\cdot\vec{n}'$ term 
with an appropriate choice of the radiation unit vector (\citealt{hb:07}), 
we obtain for the general case in three dimensions
\begin{eqnarray}
  Q^+({\nu_i{\bar {\nu}}_i}) &=& \frac{8\sigma_0\pi^2 (C_A^2 +
     C_V^2)_{\nu_i{\bar {\nu}}_i}}{3c(m_e c^2)^2} \int_0^{\infty}
     d\varepsilon_\nu \int_0^{\infty} d\varepsilon_\nu' (\varepsilon_\nu +
     \varepsilon_\nu')\,\,\times \nonumber\\
     &&J_{\nu_i} J'_{{\bar {\nu}}_i}
     (1-2\vec{h}_{\nu_i}\cdot\vec{h}'_{{\bar {\nu}}_i} + 
     \mathrm{Tr}[{\textsf
     k}_{\nu_i}:{\textsf k}'_{{\bar {\nu}}_i}]) \,\,,
\end{eqnarray}
where the term $\mathrm{Tr}[{\textsf k}_{\nu_i}:{\textsf k}'_{{\bar
{\nu}}_i}]$ is the trace of the matrix product of the two
$J_\nu$-normalized Eddington tensors.

Assuming the special choice of cylindrical coordinates in axisymmetry and neglecting 
the velocity dependence of the radiation field, we obtain
\begin{eqnarray}
  \label{eq:anni_cyl}
  Q^+({\nu_i{\bar {\nu}}_i}) &=& \frac{8\sigma_0\pi^2 (C_A^2 +
  C_V^2)_{\nu_i{\bar {\nu}}_i}}{3c(m_e c^2)^2}
    \int_0^{\infty} d\varepsilon_\nu  \int_0^{\infty} d\varepsilon_\nu' 
    (\varepsilon_\nu + \varepsilon_\nu') \,\, \times \nonumber \\
    &&JJ' \times ( 1 - 2 h_r h'_r - 2 h_z h'_z \nonumber\\
      &&+ {\textsf k}_{rr}{\textsf k}'_{rr}
      + {\textsf k}_{zz}{\textsf k}'_{zz}
      + {\textsf k}_{\phi\phi}{\textsf k}'_{\phi\phi}
    + 2 {\textsf k}_{rz}{\textsf k}'_{rz} ) \,\,,
\end{eqnarray}
where we have suppressed the $\nu_i$ 
and ${\bar {\nu}}_i$ subscripts. Note that
$k_{rz}$ is the only nonzero off-diagonal term in these coordinates.
Using the trace condition on the Eddington tensor,
${\textsf k}_{rr} + {\textsf k}_{zz} + {\textsf k}_{\phi\phi} = 1$,
the above can be recast into 
\begin{eqnarray}
  Q^+({\nu_i{\bar {\nu}}_i})&=& 
  \frac{8\sigma_0\pi^2 (C_A^2 + C_V^2)_{\nu_i{\bar {\nu}}_i}}{3c(m_e c^2)^2}
    \int_0^{\infty} d\varepsilon_\nu \int_0^{\infty} d\varepsilon_\nu'
    (\varepsilon_\nu + \varepsilon_\nu')\,\,\times\nonumber \\
    &&JJ' \times ( 1 - 2 h_r h'_r - 2 h_z
    h'_z + {\textsf k}_{rr}{\textsf k}'_{rr} + {\textsf
    k}_{zz}{\textsf k}'_{zz}\nonumber\\
    &&+ (1-{\textsf k}_{rr}-{\textsf
    k}_{zz})(1-{\textsf k}'_{rr}-{\textsf k}'_{zz}) + 2 {\textsf
    k}_{rz}{\textsf k}'_{rz} ) \,\,.
\end{eqnarray}
For completeness and comparison with Janka (1991), 
we transform to spherical coordinates and reduce
to spherically symmetric problems, which then yields
\begin{eqnarray}
Q^+({\nu_i{\bar {\nu}}_i}) &=& \frac{8\sigma_0\pi^2}{3c(m_e c^2)^2} (C_A^2
 + C_V^2)_{\nu_i{\bar {\nu}}_i} \int_0^{\infty} d\varepsilon_\nu
 \int_0^{\infty} d\varepsilon_\nu' (\varepsilon_\nu 
  + \varepsilon_\nu ')\,\,\times\nonumber\\
 &&JJ' \times ( 1 - 2
 h h' + k k' + \frac{1}{2}(1-k)(1-k') )\,\,,
\label{eq:anni_janka}
\end{eqnarray}
with scalar $h$ and $k$. Equation~(\ref{eq:anni_janka}) corresponds
to the combination of eqs. (1) and (6) of Janka (1991).

  In practice, our approach is to start from snapshots of the (converged) MGFLD radiation-hydrodynamics
simulations of the three BNS mergers presented in \S~\ref{sect:results}. 
Keeping the hydrodynamical variables frozen, the radiation variables are relaxed using the $S_n$ algorithm 
and with 8 $\vartheta$-angles. As an accuracy check, we also compare the $S_n$ results based on simulations 
performed with 16-$\vartheta$ angles. 
When we do this, the converged radiation field for the 8 $\vartheta$-angle solution is remapped into
the 16 $\vartheta$-angle run (corresponding to a total of 144 angles), 
which is then relaxed. These simulations are quite costly, and take about 4 days
with 48 processors. However, by contrast with the approach of Ruffert et al., 
the annihilation rates are then simply obtained from direct (local) integration of the moments 
of the neutrino specific intensity. 
 
  In the right half of each panel in Fig.~\ref{fig:nu_nubar_NS2}, we show the distribution in a 2D meridional slice 
for the annihilation rate $Q^+(\nu_{\rm e}\bar{\nu}_{\rm e})$ of 
electron-type neutrinos, for the BNS merger models with initially no spins ({\it left}), 
co-rotating spins ({\it middle}), and counter-rotating spins ({\it right}),
but now computed from the results of the $S_n$ calculation and the formalism presented above. 
We provide in Fig.~\ref{fig_edep_tot} volume-integral values as a function of time 
for all three models ({\it top and middle panels; solid lines}) and given in Table~\ref{tab:edep}.
By contrast with the leakage scheme results, the deposition is more evenly spread around the SMNS,
is maximum at depth (in the region where the density cut applies), and reaches peak values a factor of 
a few larger. We find that it is not only the large-angle collisions
that favor the deposition, but also the dependence on $J^2$ (and its 1/$R^4$ radial dependence
in optically-thin regions) which considerably weights the regions close to the
radiating SMNS, making the deposition large not only in the polar regions, but also all around the SMNS.
After 40\,ms, and although the associated neutrino luminosities are quite comparable between the three models, 
the (somewhat unrealistic) co-rotating spin BNS model boasts the largest annihilation rate. This stems from the very extended high-density
SMNS configuration, leading to larger neutrinosphere radii and extended regions favoring large-angle collisions. 
This constrasts with the fact that it is the least hot of all three configurations, and is a somewhat weaker emitter,
suggesting that it is not just the neutrino luminosity, but also the spatial configuration of the SMNS 
that sets the magnitude of the annihilation rate. In the new formalism based on the $S_n$ simulations, 
$\dot{E}($``$\nu_{\mu}\bar{\nu}_{\mu}$'') ({\it middle panel} of Fig.~\ref{fig_edep_tot}) 
is now only one order of magnitude smaller than $\dot{E}(\nu_{\rm e}\bar{\nu}_{\rm e})$, even for the same adopted
density cut. By contrast with the leakage scheme, and to a large extent because of the treatment of bremsstrahlung processes,
the emission from ``$\nu_{\mu}$'' neutrinos is associated with more exterior regions of the SMNS, the $\sim$25\,MeV 
``$\nu_{\mu}$'' neutrinos decoupling at $\sim$18 and $\sim$100\,km along the polar and equatorial directions, respectively,
and thus in regions with densities on the order of, or lower than, 10$^{11}$\,g\,cm$^{-3}$ (Fig.~\ref{fig_nu_sphere}).
The emission is, thus, not significantly truncated by the adopted density cut. Moreover, the subtended angle of the 
representative ``$\nu_{\mu}$'' neutrinosphere is correspondingly much bigger, favoring larger-angle collisions.

  This disagreement should not overshadow the good match between the leakage- and the $S_n$-scheme 
predictions for the $\nu_{\rm e}\bar{\nu}_{\rm e}$ annihilation rate, which are typically within 
a few tens of percent, only sometimes in disagreement by a factor of $\sim$3
(see, e.g., the no-spin BNS model at 10\,ms). The $S_n$-scheme results, thus, support the 
long-term decline of the annihilation rate predicted with the leakage scheme, but
with a $\dot{E}($``$\nu_{\mu}\bar{\nu}_{\mu}$'') value that is typically a factor of ten smaller than 
$\dot{E}(\nu_{\rm e}\bar{\nu}_{\rm e})$ at all times.
The results given here using 8 $\vartheta$-angles are already fairly converged.
For the co-rotating spin BNS model at 60\,ms after the start of the simulation,
we obtain the following differences between the 16 and 8 $\vartheta$-angle $S_n$ calculations:
$\dot{E}(\nu_{\rm e}\bar{\nu}_{\rm e})$ changes to 3.54$\times$10$^{49}$ from 
3.59$\times$10$^{49}$\,erg\,s$^{-1}$ (down by 2\%), $\dot{E}($``$\nu_{\mu}\bar{\nu}_{\mu}$'') changes 
to 1.73$\times$10$^{47}$ from 1.34$\times$10$^{47}$\,erg\,s$^{-1}$ (up by 29\%),
and $\dot{E}(\mathrm{cc})$ changes to 1.40$\times$10$^{51}$ from 1.64$\times$10$^{51}$\,erg\,s$^{-1}$ (down by 15\%)

The long time coverage of our simulations, up to $\sgreat$100\,ms, shows that the peak value, which is also coincident
with the peak neutrino luminosity at 5--10\,ms, is not sustained. The conditions at the peak of the annihilation rate 
do not correspond to a steady-state, but instead herald the steady decrease that follows as the BNS
merger radiates, contracts, and cools. In all simulations, $\dot{E}(\nu_{\rm e}\bar{\nu}_{\rm e})$ is
down by a factor 10--100 at 100\,ms compared to its peak value, while for $\dot{E}($``$\nu_{\mu}\bar{\nu}_{\mu}$'') the
decrease is by 2--3 orders of magnitude (this component is in any case sub-dominant).
The $J^2$ dependence, combined with the strong fading of all neutrino emissivities,
is at the origin of the steady decrease of the annihilation rate over the time span considered here.
In their black-hole/torus-disk configuration with $\alpha$-disk viscosity (which causes significant 
shear heating not accounted for here), \cite{setiawan:04,setiawan:06} observe that the annihilation rate 
decreases as $t^{-3/2}$, and, thus, only slightly more gradually than our prediction.
In our work, neutrino energy deposition by charge-current reactions is the dominant means 
to counteract the global cooling effect of neutrino emission.

\begin{figure}
\centering         
\plotone{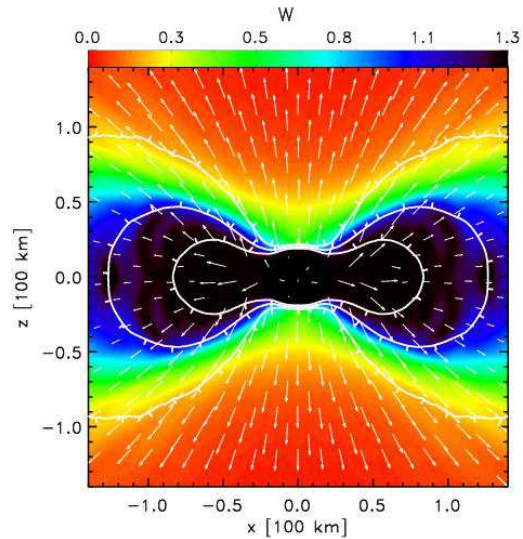}
\caption{Angular factor $W$ in the $\nu_i\bar{\nu_i}$ annihilation
rate (see eqs. \ref{eq:anni_cyl} and \ref{eq:anniW}) shown for
the $\nu_e\bar{\nu}_e$ process at $\varepsilon_\nu$ = 12.02\,MeV
for the BNS model with no initial spin. The snapshot corresponds to a time of 
60\,ms after the start of the VULCAN/2D simulation.
We superpose $\nu_e$ flux vectors at the same $\varepsilon_\nu$
and density contours corresponding to 10$^{9}$, 10$^{10}$, and 
10$^{11}$\,g\,cm$^{-3}$, indicating the local density gradient with perpendicular tick marks.
\label{fig:W}} 
\end{figure}

  We now use the results for the $S_n$ scheme to discuss the assumption
of Ruffert et al. that neutrino emission is quasi-isotropic.
More precisely, they propose that the effective 
emission for every cell is isotropic in the half space around the outward 
direction given by the density gradient $\vec{n}_\rho$.
By contrast, in the $S_n$ scheme, the angular distribution of the neutrino
radiation field is solved for, and we can determine how isotropic this emission is.
For that purpose, we study the spatial variation of the angular term in eq.~(\ref{eq:anni_cyl}),
\begin{eqnarray}
W &=& (1 - 2 h_r h^\prime_r - 2 h_z h^\prime_z +\nonumber\\ &&
k_{rr}k^\prime_{rr} +k_{zz} k^\prime_{zz} + k_{\varphi\varphi}
k^\prime_{\varphi\varphi} +2 k_{rz} k^\prime_{rz})\,\, ,
\label{eq:anniW}
\end{eqnarray}
and plot it in Fig.~\ref{fig:W} for the $\nu_e$--$\bar{\nu}_e$ annihilation rate
at a representative neutrino energy of 12.02~MeV. 
Overplotting the (representative) 12.02\,MeV $\nu_e$ flux vectors, together with 
iso-density contours, one sees that the flux is oriented \emph{everywhere} predominantly in the radial
direction, and peaks along the polar direction. 
The outward direction given by the local density gradient (given by the perpendicular 
to iso-density contours, shown here as short white tick marks), is colinear to the flux vector along
near-polar latitudes, while at mid-latitudes, they are perpendicular to each other.
Hence, the assumption by Ruffert et al. of isotropic neutrino emission in the half space around the outward 
direction given by the local density gradient is not accurate in this instance. Note that it will
likely hold more suitably once the black hole forms and only the torus disk radiates.

In Fig.~\ref{fig:W}, $W$ has a maximum of $\sim$1.33 in the
optically-thick regions of the extended (equatorial) disk-like structure.
There, the flux-factor terms and $k^\prime_{\varphi\varphi} +2 k_{rz}
k^\prime_{rz}$ are nearly zero, while $k_{rr}k^\prime_{rr} + k_{zz}
k^\prime_{zz} + k_{\varphi\varphi}k^\prime_{\varphi\varphi} = 1/3$. In
polar regions, the transition to free streaming happens at small radii
and $W$ decreases from $\sim$1.33 at $\sles$20\,km to $\sim$0.3 at
40~km.  Outside $\sim$60--100\,km, $W \approx 0$, since $h_{zz}
h^\prime_{zz} \approx k_{zz}k^\prime_{zz} \approx 1$ and all other
terms are small.  Considering now the fact that the distribution of
the mean intensity, $J$, irrespective of energy group and
species, is oblate inside the rapidly spinning supermassive neutron star 
and transitions to a prolate shape outside (see Fig.~\ref{fig_netgain}), we can
explain the spatial distribution of the pair annihilation rate
(as shown in Fig.~\ref{fig:nu_nubar_NS2}). The rate peaks inside the
oblate SMNS, since $JJ^\prime$ and $W$ are largest there. Along the
polar axis, $W$ decreases rapidly with radius, while $JJ^\prime$
becomes prolate and remains large out to large radii, compensating in
part for the small $W$.  In equatorial regions, on the other hand,
$JJ^\prime$ drops off more rapidly, but $W$ remains near 1.33 out to a
radius of $\sim$150~km, and only slowly transitions to zero as the
equatorial radiation field gradually becomes forward-peaked. This
systematic behavior sustains a significant $\nu_i\bar{\nu}_i$ annihilation
rate for equatorial radii of up to $\sim$150~km. Note, however, that
net energy deposition by pair annihilation does not occur in the equatorial 
charged-current loss regions.

The findings described in the above paragraph, indicate that
the dominant contribution to $Q^+(\nu_i\bar{\nu}_i)$ is not coming
from large-angle collisions of neutrinos emitted from the
disk, but rather is due to collisions at all angles and close to 
the high-density high-temperature SMNS surface.
This is in distinction to the annihilation rate computed using the leakage
scheme and the assumptions concerning the morphology of the neutrino radiation field
made in \cite{ruffert:97}. However, once the SMNS transitions to a black hole,
the inner contribution will vanish and the radiation will come exclusively 
from the cooling hot torus. We will address in a future paper what annihilation rates we
obtain with this BNS merger configuration.

In the future, while retaining the merits of the $S_n$ method, we will need 
to improve the computation of the annihilation
rates by accounting for GR effects. The compact and
massive configuration of BNS mergers, with $GM/Rc^2$ on the
order of 20\% at 20\,km, suggests that the numbers we present could be modified
with this improvement, although \cite{birkl:07} suggest the magnitude of the 
effect is at most a few tens of percent.

\section{Conclusion}
\label{sect:conclusion}

   We have presented multi-group flux-limited-diffusion radiation hydrodynamics
simulations of binary neutron star mergers, starting from azimuthal-averaged 2D slices 
based on the 3D SPH simulations produced with the MAGMA code
\citep{rosswog:07a}, for neutron star components
with initially no spins, co-rotating spins, and counter-rotating spins.
The main virtues of this work are 1) the solution of the radiation transport 
problem for $\nu_e$, $\bar{\nu}_e$, and ``$\nu_\mu$'' neutrinos for eight
energy groups, coupled to the hydrodynamical evolution of such mergers
over a typical timescale of $\sgreat$100\,ms after the two components come into
contact, and 2) the first quantitative assessment of baryon-pollution
by the neutrino-driven wind produced by such SMNSs.
 Interestingly, our results on neutrino signatures from such merger evens 
confirm the broad adequacy of previous work
that avoided solving the radiation transport problem by designing a neutrino-trapping, 
or leakage, scheme \citep{ruffert:96,ruffert:97,rosswog:03a}.
The only noticeable discrepancy is the much stronger ``$\nu_\mu$'' luminosities we predict
with VULCAN/2D, something we associate with nucleon-nucleon bremsstrahlung not currently 
accounted for in the above leakage schemes.

\begin{figure}
\epsscale{1.2}
\plotone{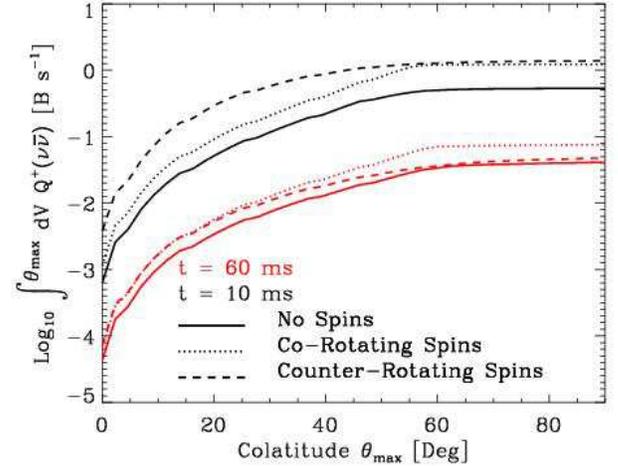}
\caption{
Variation of the log of the cumulative annihilation rate $\int^{\theta_{\rm max}} dV Q^+(\nu\bar{\nu})$ 
with polar angle $\theta_{\rm max}$ in the BNS merger models initially with  
no spin ({\it solid}), co-rotating spins ({\it dotted}), and counter-rotating spins ({\it dashed}), 
and shown here at a time of 10 (black) and 60\,ms (red) after the start of the VULCAN/2D simulation.
In other words, $\theta_{\rm max}$ is the half-opening angle of the cone that defines the 
integration volume for the quantity plotted along the ordinate axis.
We use the $S_n$ results computed with 8 $\vartheta$-angles and accounting only
for a density cut for the sites of emission and deposition. Energy deposited in the cooling
region is not subtracted off, yielding values larger by a factor of two compared with 
the corresponding values given in Table \ref{tab:edep}.
}
\label{fig:edep_vs_theta}
\end{figure}

   At 10\,ms intervals and for each BNS merger model, we select a sequence of VULCAN/2D snapshots 
computed with the MGFLD solver and post-process them with the multi-angle $S_n$ solver,
relaxing the radiation variables, but freezing the fluid variables.
Based on a knowledge of the energy-dependent, species-dependent, and angle-dependent 
neutrino specific intensity $I_\nu(\varepsilon_\nu,\vec{n})$, we compute the 
neutrino-antineutrino annihilation rate using a new formalism
that incorporates various moments of $I_\nu(\varepsilon_\nu,\vec{n})$. 
We find that the total annihilation rate computed with the
$S_n$ solution and our new formalism is larger, but at most by a factor of a few, 
compared to the results based on a combination of a leakage scheme \citep{ruffert:96} 
and paired-cell summation \citep{ruffert:97}.
With density cuts alone, the annihilation rates based on the $S_n$ solution increase 
by a factor of about two.
In our simulations, we find that all neutrino luminosities decrease after peak, 
resulting in a decrease in the annihilation rate with time, i.e., $\propto t^{-1.8}$. 
We find a cumulative total rate that decreases over 100\,ms from a few $\times$10$^{50}$ 
to $\sim$10$^{49}$\,erg\,s$^{-1}$. Adopting an average energy of $8kT\sim$40\,MeV for the 
annihilating $\nu_e\bar{\nu}_e$ \citep{brt:06}, the number of $e^-e^+$ pairs produced
varies over that time span from a few $\times$10$^{54}$ down to $\sim$10$^{53}$\,s$^{-1}$.

We show in Fig.~\ref{fig:edep_vs_theta} the cumulative annihilation rate computed with
the $S_n$ solution and our new formalism at 10 (black) and 60\,ms (red) 
after the start of the VULCAN/2D simulations.
Only $\sim$1\% of the total is deposited along the rotation axis and in a cone with an 
half-opening angle of $\sim$10$^{\circ}$, yielding a rate of $\sim$10$^{49}$\,erg\,s$^{-1}$
at 10\,ms but down to $\sim$10$^{48}$\,erg\,s$^{-1}$ at 60\,ms.
Integrating over a tenth of a second, we obtain a total energy deposition of $\sles$10$^{48}$\,erg.
This result is about an order of magnitude smaller than obtained by 
\cite{setiawan:04,setiawan:06}, although in their approach, neutrino emission is considerably
boosted by shear-heating in the differentially-rotating disk. Their annihilation 
rates are stronger and weaken slightly more gradually with time, i.e., $\propto t^{-1.5}$. 
In the VULCAN/2D simulations presented here, we have no physical viscosity to include
shear heating, nor is there adequate resolution to simulate the magneto-rotational instability. 
Our temperatures and neutrino emissions are therefore underestimates of what would occur in Nature. 
With viscous dissipation as an energy source, we anticipate that our results might 
yield more attractive powers for the generation of a relativistic $e^-e^+$-pair jet, an issue we defer to
a future study.

Baryon loading of relativistic ejecta is a recognized problem in the production of a GRB,
but this is the first time the neutrino-driven wind, the baryon polluter, has been simulated 
dynamically. All three BNS merger simulations we performed reveal the birth of a 
thermally-driven wind through neutrino energy deposition in a gain layer at the surface,
and primarily at high latitudes.
The wind blows mostly along the polar funnel, contained in a cone with an opening angle 
of $\sim$20$^{\circ}$ within $\sles$500\,km, but widening at larger distances to $\sim$90$^{\circ}$.
The electron fraction of the associated ejecta is close to 0.5, along the pole,
but is $\sim$0.1-0.2 along all lower latitudes. Low-$Y_{\rm e}$ material at near-equatorial 
latitudes and large distances has a radial velocity below the local escape speed, 
and since local pressure gradients are not negligible, it is unclear whether 
it will eventually escape. Overall, it appears that if this pre-black-hole phase lasts
for 100 milliseconds, $\sles$10$^{-4}$\,\mo of ``r-process material'' will feed the interstellar medium
through this neutrino-driven wind. 
Moreover, such baryonic pollution along the polar direction may affect any
subsequent relativistic ejecta, although the wind from the SMNS
may last for only a short time, perhaps 100\,ms. Travelling at a tenth the speed of
light, it will reach only out to 0.01 light-second, thus much shorter than the 
duration of short GRBs. This initial wind phase cannot provide the sustained confinement 
needed for the potential relativistic ejecta.
Although we focus on the phase prior to black hole formation, a neutrino-driven wind may blow
after the black hole forms too, but this time from the surrounding hot and dense torus,
and driven by charge-current neutrino absorption at the disk surface. 
Importantly, because of the centrifugal barrier felt by particles
coming in from sizable orbits in the disk, these injected baryons should remain away 
from the polar region, while neutrino-antineutrino annihilation would 
continue to contribute along the rotation axis of the SMNS, but now in an essentially
baryon-free environment. This provides an attractive mechanism for confinement, since 
neutrino emission would yield both the power source for the relativistic ejecta
(caused by neutrino-antineutrino annihilation) and the confined neutrino-driven wind (caused
by charge-current reactions). In the future,
we will investigate the neutrino signatures that obtain in the context where the compact 
object is a black hole, and only the surrounding torus material radiates neutrinos.
Such a disk wind would offer a natural confining ingredient for any relativistic
ejecta propelled along the rotation axis and powered by annihilation of neutrinos and 
antineutrinos radiated from the disk, as proposed by \cite{mochkovitch:93}.

The huge amount of free energy of rotation, on the order of 10$^{52}$\,erg in the supermassive
neutron star and its torus disk, together with millisecond orbital periods, suggest that
the magneto-rotational instability could play a major role in redistributing angular momentum,
leading to solid-body rotation, and increasing the magnetic pressure in the corresponding layers 
by orders of magnitude. Magneto-rotational effects should be strong and would lead to
a considerably stronger neutrino-driven wind, as in the AIC of white dwarfs \citep{dessart:07},
and delay the formation of a black hole. Similarly, \cite{dessart:08} found
that in collapsar candidates \citep{woosley:06,yoon:06,meynet:07}, 
the large amount of free energy of rotation in their iron core 
at the time of collapse can lead to a magnetically-driven explosion, 
associated with very large mass loss rates which may compete with the 
mass accretion rate from the torus disk and jeopardize the formation of 
a black hole. By contrast with such collapsar models,
supermassive neutron stars formed from merger events are unambiguously endowed 
with a large rotational energy budget (stored in the orbital motion), of which a large
fraction is differential and may be tapped. Hence, the phase prior to black hole formation should be modeled
at very high resolution and in combination with neutrino transport to address these issues.
The maximum neutron star mass allowed by the EOS will ultimately determine how much mass accretion
can take place from the torus-disk, and it thus represents a central question for short-duration GRBs. 
Presently, there is a lack of consensus that leaves this issue largely unsettled.

\acknowledgments

We acknowledge helpful discussions with and input from Ivan Hubeny,
Casey Meakin, Jim Lattimer, Stan Woosley, H.-Thomas Janka, 
Bernhard M\"uller, Martin Obergaulinger.
This work was partially supported by the Scientific Discovery through Advanced
Computing (SciDAC) program of the US Department of Energy under grant
numbers DE-FC02-01ER41184 and DE-FC02-06ER41452.  C.D.O. acknowleges
support through a Joint Institute for Nuclear Astrophysics
postdoctoral fellowship, sub-award no.~61-5292UA of NFS award
no.~86-6004791.  The computations were performed at the local Arizona
Beowulf cluster, on the Columbia SGI Altix machine at the Ames center
of the NASA High-End Computing Program, at the National Center for
Supercomputing Applications (NCSA) under Teragrid computer time grant
TG-MCA02N014, at the Center for Computation and Technology at
Louisiana State University, and at the National Energy Research
Scientific Computing Center (NERSC), which is supported by the Office
of Science of the US Department of Energy under contract
DE-AC03-76SF00098. 

\appendix

\section{}
\label{sect:appendix}

In this Appendix, we broaden the discussion on the $\nu_i\bar{\nu}_i$ annihilation process
presented in \S~\ref{sect:annihilate}, by moving away from BNS mergers,
and focusing instead on 100-200\,ms-old PNSs, formed from the collapse 
of their Fe or O/Ne/MG core as it exceeded the Chandrasekhar mass.
Such PNSs are hot because they are young, by contrast with BNS
mergers, whose SMNSs are ``rejuvenated'' through shear heating and shocks associated with the coalescence phase.
Specifically, we investigate the postbounce neutrino heating phase of the 20\,\mo progenitor 
model of \cite{whw:02}, adopting initally no rotation (model s20.nr) 
or differential rotation (model s20.$\pi$; the initial angular velocity at the center 
is $\pi$\,rad\,s$^{-1}$, equivalent to an initial central period of 2\,s).
These models were evolved with the MGFLD and the $S_n$ solvers using the radiation hydrodynamics
code VULCAN/2D, and the results were described in detail in \cite{ott:08}.
To investigate the effect of very fast rotation in a PNS context,
we also investigate the postbounce neutrino heating phase of the 1.92\,\mo AIC model
of \citet{dessart:06b}.
To make a meaningful comparison of the $\nu_i\bar{\nu}_i$ annihilation rate in all three models,
we take their properties at 160\,ms after bounce, thus at times when the neutrino luminosities
are comparable, and, thus, it is the shape of the radiating PNS that differs
between models (for completeness, we also include some quantities at the last computed times - see
Table~\ref{table:nunubar}). In this respect, our choice of models extends from the quasi-spherical s20.nr model, 
to the mildly oblate s20.$\pi$ model, and finally to the strongly oblate (with a disk-like extension)
AIC model. In all three models, the $S_n$ neutrino radiation field was computed with 16 $\vartheta$-angles.
Including the $\varphi$ direction, a total number of 144 angles were treated, following the
prescription described in \cite{ott:08}.

\begin{figure*}[t]
\centering
\epsscale{0.35}
\plotone{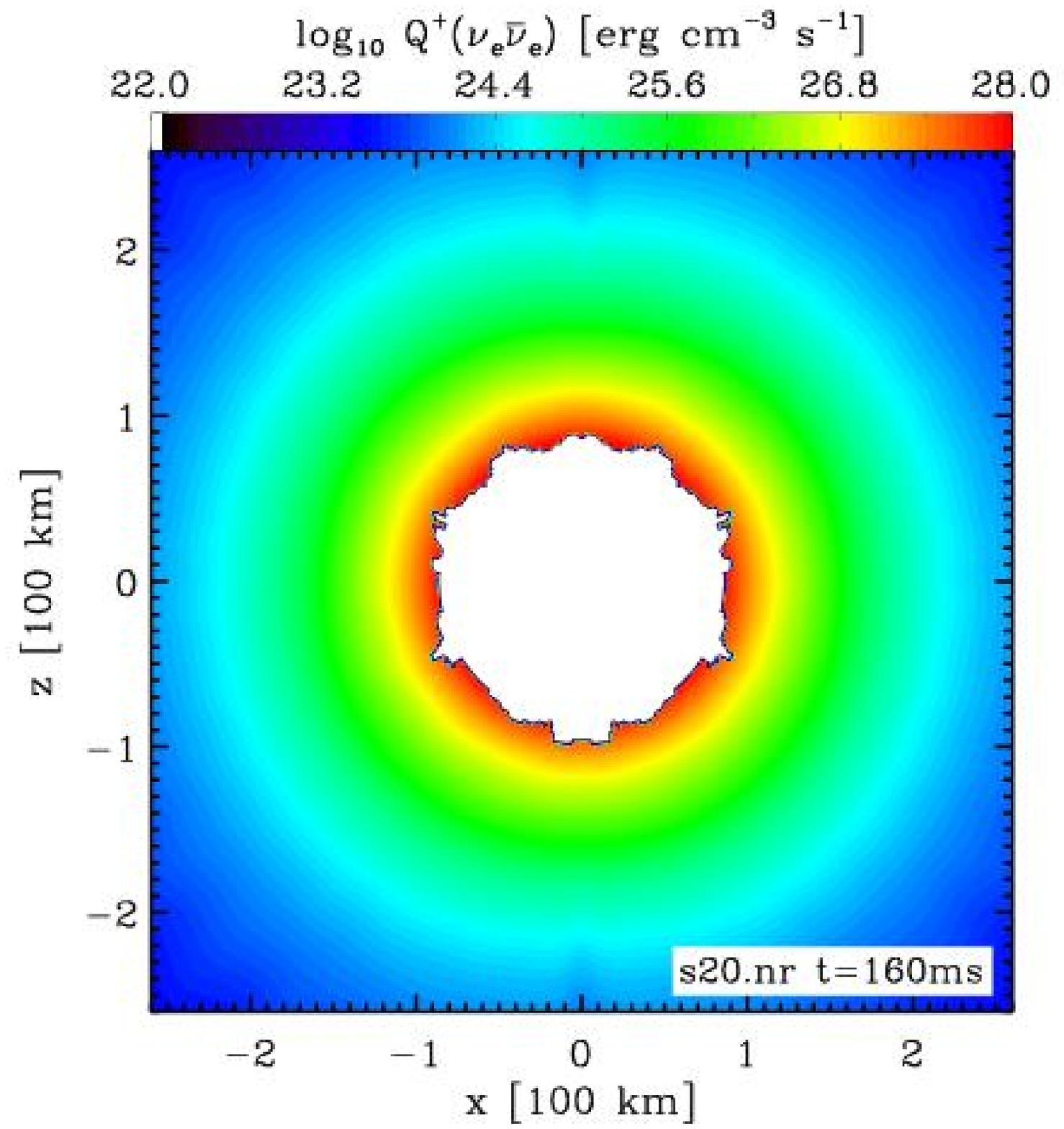}
\plotone{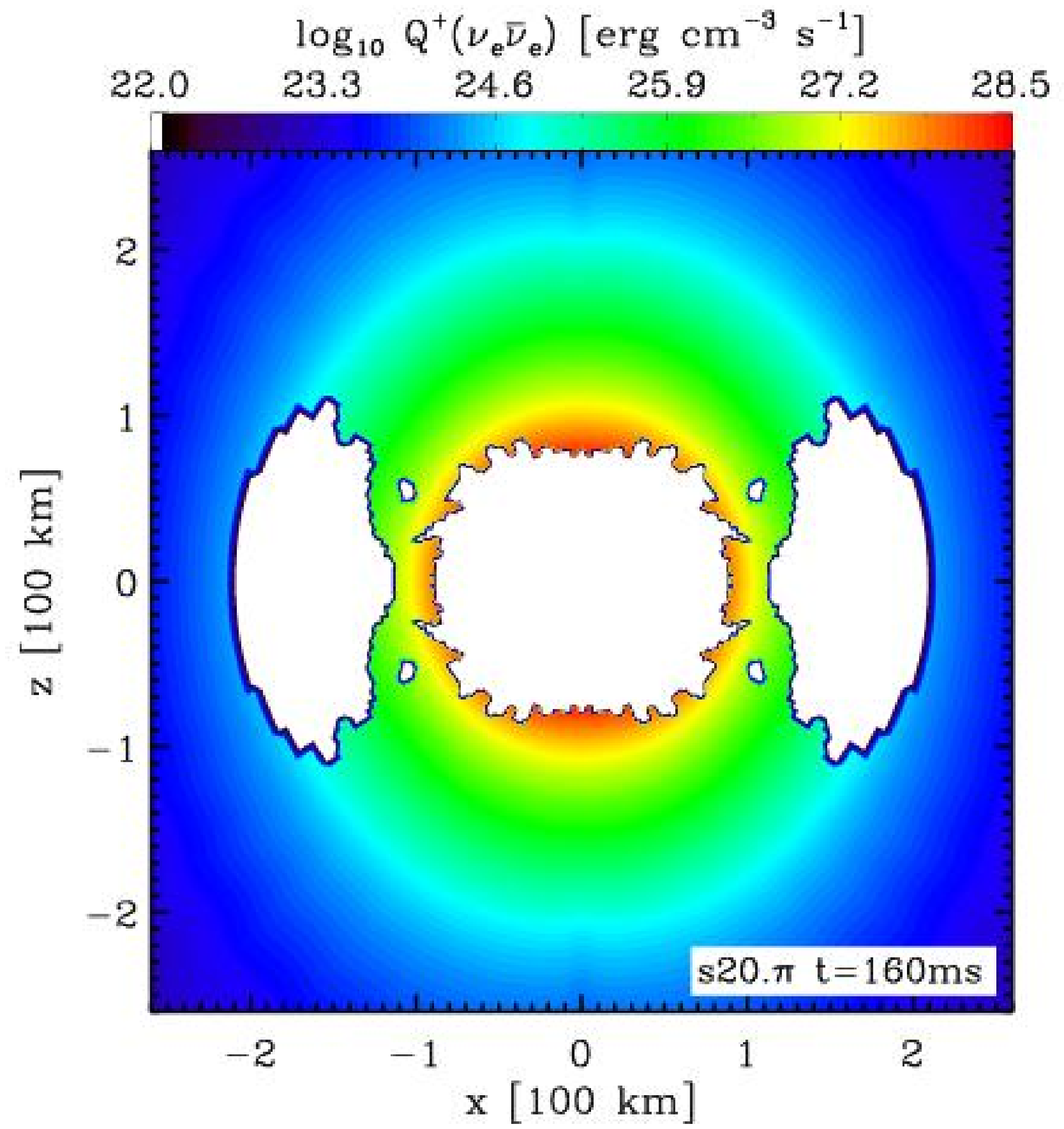}
\plotone{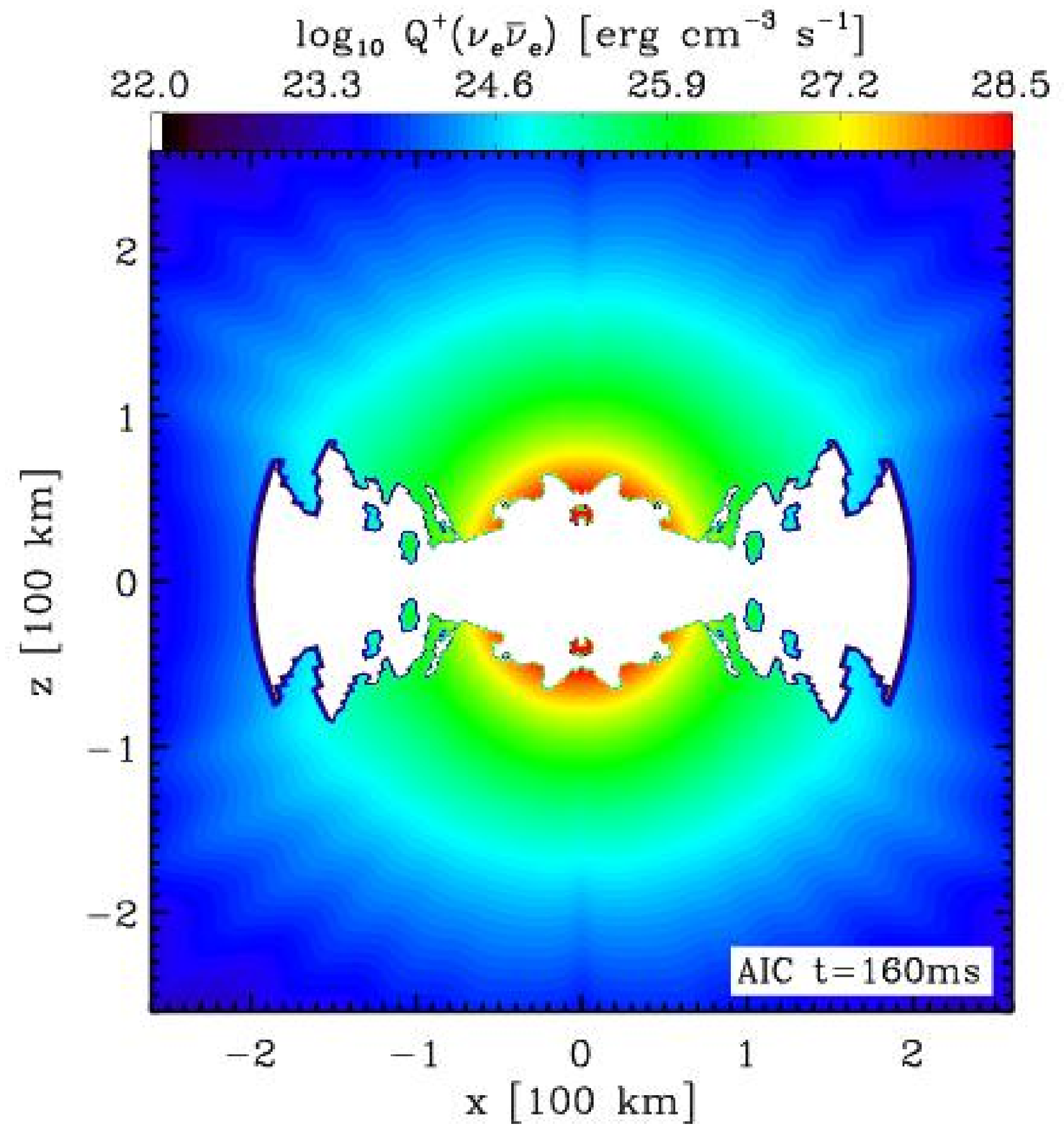}
\caption{
Colormaps of the energy deposition rate $Q^+(\nu_{\rm e}\bar{\nu}_{\rm e})$ by electron neutrino
pair annihilation in the nonrotating model s20.nr (left), the rotating
model s20.$\pi$ (center), and in the 1.92-\mo
accretion-induced collapse (AIC) model of \cite{dessart:06b} 
(right), and in each case at 160\,ms after bounce. 
The energy deposition by the ``$\nu_\mu\bar{\nu}_\mu$''
annihilation process exhibits the same qualitative spatial distribution
and is not shown here (the corresponding cumulative rates are
typically a factor of ten weaker than for $\nu_e\bar{\nu}_e$).
The black lines demarcate regions in which charged-current losses
dominate and regions deep inside the PNS where $\beta$-equilibrium
prevails. Any energy deposition from pair
annihilations are instantly re-emitted in these regions by
charged-current interactions and no net energy deposition
results. In the snapshots shown here, we find integral net
energy deposition rates by neutrino pair annihilation of
1.65$\times$10$^{49}$, 1.44$\times$10$^{49}$
and 0.59$\times$10$^{49}$\,erg\,s$^{-1}$ for models s20.nr, s20.$\pi$, and
the AIC model, respectively. In this computation, we exclude the regions with a density
larger than 10$^{11}$\,g\,cm$^{-3}$, as well as regions of negative net gain 
(shown in white). 
These numbers are one to two orders of magnitude
smaller than the corresponding rates due to charged-current neutrino
absorption. Note that all models are run with 16 $\vartheta$-angles. 
\label{fig:nu_nubar}}
\end{figure*}

In Fig.~\ref{fig:nu_nubar}, we present 2D colormaps of the spatial 
distribution of the energy deposition by $\nu_e\bar{\nu}_e$ annihilation in
the three model snapshots considered here. We draw as white the regions 
of high density ($\rho$~$\sgreat$10$^{11}$\,g\,cm$^{-3}$) 
where equilibration via charged-current
interactions is fast, as well as regions in which cooling by
charged-current interactions dominates. Any energy input by neutrino
pair annihilation in these regions is overwhelmed by cooling through
charged-current processes and does not contribute
to the net neutrino gain. Note also that we do not show the 
distribution for the ``$\nu_\mu\bar{\nu}_\mu$,'' because it is qualitatively similar
to that of $\nu_e\bar{\nu}_e$, being merely weaker, i.e., with a volume-integrated
magnitude that is a factor of about ten smaller.

In the non-rotating model s20.nr, the $Q^+(\nu_e\bar{\nu_e})$
distribution is approximately spherically symmetric, tracing regions
of high neutrino energy density in the neutrino-diffuse postshock
region, and peaking near the lower edge of the gain region around
$\sim$95\,km.  The $\nu_e\bar{\nu}_e$ process dominates over
that for ``$\nu_\mu\bar{\nu}_\mu$,'' because the benefit of the favorable 
``$\nu_\mu$'' neutrino luminosity 
is more than cancelled by the $\sim$5 times smaller ``$\nu_\mu\bar{\nu}_\mu$'' cross sections
and by the smaller decoupling radii for ``$\nu_\mu$'' neutrinos, yielding
systematically smaller collision angles.

In the 160\,ms postbounce snapshot of model s20.nr, $Q^+(\nu_i\bar{\nu}_i)$ 
falls off radially as $\sim r^{-8.3}$ (both for $\nu_e\bar{\nu}_e$ and 
the ``$\nu_\mu\bar{\nu}_\mu$'' annihilations), 
which is slightly steeper than the theoretical estimate $r^{-8}$ for
$r \gg R_\nu$ made by Goodman et al. (1987).  We find and list in
Table~\ref{table:nunubar} volume-integrated energy deposition rates of
1.30$\times$10$^{49}$ and 0.35$\times$10$^{49}$\,erg\,s$^{-1}$ for $\nu_e\bar{\nu}_e$ and
''$\nu_\mu\bar{\nu}_{\mu}$'' annihilation processes, respectively. 
These rates are more than two orders of magnitude smaller than those associated with 
the net gain from charged-current reactions, which amount to 
2.14$\times$10$^{51}$\,erg\,s$^{-1}$ at that time.  
At 500\,ms after bounce in the s20.nr model, the energy deposition rates have declined to
0.77$\times$10$^{49}$ and 0.12$\times$10$^{49}$\,erg\,s$^{-1}$ for $\nu_e\bar{\nu}_e$ and
''$\nu_\mu\bar{\nu}_{\mu}$'' annihilation, respectively. At this
time, we measure a more rapid radial fall-off $\propto r^{-9.5}$ for radii
$\sles$100\,km and somewhat shallower decline $\propto r^{-8}$ for larger radii.
These results echo the steep decrease with time of the annihilation rate found for the 
BNS mergers discussed in this paper, suggesting that neutrino-cooled compact objects are
quickly weakening neutrino-annihilation power sources in the absence of energy sources
such as shear heating in a differentially-rotating disk \citep{setiawan:04,setiawan:06}.

Our results for model s20.nr support the conclusions of previous
studies (Cooperstein et al. 1986,1987; Janka 1991) that 
argued that $\nu_i\bar{\nu_i}$ annihilation contributes little to the
neutrino heating in quasi-spherical postbounce supernova cores.  GR
effects (bending of neutrino geodesics and redshift), which yield
larger annihilation rates for very compact configurations 
\citep{salmonson:99,bhatta:07,birkl:07} are most likely irrelevant
at the large radii at which annihilation may have any significance in
our models.  $GM/Rc^2$ is already as low as $\sim$0.02 at the
inner edge of the polar gain region in models s20.nr and s20.$\pi$
and not larger than $\sim$0.06 in the AIC snapshot.

Model s20.$\pi$ is rapidly rotating and has large pole-equator neutrino flux asymmetries 
\citep{ott:08}. Though  
prolate and quickly varying with radius, the $Q^+(\nu_i\bar{\nu}_i)$
distributions in model s20.$\pi$ do not vary with angle by more than a factor of 2 
(neglecting for now the large equatorial cooling regions between $\sim$100 and 200\,km). 
Along the poles, charged-current losses negate net energy deposition by
$\nu_i\bar{\nu}_i$ annihilation at radii $\sles$80\,km.  Again,
$Q^+(\nu_e\bar{\nu}_e)$ dominates over $Q^+($``$\nu_\mu\bar{\nu}_\mu$''),
while both decrease with radius by $\sim r^{-7.8}$ at 160\,ms after
bounce and $\sim r^{-6.5}$ at 550\,ms after bounce.  As listed in
Table~\ref{table:nunubar}, we find at 160\,ms after bounce
volume-integrated energy depositions of 1.03$\times$10$^{49}$ and
0.41$\times$10$^{49}$\,erg\,s$^{-1}$, for $\nu_e\bar{\nu}_e$ and
``$\nu_\mu\bar{\nu}_\mu$,'' respectively. At 550\,ms after bounce,
these values have decreased to 0.29$\times$10$^{49}$
($\nu_e\bar{\nu}_e$) and 0.12$\times$10$^{49}$~ergs~s$^{-1}$
(``$\nu_\mu\bar{\nu}_\mu$''). The total annihilation contribution
to neutrino heating is equivalent to $\sim$1\% ($\sim$1.5\%) at
160\,ms (550\,ms). 

The right panel in Fig.~\ref{fig:nu_nubar} shows the energy
deposition by $\nu_e\bar{\nu}_e$ annihilation in the AIC model 
at 160\,ms after bounce (see \citealt{dessart:06b} for details).
In this rapidly rotating oblate PNS (the central period is 
$\sim$2.2~ms at this post-bounce time), neutrinos are emitted
from both the central object and its extended moderately hot equatorial 
disk-like structure (more specifically from the pole-facing side of these
side lobes).
Significant energy deposition by neutrino pairs occurs
at small radii in a wide polar wedge of $\sim$60$^\circ$ (widening
with radius to $\sim$120$^\circ$) and $Q^+(\nu_e\bar{\nu}_e)$ reaches
peak rates near the PNS surface at $\sim$15\,km of up to
10$^{29}$\,erg\,cm$^{-3}$~s$^{-1}$, with
$Q^+($``$\nu_\mu\bar{\nu}_\mu$''$)$ being globally smaller by an order of magnitude. 
The radial decline of $Q^+(\nu_i\bar{\nu}_i)$ along the polar direction
becomes shallower with time, with power-law exponents varying from $\sim$-8.3 to $\sim$-5.8 
from 160 to 773\,ms after bounce.

\begin{deluxetable*}{llrrrl}
\tabletypesize{\scriptsize}
\tablecaption{Cumulative $\nu_i\bar{\nu}_i$ Annihilation Rates\label{table:nunubar}}
\tablehead{
\colhead{Model Name}&
\colhead{t-t$_b$}
&\colhead{$\dot{E}(\mathrm{cc})$}
&\colhead{$\dot{E}(\nu_e\bar{\nu}_e)$}
&\colhead{$\dot{E}$(``$\nu_\mu \bar{\nu}_\mu$'')}
&\colhead{Polar}\\
&\colhead{(ms)}
&\colhead{(10$^{49}$ ergs s$^{-1}$)}
&\colhead{(10$^{49}$ ergs s$^{-1}$)}
&\colhead{(10$^{49}$ ergs s$^{-1}$)}
&\colhead{Fall-Off}}
\startdata
s20.nr       & 160 & 214.0 & 1.30 & 0.35 & $\propto \sim\! r^{-8.3}$\\
s20.nr       & 500 & 81.8 & 0.77 & 0.12 & $\propto \sim\! r^{-9.5}$--$r^{-8.0}$\\
\hline
s20.$\pi$    & 160 & 160.2 & 1.03 & 0.41 & $\propto \sim\! r^{-7.8}$\\
s20.$\pi$    & 550 & 26.2  & 0.29 & 0.12 & $\propto \sim\! r^{-6.5}$\\
\hline
AIC 1.92-\mo & 160 & 53.7 & 0.42 & 0.17 & $\propto \sim\! r^{-8.3}$ \\
AIC 1.92-\mo & 773 & 48.9 & 0.19 & 0.02 & $\propto \sim\! r^{-5.8}$ 
\enddata 
\tablecomments{Summary of the cumulative (but instantaneous) 
neutrino-antineutrino annihilation rate calculations.
t-t$_b$ corresponds to the post-bounce time of the computation.
$\dot{E}(\mathrm{cc})$ is the integrated gain from charged-current
interactions, while $\dot{E}(\nu_e\bar{\nu}_e)$ and $\dot{E}($``$\nu_\mu\bar{\nu}_\mu$'') 
denote the integral energy deposition by annihilation of $\nu_e\bar{\nu}_e$ and ``$\nu_\mu
\bar{\nu}_\mu$'' pairs (accounting for $\nu_\mu\bar{\nu}_\mu$ and
$\nu_\tau\bar{\nu}_\tau$), respectively. The rightmost column gives
the approximate power-law exponent that describes the radial fall-off
of the energy deposition rates along the polar direction (we see the same distribution
for both the $\nu_e\bar{\nu}_e$ and ``$\nu_\mu\bar{\nu}_\mu$'' cases).
For a spherically-symmetric model in which 
neutrinos irradiate isotropically from a neutrinosphere, $Q^+(\nu_i\bar{\nu}_i)\propto r^{-8}$ at radii
large compared to the neutrinosphere radius~\citep{goodman:87}. 
The ebbing annihilation rate with time results from its strong 
$JJ^\prime$ dependence (eq.~\ref{eq:anni_cyl}), and reflects PNS cooling.
[See text for discussion.]
}
\end{deluxetable*}

In this AIC model, both $Q^+(\nu_e\bar{\nu}_e)$ and
$Q^+($``$\nu_\mu\bar{\nu}_\mu$'') reach local values at small radii
that are of the same order of magnitude as the peak values in energy
deposition per unit volume by charged-current interactions, yet the
very limited volume of the high-$Q^+(\nu_i\bar{\nu}_i)$ gain regions
leads to only very modest integral values at 160\,ms after bounce of 0.42$\times$10$^{49}$ and
0.17$\times$10$^{49}$\,erg\,s$^{-1}$ for $\nu_e\bar{\nu}_e$ and
``$\nu_\mu\bar{\nu}_\mu$'' pair annihilation, respectively, in our
Newtonian model. At 773\,ms after bounce and in the same order, these values
are 0.19$\times$10$^{49}$ and 0.02$\times$10$^{49}$\,erg\,s$^{-1}$.
Compared with the charged-current interactions driving the
post-explosion wind phase of the AIC (Dessart et al. 2006b), 
even a ten times larger integral $\dot{E}(\nu_i\bar{\nu}_i)$ (e.g., via GR effects)
would still amount to only $\sles$10\% of the total charged-current
energy deposition rate in this model.


\end{document}